\documentclass[12pt]{report}
\usepackage{pstricks}
\usepackage{amsmath,amsthm,latexsym,amssymb,amsfonts,graphicx}

\makeatletter
\@addtoreset{equation}{subsection}
\@addtoreset{section}{part}
\makeatother

\newtheorem{Theorem}{Theorem}[section]
\newtheorem{Definition}{Definition}[section]

\newtheorem{Exercise}{Exercise}[section]
\def\be{\begin{equation}}
\def\ee{\end{equation}}
\def\ba{\begin{eqnarray}}
\def\ea{\end{eqnarray}}
\def\a{{\cal A}}
\def\ab{\overline{\a}}

\def\Nl{{\mathchoice
{\setbox0=\hbox{$\displaystyle\rm N$}\hbox{\hbox to0pt
{\kern0.4\wd0\vrule height0.9\ht0\hss}\box0}}
{\setbox0=\hbox{$\textstyle\rm N$}\hbox{\hbox to0pt
{\kern0.4\wd0\vrule height0.9\ht0\hss}\box0}}
{\setbox0=\hbox{$\scriptstyle\rm N$}\hbox{\hbox to0pt
{\kern0.4\wd0\vrule height0.9\ht0\hss}\box0}}
{\setbox0=\hbox{$\scriptscriptstyle\rm N$}\hbox{\hbox to0pt
{\kern0.4\wd0\vrule height0.9\ht0\hss}\box0}}}}
\def\Zl{{\mathchoice
{\setbox0=\hbox{$\displaystyle\rm Z$}\hbox{\hbox to0pt
{\kern0.4\wd0\vrule height0.9\ht0\hss}\box0}}
{\setbox0=\hbox{$\textstyle\rm Z$}\hbox{\hbox to0pt
{\kern0.4\wd0\vrule height0.9\ht0\hss}\box0}}
{\setbox0=\hbox{$\scriptstyle\rm Z$}\hbox{\hbox to0pt
{\kern0.4\wd0\vrule height0.9\ht0\hss}\box0}}
{\setbox0=\hbox{$\scriptscriptstyle\rm Z$}\hbox{\hbox to0pt
{\kern0.4\wd0\vrule height0.9\ht0\hss}\box0}}}}
\def\Ql{{\mathchoice
{\setbox0=\hbox{$\displaystyle\rm Q$}\hbox{\hbox to0pt
{\kern0.4\wd0\vrule height0.9\ht0\hss}\box0}}
{\setbox0=\hbox{$\textstyle\rm Q$}\hbox{\hbox to0pt
{\kern0.4\wd0\vrule height0.9\ht0\hss}\box0}}
{\setbox0=\hbox{$\scriptstyle\rm Q$}\hbox{\hbox to0pt
{\kern0.4\wd0\vrule height0.9\ht0\hss}\box0}}
{\setbox0=\hbox{$\scriptscriptstyle\rm Q$}\hbox{\hbox to0pt
{\kern0.4\wd0\vrule height0.9\ht0\hss}\box0}}}}
\def\Rl{{\mathchoice
{\setbox0=\hbox{$\displaystyle\rm R$}\hbox{\hbox to0pt
{\kern0.4\wd0\vrule height0.9\ht0\hss}\box0}}
{\setbox0=\hbox{$\textstyle\rm R$}\hbox{\hbox to0pt
{\kern0.4\wd0\vrule height0.9\ht0\hss}\box0}}
{\setbox0=\hbox{$\scriptstyle\rm R$}\hbox{\hbox to0pt
{\kern0.4\wd0\vrule height0.9\ht0\hss}\box0}}
{\setbox0=\hbox{$\scriptscriptstyle\rm R$}\hbox{\hbox to0pt
{\kern0.4\wd0\vrule height0.9\ht0\hss}\box0}}}}
\def\Cl{{\mathchoice
{\setbox0=\hbox{$\displaystyle\rm C$}\hbox{\hbox to0pt
{\kern0.4\wd0\vrule height0.9\ht0\hss}\box0}}
{\setbox0=\hbox{$\textstyle\rm C$}\hbox{\hbox to0pt
{\kern0.4\wd0\vrule height0.9\ht0\hss}\box0}}
{\setbox0=\hbox{$\scriptstyle\rm C$}\hbox{\hbox to0pt
{\kern0.4\wd0\vrule height0.9\ht0\hss}\box0}}
{\setbox0=\hbox{$\scriptscriptstyle\rm C$}\hbox{\hbox to0pt
{\kern0.4\wd0\vrule height0.9\ht0\hss}\box0}}}}
\def\Hl{{\mathchoice
{\setbox0=\hbox{$\displaystyle\rm H$}\hbox{\hbox to0pt
{\kern0.4\wd0\vrule height0.9\ht0\hss}\box0}}
{\setbox0=\hbox{$\textstyle\rm H$}\hbox{\hbox to0pt
{\kern0.4\wd0\vrule height0.9\ht0\hss}\box0}}
{\setbox0=\hbox{$\scriptstyle\rm H$}\hbox{\hbox to0pt
{\kern0.4\wd0\vrule height0.9\ht0\hss}\box0}}
{\setbox0=\hbox{$\scriptscriptstyle\rm H$}\hbox{\hbox to0pt
{\kern0.4\wd0\vrule height0.9\ht0\hss}\box0}}}}
\def\Ol{{\mathchoice
{\setbox0=\hbox{$\displaystyle\rm O$}\hbox{\hbox to0pt
{\kern0.4\wd0\vrule height0.9\ht0\hss}\box0}}
{\setbox0=\hbox{$\textstyle\rm O$}\hbox{\hbox to0pt
{\kern0.4\wd0\vrule height0.9\ht0\hss}\box0}}
{\setbox0=\hbox{$\scriptstyle\rm O$}\hbox{\hbox to0pt
{\kern0.4\wd0\vrule height0.9\ht0\hss}\box0}}
{\setbox0=\hbox{$\scriptscriptstyle\rm O$}\hbox{\hbox to0pt
{\kern0.4\wd0\vrule height0.9\ht0\hss}\box0}}}}

\title{Lectures on Loop Quantum Gravity}
\author{T. Thiemann\thanks{thiemann@aei-potsdam.mpg.de} \\
       MPI f. Gravitationsphysik, Albert-Einstein-Institut, \\
           Am M\"uhlenberg 1, 14476 Golm near Potsdam, Germany}
\date{{\small Preprint AEI-2002-087}}

\begin{document}

\maketitle

\begin{abstract}
Quantum General Relativity (QGR), sometimes called Loop Quantum Gravity,
has matured over the past fifteen years to a mathematically rigorous
candidate quantum field theory of the gravitational field. The features
that distinguish it from other quantum gravity theories are 
1) {\it background independence} and 2) {\it minimality of structures}. 

Background 
independence means that this is a non-perturbative approach in which one 
does not perturb around a given, distinguished, classical background 
metric, rather arbitrary fluctuations are allowed, thus precisely encoding
the quantum version of Einstein's radical perception that {\it gravity is 
geometry}.  

Minimality here means that one explores the logical consequences 
of bringing together the two fundamental principles of modern physics,
namely general covariance and quantum theory, without adding any 
experimentally unverified additional structures such as extra dimensions,
extra symmetries or extra particle content beyond the standard model.
While this is a very conservative approach and thus maybe not very 
attractive to many researchers, it has the advantage that pushing the 
theory to its logical frontiers will undoubtedly either result in a 
successful theory or derive exactly which extra structures are 
required, if necessary. Or put even more radically, it may
show which basic principles of physics have to be given up and 
must be replaced by more fundamental ones. 

QGR therefore is, by definition, not a unified theory of 
all interactions in the standard sense since such a theory would require a 
new symmetry principle. 
However, it unifies all presently known 
interactions in a new sense by quantum mechanically implementing their 
common symmetry group, the four-dimensional diffeomorphism group, which is 
almost completely broken in perturbative approaches. 
\end{abstract}

\newpage

{\small \tableofcontents}

\newpage

~~~~~\\
This is the expanded version of a talk given by the author 
at the 271st WE Heraeus Seminar ``Aspects of Quantum Gravity: From Theory
to Experimental Search'', Bad Honnef, Germany, February 25th -- March 1st 
2002, http://www.uni-duesseldorf.de/QG-2002.

Basically, we summarize the present status of Canonical Quantum General 
Relativity (QGR), also known as ``Loop Quantum Gravity". Our presentation 
tries to be precise and
at the same time technically not too complicated by skipping the proofs 
of all the statements made. These many missing details, which are 
relevant to the serious reader, can be found in the notation used here 
in this overview e.g. in the recent, close to 
exhaustive review \cite{0} and references therein. Of course, in order 
to be useful as a text for first reading we did not include all the 
relevant references here. We apologize for that to the researchers 
in the field but we hope that a close to complete list of their work 
can be found in \cite{0}. Nice reports, treating complementary subjects 
of the field and more general aspects of quantum gravity can be found in 
\cite{0a}.

The text is supplemented by numerous exercises 
of varying degree of difficulty whose purpose is to cut the length of 
the exposition and to arouse interest in further studies. Solving the 
problems is not at all mandatory for an undertsanding of the 
material,
however, the exercises contain further information and thus should be 
looked at even on a first reading. 

On the other hand, if one solves the problems then one should get a fairly
good insight into the techniques that are important in QGR and in 
principle could serve as a preparation for a diploma thesis or a 
dissertation in this field. The problems sometimes involve 
mathematics that may be unfamiliar to students, however, this should not 
scare off but rather encourage the serious reader to learn the necessary 
mathematical background material. Here is a small list of mathematical texts,
from the author's own favourites, geared at theoretical and mathematical 
physicists, that might be helpful: 
\begin{itemize}
\item {\it General}\\
A fairly good encyclopedia is\\ 
Y. Choquet-Bruhat, C. DeWitt-Morette, ``Analysis, Manifolds
and Physics", North Holland, Amsterdam, 1989
\item {\it General Topology} \\
A nice text, adopting almost no prior knowledge is\\
J. R. Munkres, ``Toplogy: A First Course", Prentice Hall Inc., Englewood 
Cliffs (NJ), 1980 
\item {\it Differential and Algebraic Geometry}\\
A modern exposition of this classical material can be found in\\
M. Nakahara, ``Geometry, Topology and Physics",
Institute of Physics Publishing, Bristol, 1998
\item {\it Functional Analysis}\\
The number one, unbeatable and close to complete exposition is \\
M. Reed, B. Simon, ``Methods of Modern Mathematical Physics",
vol. 1 -- 4, Academic Press, New York, 1978\\
especially volumes one and two.
\item {\it Measure Theory}\\
An elementary introduction to measure theory can be found in the beautiful
book\\
W. Rudin, ``Real and Complex Analysis", McGraw-Hill, New York, 1987
\item {\it Operator Algebras}\\
Although we do not really make use of $C^\ast-$algebras in this review,
we hint at the importance of the subject, so let us include\\ 
O. Bratteli, D. W. Robinson, ``Operator Algebras and Quantum Statistical
Mechanics", vol. 1,2, Springer Verlag, Berlin, 1997
\item {\it Harmonic Analysis on Groups}\\
Although a bit old, it still contains a nice collection of the material
around the Peter \& Weyl theorem:\\
N. J. Vilenkin, ``Special Functions and the Theory of Group
Representations", American Mathematical Society, Providence, Rhode
Island, 1968
\item {\it Mathematical General Relativity}\\
The two leading texts on this subject are \\
R. M. Wald, ``General Relativity", The University of Chicago
Press, Chicago, 1989\\
S. Hawking, Ellis, ``The Large Scale Structure of Spacetime",
Cambridge University Press, Cambridge, 1989
\item {\it Mathematical and Physical Foundations of Ordinary QFT}\\
The most popular books on axiomatic, algebraic and constructive quantum
field theory are \\
R. F. Streater, A. S. Wightman, ``PCT, Spin and Statistics, and
all that", Benjamin, New York, 1964\\
R. Haag, ``Local Quantum Physics", 2nd ed., Springer Verlag,
Berlin, 1996\\
J. Glimm, A. Jaffe, ``Quantum Physics", Springer-Verlag, New York, 1987
\end{itemize}
In the first part we motivate the particular approach to a quantum theory
of gravity, called (Canonical) Quantum General Relativity, and develop the 
classical
foundations of the theory as well as the goals of the quantization programme.

In the second part we list the solid results that have been obtained so 
far within QGR. Thus, we will apply step by step the quantization 
programme outlined at the end of section \ref{s1.3} to the classical 
theory that we defined in section \ref{s1.2}. Up to now, these steps have 
been completed approximately until step vii) at least with respect to 
the Gauss -- and the spatial diffeomorphism constraint. The analysis
of the Hamiltonian constraint has also reached level vii) already, however,
its classical limit is presently under little control which is why we 
discuss it in part three where current research topics are listed.

In the third part we discuss a selected number of active research 
areas.
The topics that we will describe already have produced a large number of 
promising results, however, the analysis is in most cases not even close 
to being complete and therefore the results are less robust than those 
that we have obtained in the previous part.

Finally, in the fourth part we summarize and list the most important 
open problems that we faced during the discussion in this report.

\part{Motivation and Introduction}
\label{s1}

\section{Motivation}
\label{s1.1}

\subsection{Why Quantum Gravity in the 21'stCentury ?}
\label{s1.1.1}

Students that plan to get involved in quantum gravity research should 
be aware of the fact that in our days, when financial resources for 
fundamental research are more and more cut and/or more and more absorbed 
by research
that leads to practical apllications on short time scales, one should
have a good justification for why tax payers should support any quantum 
gravity research at all. 
This seems to be difficult at first due to the fact that even 
at CERN's LHC we will be 
able to reach energies of at most $10^4$ GeV which is {\bf fifteen
orders of magnitude below the Planck scale} which is the energy scale 
at which quantum gravity is believed to become important. Therefore
one could argue that quantum gravity research in the 21'st century is 
of purely academic interest only.

To be sure, it is a shame that one has to justify fundamental research at 
all, a situation unheard of in the beginning of the 20'th century which
probably was part of the reason for why so many breakthroughs especially 
in fundamental physics have happened in that time. Fundamental research
can work only in absence of any pressure to produce (mainstream) results,
otherwise new, radical and independent thoughts are no longer produced.   
To see the time scale on which fundamental research leads to practical
results, one has to be aware that
General Relativity (GR) and Quantum Theory (QT) were discovered 
in the 20's and 30's already but it took some 70 years before 
quantum mechanics through, e.g. computers, mobile phones, the internet,
electronic devices or general relativity through e.g. space travel or the 
global 
positioning system (GPS) became an integral part of life of a large 
fraction of 
the human population. Where would we be today if the independent thinkers 
of those times were forced to do practical physics due to lack of 
funding for analyzing their fundamental questions ?

Of course, in the beginning of the 20'th century, one could say that 
physics had come to some sort of crisis, so that there was urgent need 
for some revision of the fundamental concepts: Classical 
Newtonian mechanics, classical
electrodynamics and thermodynamics were so well understood that Max Planck
himself was advised not to study physics but engineering. However, 
although from a practical point of view all seemed well, there were
subtle inconsistencies among these theories if one drove them to their
logical frontiers. We mention only three of them:\\
1) Although the existence 
of atoms was by far not widely accepted at the end of the 19th century 
(even Max Planck denied them), if they existed then there was a serious 
flaw, namely, how should atoms be stable ? Acceralated charges radiate 
Bremsstrahlung according to Maxwell's theory, thus an electron should fall 
into the nucleus after a finite amount of time. \\
2) If Newton's theory of absolute space and time was correct then 
the speed of light should depend on the speed of the inertial
observer. The fact that such velocity dependence was ruled out 
to quadratic order in $v/c$ in the famous Michelson-Morley experiment
was explained by postulating an unknown medium, called ether, with
increasingly (as measurement precision was refined) bizarre properties
in order to conspire to a negative outcome of the interferometer
experiment and to preserve Newton's notion of space and time.\\
3) The precession of mercury around the sun contradicted the ellipses
that were predicted by Newton's theory of gravitation.

Today we easily resolve these problems by 1) quantum mechanics,
2) special relativity and 3) general relativity. Quantum mechanics 
does not allow for continuous radiation but predicts a discrete energy 
spectrum of the atom, special relativity removed the absolute notion of 
space and time and general relativity generalizes the static Minkowski metric
underlying special relativity to a dynamical theory of a metric field 
which revolutionizes our understanding of gravity not as a force but
as geometry. Geometry is curved at each point in a manifold proportional
to the matter density at that point and in turn curvature tells matter
what are the straightest lines (geodesics) along which to move. The 
ether became completely unnecessary by changing the foundation of 
physics and beautifully demonstrates that driving a theory to 
its logical frontiers can make extra structures redundant, what one 
had to change is the basic principles of physics.\footnote{
Notice, that the stability of atoms is still not
satisfactorily understood even today because the full problem also treats 
the radiation field, the nucleus and the electron as quantum objects
which ultimately results in a problem in QED, QFD and QCD for which we 
have no entirely satisfactory description today, see below.} 
 
This historic digression brings us back to the motivation for studying
quantum gravity in the beginning of the 21st century. The question is 
whether fundamental physics also today is in a kind of crisis.
We will argue that indeed
we are in a situation not unsimilar to that of the beginning of the 
20th century:\\
Today we have very successful theories of all interactions.
Gravitation is described by general relativity, matter interactions  
by the standard model of elementary particle physics. As classical 
theories, their dynamics is summarized in the classical Einstein equations. 
However, there are several problems with these theories, some of 
which we list below:
\begin{itemize}
\item[i)] {\it Classical -- Quantum Inconsistency}\\
The fundamental principles collide in the classical Einstein equations
\be \nonumber
\fbox{$\displaystyle
\underbrace{R_{\mu\nu}\;\;-\;\;\frac{1}{2}R g_{\mu\nu}}_{{\blue 
\mbox{{\bf Geometry (GR, gen. covariance)}}}}
\;\;=\;\; 
\underbrace{\kappa\; T_{\mu\nu}(g)}_{{\green 
\mbox{{\bf Matter (Stand.model, QT)}}}}
$} 
\ee
These equations relate matter density in form of the energy momentum
tensor $T_{\mu\nu}$ and geometry in form of the Ricci curvature tensor 
$R_{\mu\nu}$. Notice that the metric tensor $g_{\mu\nu}$ enters also the 
definition of the energy momentum tensor. However,
while the left hand side is described until today only by a classical 
theory,
the right hand side is governed by a quantum field theory (QFT). Since
complex valued functions and operators on a Hilbert space are two 
completely different mathematical objects, the only way to make sense
out of the above equations while keeping the classical and quantum nature 
of 
geometry and matter respectively is to take expectation values of the 
right hand side, that is, 
\be \nonumber 
R_{\mu\nu}-\frac{1}{2}R g_{\mu\nu}\;\;=\;\;\kappa\; 
<\hat{T}_{\mu\nu}(g_0)>,
\;\;\;\;\kappa=8\pi G_{\mbox{Newton}}/c^3
\ee
Here $g_0$ is an arbitrary background metric, 
say the Minkowski metric $\eta=\mbox{diag}(-1,1,1,1)$.
However, even if the state with respect to which the expectation value is 
taken is 
the vacuum state $\psi_{g_0}$ with respect to $g_0$ (the notion of vacuum
depends on the background metric itself, see below), 
the right hand side is generically non-vanishing due to the vacuum 
fluctuations, enforcing $g=g_1\not=g_0$. Hence, 
in order to make this system of equations consistent, one could iterate 
the procedure by computing the vacuum state $\psi_{g_1}$ and reinserting 
$g_1$ into $\hat{T}_{\mu\nu}(.)$, resulting in $g_2\not=g_1$ etc.
hoping that the procedure converges. However, this is generically not the 
case and results in ``run -- away solutions" \cite{1}.

Hence, we are enforced to quantize the metric itself, that is, we need
a quantum theory of gravity resulting in the
\fboxsep1cm
\boldmath
\be \nonumber
\fbox{$^{\prime\prime}
\left. \begin{array}{c}
{\large {\blue \widehat{R}_{\mu\nu} \;\; -\;\; 
\frac{1}{2}\widehat{R}\widehat{g}_{\mu\nu}} \;\;
{\red =} \;\;
{\green \kappa\; \widehat{T}_{\mu\nu}(}{\blue \widehat{g}}{\green )}} \\
\\
\mbox{{\bf\large {\green Quantum}  {\blue - Einstein - }  {\red 
Equations}}}
\end{array} \right. \;\;^{\prime\prime}
$}
\ee
\unboldmath
~~~~~~~\\
The inverted commas in this equation are to indicate that this 
equation is to be made rigorous in a Hilbert space context. QGR is
designed to exactly do that, see section \ref{s3.1}. 
\item[ii)] {\it General Relativity Inconsistencies}\\
It is well-known that classical general relativity is an incomplete theory
because it predicts the existence of so-called spacetime singularities, 
regions 
in spacetime where the curvature or equivalently the matter density 
becomes
infinite \cite{2}. The most prominent singularities of this kind are 
black hole and big bang singularities and such singularities are generic
as shown in the singularity theorems due to Hawking and Penrose. When
a singularity appears it means that the theory has been pushed beyond its 
domain of 
validity, certainly when matter collapses it reaches a state of extreme 
energy density at which quantum effects become important. A quantum theory
of gravity could be able to avoid these singularities in a similar way
as quantum mechanics explains the stability of atoms. We will see that
QGR is able to achieve this, at least in the simplified context of 
``Loop Quantum Cosmology", see section \ref{s3.2}.
\item[iii)] {\it Quantum Field Theory Inconsistencies}\\
As is well-known, QFT is plagued by UV (or short distance) divergences. 
The fundamental 
operators of the theory are actually not operators but rather operator
valued distributions and usually interesting objects of the theory
are (integrals of) polynomials of those evaluated in the same point. 
However, the product of distributions is, by definition, ill-defined.
The 
appearance of these 
divergences is therefore, on the one hand, not surprising, on the other 
hand it indicates again that the theory is incomplete: In a complete 
theory there is no room for infinities. Thus, either the appropriate 
mathematical framework has not been found yet, or they arise because one
neglected the interaction with the gravitational field. In fact, 
in renormalizable theories one can deal with these infinities by 
renormalization, that is, one introduces a short distance cut-off (e.g. by 
point splitting the operator valued distributions) and then redefines 
masses and coupling constants of the theory in a cut-off dependent 
way such that they stay finite as the cut-off is sent to zero. 
This redefinition is done in the framework of perturbation theory 
(Feynman diagrammes) by subtracting counter terms from the original
Lagrangean which are formally infinite and a theory is said to be 
renormalizable if the number of algebraically different such counter 
terms is finite.

The occurance of UV singularities is in deep conflict with general
relativity due to the following reason: In perturbation theory, the 
divergences have their origin in Feynman loop integrals in momentum space
where the inner loop 4-momentum $k=(E,P)$ can become arbitrarily large,
see figure \ref{f1} for an example from QED (mass renormalization). Now 
such virtual (off-shell) particles with energy $E$ and momentum $P$
have a spatial extension of the order of the Compton radius 
$\lambda=\hbar/P$
and a mass of the order of $E/c^2$. Classical general relativity 
predicts that this lump of energy turns into a black hole once  
$\lambda$ reaches the Schwarzschild radius of the order of $r=G E/c^4$. In 
a Lorentz frame where $E\approx Pc$ this occurs at the Planck energy
$E=E_P=\sqrt{\hbar/\kappa}c\approx 10^{19}$GeV or at the Planck length 
Compton radius $\ell_P=\sqrt{\hbar\kappa}\approx 10^{-33}$cm. 
However, when a (virtual) particle turns into a black hole it completely
changes its properties. For instance, if the virtual particle is an 
electron then it is able to interact only 
electroweakly and thus can radiate only particles of the electrowak 
theory. However, once a black hole has formed, also Hawking processes
are possible and now any kind of particles can be emitted, but at a 
different production rate. 
\begin{figure}
\includegraphics[width=10cm,height=7cm]{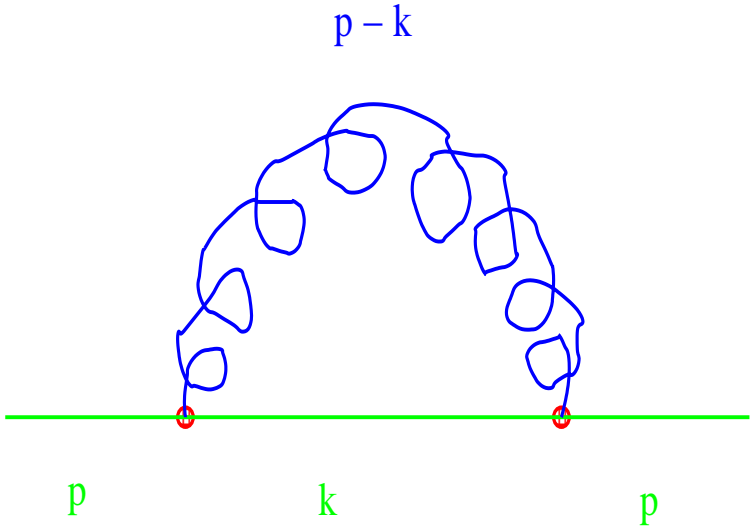}
\caption{One loop correction to the electron propagator in QED}
\label{f1}
\end{figure}
Of course, this is again an energy regime at 
which quantum gravity must be important and these qualitative pictures 
must be fundamnetally wrong, however, they show that there is a problem
with integrating virtual loops into the UV regime.
In fact, these qualitative thoughts suggest that gravity could serve 
as a {\it natural cut-off} because a black hole of Planck mass size
$\ell_P$ should decay within a Planck time unit $t_P=\ell_P/c\approx
10^{-43}s$ so that one has to integrate $P$ only until $E_P/c$. Moreover,
it indicates that spacetime geometry itself acquires possibly a {\it 
discrete structure} since arguments of this kind make it plausible that it 
is impossible to resolve 
spacetime distances smaller than $\ell_P$, basically because the 
spacetime behind an event horizon is in some sense ``invisible". These 
are, of course, only 
hopes and must be demonstrated within a concrete theory. We will see 
that QGR is able to precisely do that and its fundamental discreteness 
is in particular responsible for why the Bekenstein Hawking entropy
of black holes is finite, see sections \ref{s2.2.2}, \ref{s3.1.1} and 
\ref{s3.4}. 
\end{itemize}
So we see that there is indeed a fundamental inconsisteny within the 
current description of fundamental physics comparable to the time 
before the discovery of GR and QT and its resolution, Quantum
Gravity, will revolutionize not only our understanding of nature but will
also drive new kinds of technology that we do not even dare to dream 
of today.

\subsection{The Role of Background Independence}
\label{s1.1.2}

\begin{figure}
\includegraphics[width=10cm,height=7cm]{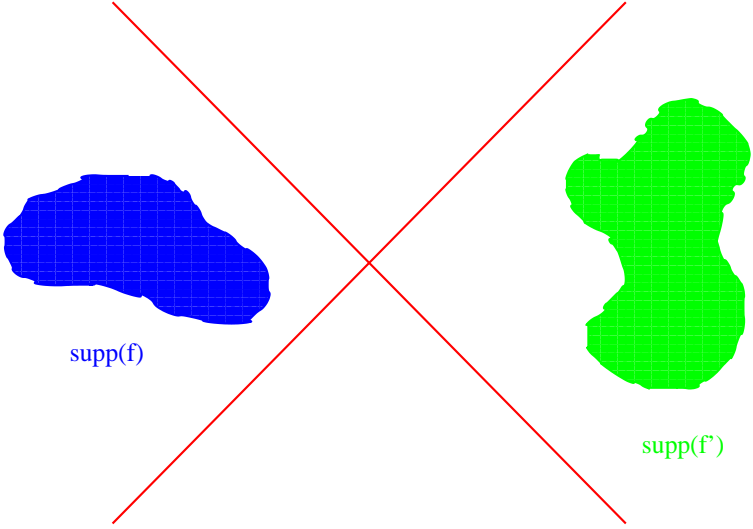}
\caption{Spacelike separated regions in Minkowski space}
\label{f2}
\end{figure}

Given the fact that both QT and GR were discovered already more than 70
years ago and that people have certainly thought about quantizing GR
since then and that matter interactions are more or less succesfully
described by ordinary quantum field theories (QFT), it is somewhat 
surprising that 
we do
not yet have a quantum gravity theory. Why is it so much harder to
combine gravity with the principles of quantum mechanics than for the 
other interactions ? The short answer is that \\
\\
{\bf Ordinary QFT only incorporates Special Relativity.}\\ 
\\
To see why, we just have to remember that ordinary QFT has an axiomatic
definition, here for a scalar field for simplicity:\\
\\
{\bf\blue\large WIGHTMAN AXIOMS} (Scalar Fields on Minkowsk Space)
\begin{itemize}
\item[{\green W1}] ~{\red\bf Poincar\'e Group ${\cal P}$}:\\
$\exists$ continuous, unitary representation $\hat{U}$ of 
{\red\bf ${\cal P}$} on a Hilbert space $\cal H$.
\item[{\green W2}] Forward Lightcone {\bf\red Spektral Condition}:\\
For the generators {\bf\red$\hat{P}^\mu$} of the translation subgroup
of {\bf\red$\cal P$} holds
{\bf\red $\eta_{\mu\nu}\hat{P}^\mu\hat{P}^\nu\le 0,\;\;\hat{P}^0\ge 0$}.
\item[{\green W3}] Existence and  Uniqueness of a {\bf\red 
${\cal P}-$inariant Vacuum $\Omega$}:\\
$\hat{U}(p)\Omega=\Omega\;\;\forall p\in{\cal P}$.
\item[{\green W4}] ~{\bf\red ${\cal P}-$Covariance}:\\
$$
\hat{\phi}(f):=\int d^{D+1}x f(x)\hat{\phi}(x),\;\;f\in 
{\cal S}(R^{D+1})$$
$$
\hat{\phi}(f_1)..\hat{\phi}(f_n){\red \Omega} \mbox{ dense in } {\cal H}
\mbox{ and } \hat{U}(p)\hat{\phi}(f)\hat{U}(p)^{-1}=\hat{\phi}(f\circ p)
$$
\item[{\green W5}] ~{\bf\red Locality (Causality):}\\
If supp$(f)$, supp$(f')$ {\bf\red spacelike separated} (see figure
\ref{f2}), then 
$[\hat{\phi}(f),\hat{\phi}(f')]=0$.
\end{itemize}
It is obvious that due to the presence of the Minkowski backgrounde metric
$\eta$ we have available a large amount of structure which forms the 
fundament on which ordinary QFT is built. Roughly, we have the following 
scheme:
\be
\left. \begin{array}{ccccc}
\eta_{\mu\nu} & \Rightarrow & {\cal P} & \Rightarrow & \hat{H}=\hat{P}^0\\
\mbox{b.-metric} & & \mbox{symm.-group} & & \mbox{Ham.-operator} \\
& & & &\\
\Downarrow & & & & \Downarrow\\
& & & &\\
(x-y)^2=0 & \Rightarrow & \mbox{commutation} &  & 
\Omega\\
\mbox{lightcone} & & \mbox{relations} & & \mbox{ground state} 
\end{array} \right. \nonumber
\ee
Notice that a generic background metric has no symmetry group at all
so that it is not straightforward to generalize these axioms to QFT
on general curved backgrounds, however, since any metric is locally 
diffeomorphic to the Minkowski metric, a local generalization is 
possible and results in the so-called microlocal analysis in which
the role of vacuum states is played by Hadamard states, see e.g. 
\cite{3}.\\
\\
{\bf\large The fundamental, radically new feature of Einstein's theory is 
that there is no background metric at all: The theory is background 
independent. The lightcones themselves 
are fluctuating, causality and locality become empty notions. The dome of 
ordinary QFT completely collapses.}\\
\\
Of course, there must be a regime in any quantum gravity theory where
the quantum fluctuations of the metric operator are so tiny that 
we recover the well established theory of free ordinary quantum fields
on a given background metric, however, the large fluctuations of the
metric operator can no longer be ignored in extreme astrophysical
or cosmological situations, such as near a black 
hole or big bang singularity.

People have tried to rescue the framework of ordinary QFT by splitting
the metric into a background piece and a fluctuation piece
\be \label{1.1.1}
\left. \begin{array}{ccccc}
g_{\mu\nu} & = & \eta_{\mu\nu} & + & h_{\mu\nu} \\
& & & &\\
\uparrow & & \uparrow & & \uparrow \\
\mbox{full metric} & & \mbox{background (Minkowski)} & & 
\mbox{perturbation (graviton)}
\end{array} \right. 
\ee
which results in a Lagrangean for the graviton field $h_{\mu\nu}$
and could in principle be the definition of a graviton QFT on 
Minkowski space. However, there are serious drawbacks:
\begin{itemize}
\item[i)] {\it Non-Renormalizability}\\
The resulting theory is perturbatively non-renormalizable 
\cite{4} as could have 
been expected from the fact that the coupling constant of the theory,
the Planck area $\ell_P^2$, has negative mass dimension (in Planck units).
Even the supersymmetric extension of the theory, in any possible dimension 
has this bad feature \cite{5}. It could be that the theory is 
non-perturbatively renormalizable, meaning that it has a non-Gaussian 
fix point in the language of Wilson, a possibility that has recently
regained interest \cite{6}.
\item[ii)] {\it Violation of Background Independence}\\
The split of the metric performed above again distinguishes the Minkowski
metric among all others and reintroduces therefore a background 
dependence. This violates the {\bf key feature} of Einstein's theory
and thus somehow does not sound correct, we better keep background 
independence if we want to understand how quantum mechanics can possibly
work together with general covariance.
\item[iii)] {\it Violation of Diffeomorphism Covariance}\\
The split of the metric performed above is certainly not diffeomorphism
covariant, it breaks the diffeomorphism group down to Poincar\'e group.
Violation of fundamental, local gauge symmetries is usually considered 
as a bad feature in Yang-Mills theories on which all the other 
interactions are based, thus already from this point of view perturbation
theory looks dangerous. As a side remark we see that background dependence
and violation of general covariance are synonymous.
\item[iv)] {\it Gravitons and Geometry}\\
Somehow the whole idea of the gravitational interaction as a result 
of graviton exchange on a background metric contradicts Einstein's 
original and fundamental
idea that gravity is geometry and not a force in the usual sense.
Therefore such a perturbative description of the theory is very unnatural
from the outset and can have at most a semi-classical meaning when 
the metric fluctuations are very tiny.
\item[v)] {\it Gravitons and Dynamics}\\
All that classical general relativity is about is how a metric evolves in 
time in an interplay with the matter present. It is clear that an 
initially (almost) Minkowskian metric can evolve to something that is far 
from Minkowskian at other times, an example being cosmological big bang
situations or the collapse of initially dilluted matter (evolved 
backwards). In 
such situations the assumption being made in (\ref{1.1.1}), namely that
$h$ is ``small" as compared to $\eta$ is just not dynamically stable.
In some sense it is like trying to use Cartesian coordinates for a sphere
which can work at most locally. 
\end{itemize}
All these points just naturally ask for a non-perturbative approach to
quantum gravity. This, in turn, could also cure another unpleasant feature 
about ordinary QFT: Today we do not have a single example of a rigorously
defined interacting ordinary QFT in four dimensions, in other words, the 
renormalizable theories that we have are only defined order by order in 
perturbation theory but the perturbation series diverges. A 
non-perturbative definition, to which we seem to be forced when coupling 
gravity 
anyway, might change this unsatisfactory situation. \\ 
\\ 
It should be noted here that there is in fact a 
consistent perturbative description of a candidate quantum gravity theory, 
called string
theory (or M -- Theory nowadays) \cite{7}.\footnote{String theory
is an ordinary QFT but not in the usual sense: It is an ordinary
scalar QFT on a 2d Minkowski space, however, the scalar fields themselves
are coordinates of the ambient target Minkowski space which in this case 
is 10 dimensional. Thus, it is similar to a first quantized theory
of point particles. The theory is renormalizable and presumably 
even finite order by order in 
perturbation theory but the perturbation series does not converge.} 
However, in order to achieve 
this celebrated rather non-trivial result, expectedly one must introduce 
extra 
structure: The theory lives in 10 (or 11) rather than 4 dimensions, it is 
necessarily supersymmetric and it has an infinite number of extra 
particles besides those that are needed to make the theory compatible 
with the standard model. Moreover, at least as presently understood, 
again the fundamental new 
ingredient of Einstein's theory, background independence, is 
violated in string theory. This current background dependence of string 
theory is supposed to be overcome once M -- Theory has been rigorously 
defined.

At present only string theory
has a chance to explain the matter content of our universe. The 
unification of symmetries is a strong guiding principle in physics 
as well and has been pushed also by Einstein in his programme of 
geometrization of physics attemting to unify electromagnetism and gravity
in a five-dimensional Kaluza -- Klein theory. 
The unification of the electromagnetic and the weak force in the 
electroweak theory is a prime example for the success of such ideas.
However, unification of forces is an additional principle completely 
independent of background independence and is not necessarily 
what a {\it quantum theory} of gravity must achieve: Unification
of forces can be analyzed at the purely 
{\it classical level}\footnote{In fact, e.g. the unified electroweak 
$SU(2)_L\times U(1)$
theory with its massless gauge bosons can be perfectly described by a 
classical Lagrangean. The symmetry broken, massive $U(1)$ theory can be 
derived 
from it, also classically, by introducing a constant background Higgs field
(Higgs mechanism) and expanding the symmetric Lagrangean around it. 
It is true that the search for a massless, symmetric theory was inspired 
by the fact that a theory with massive gauge bosons is not renormalizable
(so the motivation comes from quantum theory) and, given the 
non-renormalizability of general relativity, many take this as an 
indication that one must unify gravity with matter, one incarnation of 
which is 
string theory. However, the argument obviously fails should it be 
possible to quantize gravity {\it non-perturbatively}.}.
Thus, the only question is whether the theory can be quantized before
unification or not (should unification of geometry and matter be 
realized in nature at all).\\
\\
We are therefore again in a situation, similar to that
before the discovery of special relativity, where we have the choice 
between a) preserving an old principle, here renormalizability of 
perturbative QFT on background 
spacetimes $(M,\eta)$, at the price of introducing 
extra structure (extra unification symmetry), or b) replacing the old 
principle by a new principle, here 
non-perturbative QFT on a differentiable manifold $M$, 
without new hypothetical structure. At this point it unclear which 
methodology has more chances for success, historically there is evidence 
for either of them (e.g. the unification of electromagnetism and the 
massive Fermi model is evidence for the former, the replacement of 
Newton's notion of spactime by special relativity is evidence for the 
latter)
and it is quite possible that we actually need both ideas. 
In QGR we take 
the latter point of view to begin with since there maybe zillions 
of ways to unify forces and it is hard to judge whether there is a 
``natural one", therefore the approach is {\it purposely conservative}
because we actually may be able to {\it derive} a natural way of 
unification, if necessary, if we drive the theory to its logical frontiers.  
Among the various 
non-perturbative approaches available we will choose the canonical one. 

Pictorially, one could illustrate the deep difference between a 
background dependent QFT and background independent QFT as follows:
\begin{figure} 
\includegraphics[width=10cm,height=5cm]{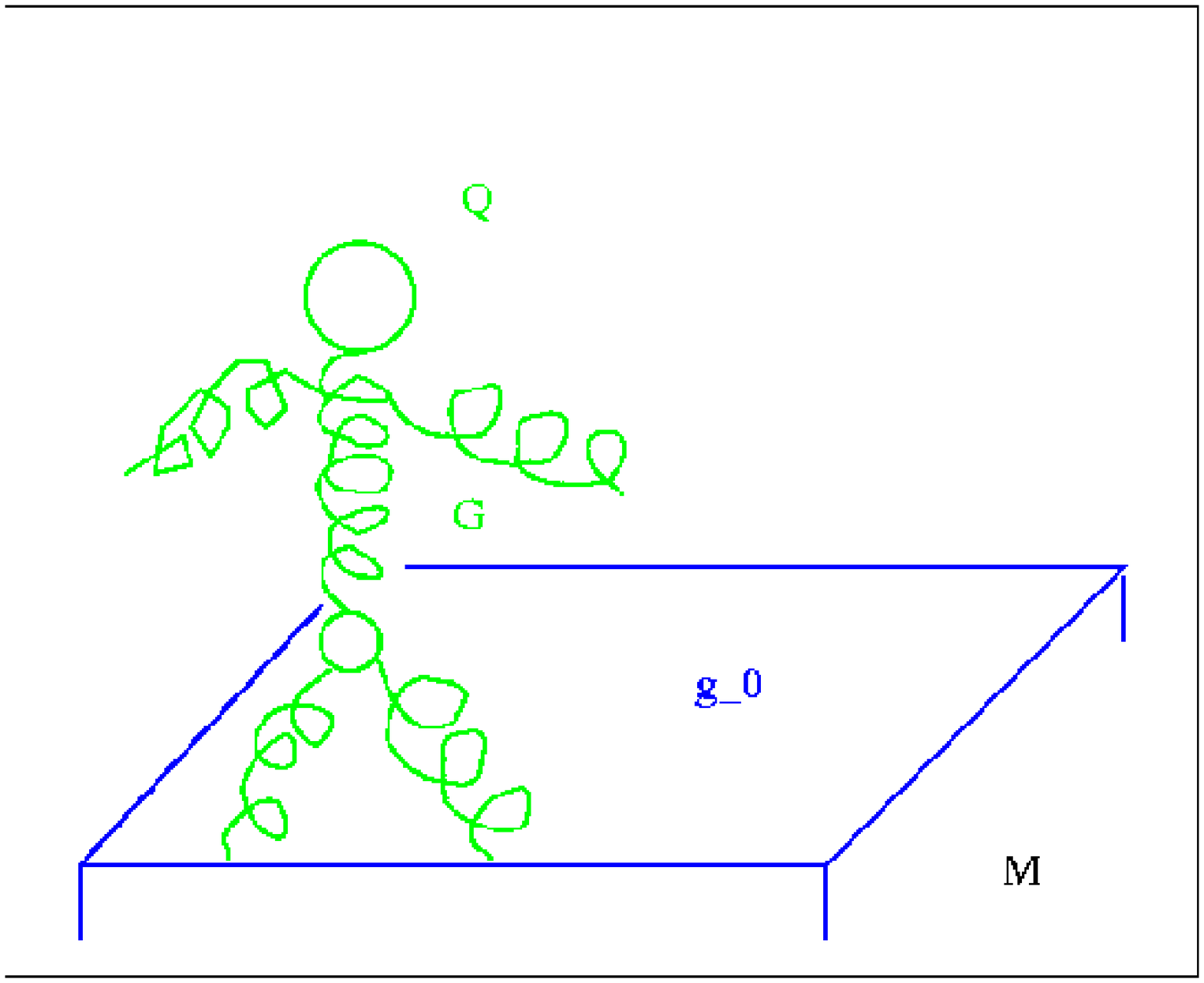}
\caption{QFT on Background Spacetime $(M,g_0)$:
Actor: {\bf\green Matter}, Stage: {\bf\blue Geometry} + Manifold $M$.}
\label{f3}
\end{figure}
\begin{figure} 
\includegraphics[width=10cm,height=5cm]{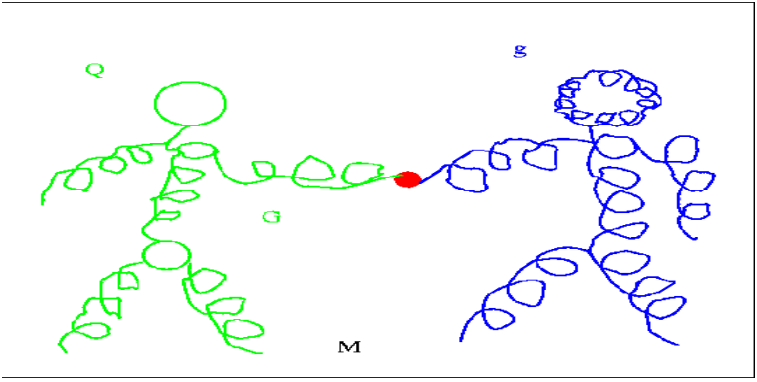}
\caption{QGR on Differential Manifold  $M$:
Actor: {\bf\green Matter} + {\bf\blue Geometry}, Stage: Manifold $M$}
\label{f4}
\end{figure}
In figure \ref{f3} we see matter in the form of QCD (notice the quark 
($Q$)
propagators, the quark-gluon vertices and the three -- and four point
gluon ($G$) vertices) displayed as an actor in green. Matter propagates on 
a fixed background spacetime $g_0$ according to well-defined rules, 
particles know exactly what timelike geodesics are etc. This fixed  
background spacetime $g_0$ is displayed as a firm stage in blue.
This is the situation of a QFT on a {\it Background Spacetime}. 

In contrast, in figure \ref{f4} the stage has evaporated, it has become
itself an actor (notice the arbitrarily high valent graviton ($g$) 
vertices) displayed in blue as well. Both matter and geometry are now
dynamical entities and interact as displayed by the red vertex. 
There are no light cones any longer, rather the causal structure 
is a semiclassical concept only. This is the situation of a QFT
on a {\it Differential Manifold} and this is precisely what QGR aims
to rigorously define. 

It is clear from these figures that the passage from a 
QFT on a background spacetime to a QFT on a differential manifold is
a very radical one: It is like removing the chair on which 
you sit and trying to find a new, yet unknown, mechanism that keeps 
you from falling down.
We should mention here that for many researchers in quantum gravity
even that picture is not yet radical enough, some proposals require 
not only to get rid of the background metric $g_0$ but also 
of the differential manifold, allowing for topology change. 
This is also very desired in QGR but considered as a second step.
In 3d QGR also this step could be completed and the final picture 
is completely combinatorial.\\
\\
Let us finish this section by stating once more what we mean by 
Quantum General Relativity (QGR).

\newpage 
~~\\
{\bf\magenta\large  DEFINITION}\\
\\
\fbox{\fbox{\parbox{14cm}{{\bf
(Canonical) {\red 
Quantum General Relativity (QGR)} is an attempt to construct a
{\red mathematically rigorous}, {\green non -- perturbative}, 
{\blue background independent} Quantum Field Theory of four -- dimensional,
Lorentzian General Relativity and all known matter in the continuum.\\
\\
No additional, experimentally unverified structures are introduced.
The fundamental principles of {\blue General Covariance} and 
{\green Quantum Theory} are brought together and driven to their 
logical frontiers guided by mathematical consistency.\\
\\  
QGR is not a unified theory of all interactions in the standard sense 
since unification of gauge symmetry groups 
is not necessarily required in a non-perturbative approach.
However, 
{\blue Geometry} and {\green Matter} are unified in a non -- standard 
sense by making them both transform covariantly under the 
{\red Diffeomorphism Group} at the quantum level. 
}}}}

\section{Introduction: Classical Canonical Formulation
of General Relativity}
\label{s1.2}

In this section we sketch the classical Hamiltonian formulation of general
relativity in terms of Ashtekar's new variables. There are many ways 
to arrive at this new formulation and we will choose the one that is the most
convenient one for our purposes.

The Hamiltonian formulation by definition requires some kind of split 
ot the spacetime variables into time and spatial variables. This seems 
to contradict the whole idea of general covariance, however, quantum 
meachanics as presently formulated requires a notion of time because 
we interpret expection values of operators as instantaneous 
measurement values averaged over a large number of measurements. In order 
to avoid this one has to ``covariantize" the interpretation of quantum 
mechanics, in particular the measurement process, see e.g. \cite{8a}
for a discussion. There are a number of proposals to make the canonical
formulation more covariant, e.g.\footnote{Path integrals \cite{8} use 
the Lagrangean rather than the Hamiltonian and therefore seem to be better 
suited to a covariant formulation than the canonical one, however, 
usually the path integral is interpreted as some sort
of propagator which makes use of instantaneous time Hilbert spaces 
again which therefore cannot be completely discarded with. 
At present, this connection with the canonical 
formulation is not very transparent, part of the reason being that
the path integral is usually only defined in its Euclidean formulation, 
however
the very notion of analytic continuation in time is not very meaningful
in a theory where there is no distinguished choice of time, see however
\cite{14} for recent progress in this direction.}:
Multisymplectic Ans\"atze \cite{9}
in which there are multimomenta, one for each spacetime dimension, 
rather
than just one for the time coordinate; Covariant phase space formulations    
\cite{10} 
where one works on the space of solutions to the field equations rather 
than on the intial value instantaneous phase space; Peierl's bracket 
formulations \cite{11} which covariantize the notion of the usual Poisson 
bracket;
history bracket formulations \cite{12}, which grew out of the consistent
history formulation of quantum mechanics \cite{13}, and which 
extends the usual spatial Poisson bracket to spacetime.

At the classical level all these formulations are equivalent. 
However, at the quantum level, one presently gets farthest 
within the the standard canonical formulation:
The quantization of the multisymplectic approach is still in
its beginning, see \cite{15} for the most advanced results in this respect;
The covariant phase space formulation is not only very implicit because 
one usually does not know the space of solutions to the classical field 
equations, but even if one manages to base a quantum theory on it, it
will be too close to the classical theory since certainly the 
singularities of the classical theory are also built into the quantum 
theory; The Peierl's bracket also needs the explicit space of solutions 
to the classical field equations; Also the quantization of the history 
bracket formulation just has started, see \cite{16} for first steps in that
direction.

Given this present status of affairs, we will therefore proceed with the 
standard canonical quantization and see how far we get. Notice that there
is no obvious problem with general covariance: 
For instance, standard Maxwell theory
can be quantized canonically without any problem and one can show that
the theory is Lorentz covariant although the spacetime split into space 
and time seems to break the Lorentz group down to the rotation group. 
{\it This is not at all the case} ! It is just that Lorentz covariance is 
not manifest, one has to do some work in order to establish Lorentz 
covariance.
Indeed, as we will see, at least at the classical level we will explicitly
recover the four-dimensional diffeomorphism group in the formalism, although
it is admittedly deeply hiddeen in the canonical formalism.

With these cautionary remarks out of the way, we will thus assume that 
the four dimensional spacetime manifold has the topology $\Rl\times \sigma$,
where $\sigma$ is a three dimensional manifold of arbitrary topology, in 
order to perform the $3+1$ split. This assumption about the topology of
$M$ may seem rather restrictive, however, it is not due to the following 
reasons: (1) According to a theorem due to Geroch any globally hyperbolic 
manifold (roughly those that admit a smooth metric with everywhere 
Lorentzian 
signature) are necessarily of that topology. Since Lorentzian metrics are 
what we are interested in, at least classically the assumption about the 
topology of $M$ is forced on us. (2) Any four manifold $M$ has the topology  
of a countable disjoint union $\cup_\alpha I_\alpha\times \sigma_\alpha$ 
where either 
$I_\alpha$ are open intervals and $\sigma_\alpha$ is a three manifold or 
$I_\alpha$ is a one point set and $\sigma_\alpha$ is a two manifold (the 
latter are the intersections of the closures of the former). In this 
most generic situation we thus allow topology change between different
three manifolds and it is even classically an open question how to make this
compatible with the action principle. We take here a practical point of 
view and try to understand the quantum theory first for a single copy 
of the form $\Rl\times \sigma$ and later on worry how we glue the theories
for different $\sigma's$ together.

\subsection{The ADM Formulation}
\label{s1.2.1}

\begin{figure} 
\includegraphics[width=10cm,height=5cm]{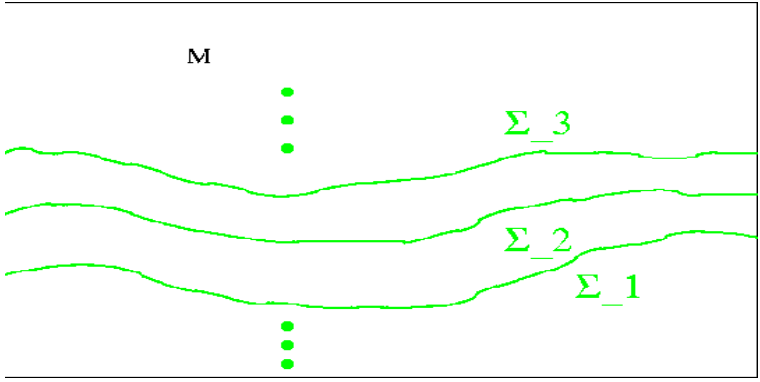}
\caption{Foliation of $M$} 
\label{f5}
\end{figure}
In this nice situation the $3+1$ split is well known as the 
Arnowitt -- Deser -- Misner (ADM) formulation of general relativity,
see e.g. \cite{2} and we briefly sketch how this works. Since $M$ is 
diffeomorphic to $\Rl\times \sigma$
we know that $M$ foliates into hypersurfaces $\Sigma_t,\;t\in\Rl$ as in
figure \ref{f5}, where $t$ labels the hypersurface and will play the role 
of our time coordinate. If we denote 
the four dimensional coordinates by $X^\mu,\; \mu=0,1,2,3$ and the three 
dimensional coordinates by $x^a,\; a=1,2,3$ then we know that there 
is a diffeomorphism $\varphi:\;\Rl\times \sigma\to M;\; (t,x)\mapsto
X=\varphi(t,x)$ where $\Sigma_t=\varphi(t,\sigma)$. We stress that the 
four diffeomorphism $\varphi$ is {\it completely arbitrary} 
until this point and thus {\it the foliation of $M$ is not at all fixed},
in other words, the set of foliations {\it is in one to one 
correspondence with Diff$(M)$, the four dimensional diffeomorphism group}. 
Consider the 
tangential vector fields to $\Sigma_t$ given by 
\be \label{1.2.1}
S_a(X):=(\partial_a)_{\varphi(t,x)=X}=
(\varphi^\mu_{,a}(t,x))_{\varphi(t,x)=X}\;\;\partial_\mu
\ee
Denoting the four metric by $g_{\mu\nu}$ we define a normal vector field 
$n^\mu(X)$ by $g_{\mu\nu} n^\mu S_a^\nu=0,\;g_{\mu\nu} n^\mu n^\nu=-1$.
Thus, while the tangential vector fields depend only on the foliation, 
the normal vector field depends also on the metric. Let us introduce the 
foliation vector field
\be \label{1.2.1a}
T(X):=(\partial_t)_{\varphi(t,x)=X}=
(\varphi^\mu_{,t}(t,x))_{\varphi(t,x)=X}\;\;\partial_\mu
\ee
and let us decompose it into the basis $n,S_a$. This results in 
\be \label{1.2.2}
T=Nn+U^a S_a
\ee
where $N$ is called the {\it lapse function} while $U^a S_a$ is called the 
{\it shift vector} field. The arbitrariness of the foliation is expressed 
in the arbitrariness of the fields $N,U^a$.
We can now introduce two 
symmetric spacetime tensor fields ($\nabla$ is the unique, torsion free 
covariant differential compatible with $g_{\mu\nu}$)
\be \label{1.2.3}
q_{\mu\nu}=g_{\mu\nu}+n_\mu n_\nu,\;\;
K_{\mu\nu}=
q_{\mu\rho} q_{\nu\sigma} \nabla^\rho n^\nu
\ee
called the intrinsic metric and the extrinsic curvature respectively which
are spatial, that is, their contraction with $n$ vanishes. Thus, their 
full information is contained in their components with respect to the spatial
fields $S_a$, e.g. $q_{ab}(t,x)=[q_{\mu\nu} S^\mu_a S^\nu_b](X(t,x))$.
In particular,
\be \label{1.2.4}
K_{ab}(t,x)=\frac{1}{2N}[\dot{q}_{ab}-{\cal L}_U q_{ab}]
\ee
contains information about the {\it velocity} of $q_{ab}$.
Here $\cal L$ the Lie 
derivative. The metric $g_{\mu\nu}$ is completely specified in terms 
of $q_{ab},N,U^a$ as one easily sees by expressing the line 
element $ds^2=g_{\mu\nu} dX^\mu\; dX^\nu$ in terms of $dt,dx^a$.
\begin{Exercise} \label{ex1.2.0} ~\\
Recall the definition of the Lie derivative and 
verify that $K_{\mu\nu}$ is indeed symmetric and that formula (\ref{1.2.4})
holds.\\
Hint:\\
A hypersurface $\Sigma_t$ can be defined by the solution of an equation of 
the form $\tau(X)=t$. Conclude that $n_\mu\propto \nabla_\mu\tau$ and 
use torsion -- freness of $\nabla$.
\end{Exercise}
The Legendre transformation of the Einstein-Hilbert action
\be \label{1.2.5}
S=\frac{1}{\kappa}\int_M d^4X \sqrt{|\det(g)|} R^{(4)}
\ee
with $q_{ab},N,U^a$ considered as configuration coordinates in an 
infinite dimensional phase space is standard and we will not repeat 
the analysis here, which uses the so - called Gauss - Codacci equations. 

Here we are considering for 
simplicity only the case that $\sigma$ is compact, otherwise (\ref{1.2.5})
would contain boundary terms. The end result is 
\be \label{1.2.6}
S=\frac{1}{\kappa}\int_\Rl dt \int_\sigma 
d^3x \{\dot{q}_{ab} P^{ab}+\dot{N} P+\dot{N}^a P_a-
[\lambda P+\lambda^a P_a+U^a V_a+N C]\}
\ee
where 
\be \label{1.2.7}
P^{ab}=\frac{\kappa\delta S}{\delta q_{ab}}=\sqrt{\det(q)}
[q^{ac} q^{bd}-q^{ab} q^{cd}] K_{cd}
\ee
and $P,P_a$ are the momentua conjugate to $q_{ab},N,U^a$ respectively.
Thus, we have for instance the equal time Poisson brackets
\be \label{1.2.7a}
\{P^{ab}(t,x),P^{cd}(t,y)\}=\{q_{ab}(t,x),q_{cd}(t,y)\}=0,\;\;
\{P^{ab}(t,x),q_{cd}(t,y)\}=\kappa 
\delta^a_{(c} \delta^b_{d)} \delta(x,y)
\ee
where $(.)_{(ab)}:=[(.)_{ab}+(.)_{ba}]/2$ denotes symmetrization.
The functions $C,V_a$ which depend only on $q_{ab},P^{ab}$ are called the 
{\it Hamiltonian and Spatial Diffeomorphism constraint} respectively for 
reasons 
that will become obvious in a moment. Their explicit form is given by
\ba \label{1.2.7b}
V_a &=& -2 q_{ac} D_b P^{bc} \nonumber\\
C &=& \frac{1}{\sqrt{\det(q)}}
[q_{ac} q_{bd}-\frac{1}{2} q_{ab} q_{cd}] P^{ab} P^{cd}-
\sqrt{\det(q)} R
\ea
where $D$ is the unique, torsion -- free covariant differential 
compatible with $q_{ab}$ and $R$ is the curvature scalar associated with 
$q_{ab}$.

The reason for the occurence of the Lagrange multipliers $\lambda,\lambda^a$
is that the Lagrangean (\ref{1.2.5}) is singular, that is, one cannot 
solve all the velocities in terms of momenta and therefore one must
use Dirac's procedure \cite{17} for the Legendre transform of singular 
Lagrangeans.
In this case the singularity structure is such that the momenta conjugate 
to $N,U^a$ vanish identically, whence the Lagrange multipliers which when
varied give the equations of motion $P=P_a=0$. The equations of motion 
with respect to the Hamiltonian (i.e. $\dot{F}:=\{H,F\}$ for any 
functional $F$ of the canonical coordinates)
\be \label{1.2.8}
H=\int d^3x [\lambda P+\lambda^a P_a+U^a V_a+N C] 
\ee
for
$N,U^a$ reveal that $N,U^a$ are themselves Lagrange multipliers, i.e.
completely unspecified functions (proportional to $\lambda,\lambda^a$) while 
the equations of motion for 
$P,P_a$ give $\dot{P}=-C,\;\dot{P}_a=-V_a$. Since $P,P_a$ are supposed to 
vanish, this requires $C=V_a=0$ as well. Thus we see that the 
Hamiltonian 
{\it is constrained to vanish in GR} ! We will see that this is a direct 
consequence of the four dimensional diffeomorphism invariance of the 
theory.

Now the equations of motion for $q_{ab},P^{ab}$ imply the 
so-called {\it Dirac (or hypersurface deformation) algebra}
\ba \label{1.2.9}
\{V(U),V(U')\}&=&\kappa V({\cal L}_U U') \nonumber\\
\{V(U),C(N)\}&=&\kappa C({\cal L}_U N) \nonumber\\
\{C(N),C(N')\}&=&\kappa V(q^{-1}(N dN'-N' dN))
\ea
where e.g. $C(N)=\int d^3x N C$. These equations tell us that
the condition $H=V_a=0$ is preserved under evolution, in other words,
the evolution is consistent ! This is a non-trivial result. One says,
the Hamiltonian and vector constraint form a {\it first class constraint 
algebra}.
This algebra is much more complicated than the more familiar Kac-Moody
algebras due to the fact that it is not an (infinite) dimensional Lie algebra
in the true sense of the word because the ``structure constants" on the 
right hand side of the last line in (\ref{1.2.9}) are not really constants,
they depend on the phase space. Such algebras are open in the the 
terminology of BRST \cite{18} and about their representation theory
only very little is known.
\begin{Exercise} \label{ex1.2.1.0} ~~~\\
Derive (\ref{1.2.9}) from (\ref{1.2.7a}).\\
Hint:\\
Show first that the Poisson bracket between local functions which contain
spatial derivatives is simply the spatial derivatives applied to the
Poisson bracket. Since the Poisson bracket of local functions is
distributional recall that derivatives of distributions are defined 
through an integration by parts.    
\end{Exercise}

Since the 
variables $P,P_a$ drop out completely from the analysis and $N,U^a$ 
are Lagrange multipliers, we may replace (\ref{1.2.6}) by
\be \label{1.2.10}
S=\frac{1}{\kappa}\int_\Rl dt \int_\sigma 
d^3x \{\dot{q}_{ab} P^{ab}-[U^a V_a+N H]\}
\ee
with the understanding that $N,U^a$ are now completely arbitrary 
functions which parameterize the freedom in choosing the foliation.
Since the Hamiltonian of GR depends on the completely unspecified functions
$N,U^a$, the motions that it generates in the phase space $\cal M$
coordinatized by $(P^{ab},q_{ab})$ subject to the Poisson brackets 
(\ref{1.2.7a}) are to be considered as pure 
gauge transformations. The infinitesimal flow (or motion) of the canonical 
coordinates generated by 
the corresponding
Hamiltonian vector fields on $\cal M$ has the following form for an 
arbitrary tensor $t_{ab}$ built from $q_{ab}, P^{ab}$
\ba \label{1.2.11}
\{V(U),t_{ab}\}_{EOM}=\kappa ({\cal L}_U 
t_{ab})
\nonumber\\
\{C(N),t_{ab}\}_{EOM}=\kappa
({\cal L}_{Nn} t_{ab})
\ea
where the subscript $EOM$ means that these relations hold for generic
functions on $\cal M$ only when the vacuum equations of motion (EOM)
$R^{(4)}_{\mu\nu}-R^{(4)} g_{\mu\nu}/2=0$ hold. Equation (\ref{1.2.11})
reveals that Diff$(M)$ is implemented also in the canonical formalism, 
however, in a rather non-trivial way: The gauge motions generated by the 
constraints can be interpreted as four-dimensional diffeomorphisms {\it
only when the EOM hold}. This was to be expected because a diffemorphism
orthogonal to the hypersurface means evolution in the time 
parameter, what is surprising though is that this evolution is considered 
as a 
gauge transformation in GR. Off the solutions, the constraints generate 
different motions, in other words, the set of gauge symmetries is 
not Diff$(M)$ everywhere in the phase space. This is not unexpected: The 
action (\ref{1.2.5})
is obviously Diff$(M)$ invariant, but so would be any action that is an 
integral over a four-dimensional scalar density of weight one formed
from polynomials in the curvature tensor and its covariant derivatives.
This symmetry is completely insensitive to the specific Lagrangean in 
question, it is kinematical. The dynamics generated by a specific 
Lagrangean must depend on that Lagrangean, otherwise all 
Lagrangeans underlying four dimensionally
diffeomorphism invariant actions would equal each other up to a 
diffeomorphism which is certainly not the case (consider for instance 
higher derivative theories). In particular, that dynamics is, a 
priori, completely independent of Diff$(M)$. As a consequence, Dirac 
observables, that 
is, functions on $\cal M$ which are gauge invariant (have vanishing 
Poisson brackets with the constraints), {\it 
are not simply
functionals of the four metric invariant under four diffeomorphisms} because
they must depend on the Lagrangean. The set of 
these dynamics dependent gauge transformations does not obviously form a 
group 
as has been investigated by Bergmann and Komar \cite{19}. The geometrical 
origin of the hypersurface deformation algebra has been investigated in 
\cite{20}. Torre and Anderson have shown that for compact $\sigma$ there 
are no Dirac observables
which depend on only a finite number of spatial derivatives of the canonical
coordinates \cite{21} which means that Dirac observables will be highly
non-trivial to construct. 

Let us summarize the gauge theory of GR in figure \ref{f6}:
\begin{figure} 
\includegraphics[width=10cm,height=5cm]{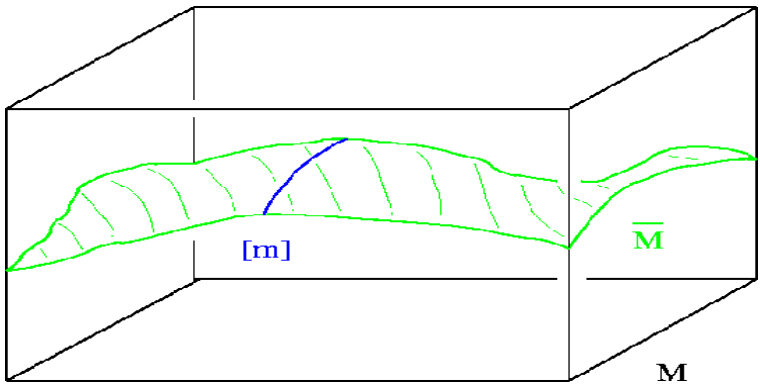}
\caption{Constraint hypersurface $\overline{{\cal M}}$ and gauge orbit 
$[m]$ of $m\in \overline{{\cal M}}$ in $\cal M$}. 
\label{f6}
\end{figure}
The constraints $C=V_a=0$ define a constraint hypersurface 
$\overline{{\cal M}}$ within the 
full phase space $\cal M$. The gauge motions are defined on all of 
$\cal M$ but they have the feature that they leave the constraint 
hypersurface invariant, and thus the orbit of a point $m$ in the 
hypersurface under gauge transformations will be a curve or gauge 
orbit $[m]$ entirely 
within it. The set of these curves defines the so-called reduced phase 
space  and Dirac observables restricted to $\overline{{\cal M}}$ depend 
only on these orbits.
Notice that as far as the counting is concerned we have twelve phase 
space coordinates $q_{ab},P^{ab}$ to begin with. The four constraints 
$C,V_a$ can be solved to eliminate four of those and there are still
identifications under four independent sets of motions among the remaining
eight variables leaving us with only four Dirac observables. The 
corresponding so-called reduced phase space has therefore precisely the two
configuration degrees of freedom of general relativity.

\subsection{Gauge Theory Formulation}
\label{s1.2.2}

We can now easily introduce the shift from the ADM variables 
$q_{ab},P^{ab}$ to the connection variables introduced first by Ashtekar
\cite{22} and later somewhat generalized by Immirzi \cite{23} and Barbero
\cite{24}.
We introduce $su(2)$ indices $i,j,k,..=1,2,3$ and co-triad variables $e_a^j$
with inverse $e^a_j$ whose relation with $q_{ab}$ is given by 
\be \label{1.2.12}
q_{ab}:=\delta_{jk} e^j_a e^k_b
\ee
Defining the spin connection $\Gamma_a^j$ through the equation
\be \label{1.2.13}
\partial_a e_b^j-\Gamma^c_{ab} e_c^j+\epsilon_{jkl} \Gamma_a^k e_b^l=0
\ee
where $\Gamma_{ab}^c$ are the Christoffel symbols associated with $q_{ab}$
we now define 
\be \label{1.2.14}
A_a^j=\Gamma_a^j+\beta K_{ab} e^b_j,\;\;E^a_j=\sqrt{\det(q)} e^a_j/\beta
\ee
where $\beta\in \Cl-\{0\}$ is called the {\it Immirzi} parameter.
In this article we only consider real valued and positive $\beta$.
Finally we introduce the $SU(2)$ {\it Gauss constraint}
\be \label{1.2.15}
G_j:=\partial_a E^a_j+\epsilon_{jkl} A_a^k E^b_l
\ee
with $\epsilon_{jkl}$ the structure constants of $su(2)$
which we would encounter in the canonical formulation of any $SU(2)$ 
gauge theory.
As one can check, modulo $G_j=0$ one can then write $C,V_a$ in terms of 
$A,E$ as follows 
\ba \label{1.2.16}
V_a &=& F_{ab}^j E^b_j, \nonumber\\
C &=& \frac{F_{ab}^j \epsilon_{jkl} E^a_j E^b_l}{\sqrt{|\det(E)|}}\;\;+\;\;
\mbox{ More }
\ea
where $F=2(dA+[A,A])$ is the curvature of $A$ and ``More" is an additional 
term which is more complicated but can be treated by similar methods as 
the one displayed.
 
We then have the following theorem \cite{22}.
\begin{Theorem} \label{th1.2.0}
~~~\\
Consider the phase space $\cal M$ coordinatized by $(A_a^j,E^b_j)$
with Poisson brackets
\be \label{1.2.17}
\{E^a_j(x),E^b_k(y)\}=
\{A_a^j(x),A_b^k(y)\}=0,\;\;
\{E^a_j(x),A_b^k(y)\}=\kappa\delta^a_b\delta_j^k\delta(x,y)
\ee
and constraints $G_j,C,V_a$. Then, solving only the constraint $G_j=0$ and 
determining the Dirac observables with respect to it leads us back
to the ADM phase space with constraints $C,V_a$.
\end{Theorem}
The proof of the theorem is non-trivial and tedious and can be found 
in the notation used here in \cite{0}. Alternatively one can find 
directions for a proof in the subsequent exercise. In 
particular, this works only 
because the Gauss constraint is in evolution with itself and the other 
constraints, specifically $\{G,G\}\propto G,\;\{G,V\}=\{G,H\}=0$.
\begin{Exercise} \label{ex1.2.1} ~~~~\\
i)\\
Prove theorem \ref{th1.2.0}.\\
Hint:\\
Express $q_{ab},P^{ab}$ in terms of $A_a^j,E^a_j$ by using 
(\ref{1.2.12}), (\ref{1.2.13}), (\ref{1.2.14}), and (\ref{1.2.15})
and check that the Poisson brackets, with respect to (\ref{1.2.17}),
among the solutions 
$q_{ab}=s_{ab}[A,E],P^{ab}=S^{ab}[A,E]$ equal precisely
(\ref{1.2.7a}) modulo terms proportional to $G_j$.\\
ii)\\ 
Define $G(\Lambda):=\int_\sigma d^3x \Lambda^j G_j$ and 
$D(U):=\int_\sigma U^a[V_a-A_a^j G_j]$ and $[\Lambda,\Lambda']_j
=\epsilon_{jkl} \Lambda^k(\Lambda')^l$. Verify the following 
Poisson brackets
\ba \label{ex1.2.1a}
\{G(\Lambda),G(\Lambda')\} &=& \kappa G([\Lambda,\Lambda'])
\nonumber\\
\{G(\Lambda),V(U)\} &=& 0 \nonumber\\
\{D(U),D(U')\}&=&\kappa D([U,U'])
\ea
and conclude that the Hamiltonian vector fields of $G(\Lambda)$ and $D(U)$ 
respectively generate $SU(2)$ gauge transformations and spatial
diffeomorphisms of $\sigma$ respectively. \\
Hint:\\
Show first that 
\ba \label{ex1.2.1b}
\{G(\Lambda/\kappa),A_a^j(x)\} &=&-\Lambda^j_{,a}+\epsilon_{jkl} 
\Lambda^k A_a^l \nonumber\\
\{D(U/\kappa),A_a^j(x)\} &=& U^b A^j_{a,b}+ U^b_{,a} A^j_b
\ea
to conclude that $A$ transforms as a connection under infinitesimal
gauge transformations and as a one-form under infinitesimal 
diffeomorphisms. Consider then $g_t(x):=\exp(t\Lambda^j \tau_j/(2\kappa))$
and $\varphi_t(x):=c_{U,x}(t)$ where $t\mapsto c_{U,x}(t)$ is the unique 
integral curve of $U$ through $x$, that is, 
$\dot{c}_{U,x}(t)=U(c_{U,x}(t)),\;c_{U,x}(0)=x$. Recall
that the usual transformation behaviour of connections and one-forms under
finite gauge transformations and diffeomorphisms respectively is given by
(e.g. \cite{24a})
\ba \label{ex1.2.1c}
A^g &=& -dg g^{-1}+\mbox{Ad}_g (A) \nonumber\\
A^\varphi &=& \varphi^\ast A 
\ea
where $A=A_a^j dx^a \tau_j/2$, Ad$_g(.)=g(.)g^{-1}$ denotes the 
adjoint representation of $SU(2)$ on $su(2)$ and $\varphi^\ast$ denotes 
the pull-back map of $p-$forms and $i\tau_j$ are the Pauli matrices so 
that  $\tau_j \tau_k=-\delta_{jk} 1_2+\epsilon_{jkl} \tau_l$. Verify then 
that (\ref{ex1.2.1b}) is the derivative at $t=0$ of (\ref{ex1.2.1c})
with $g:=g_t,\;\varphi:=\varphi_t$. Similarly, derive that $E$ transforms
as an $su(2)-$valued vector field of density weight one. (Recall that a 
tensor field $t$ of some type is 
said to be of density weight $r\in \Rl$ if $t\sqrt{|\det(s)|}^{-r}$ is an 
ordinary tensor field of the same type where $s_{ab}$ is any 
non-degenerate symmetric tensor field). 
\end{Exercise}
From the point of view of the 
classical theory we have made things more complicated: Instead of twelve 
variables $q,P$ we now have eighteen $A,E$. However, the additional six 
degrees of freedom are removed by the first class Gauss constraint which 
shows that
working on our gauge theory phase space is equivalent to working on the 
ADM phase space. The virtue of this extended phase space is that 
{\it canonical GR can be formulated in the language of a canonical gauge 
theory} where $A$ plays the role of an $SU(2)$ connection with 
canonically conjugate electric field $E$. Besides the remark that this fact
could be the starting point for a possible gauge group unification of all
four forces we now have access to a huge arsenal of techniques that have been
developed for the canonical quantization of gauge theories. It is 
precisely this fact that has enabled steady progress in this field in the 
last fifteen years while one was stuck with the ADM formulation for almost
thirty years.

\section{Canonical Quantization Programme for Theories with
Constraints}
\label{s1.3}

\subsection{Refined Algebraic Quantization (RAQ)}
\label{s1.3.1}

As we have seen, GR can be formulated as a constrained Hamiltonian system 
with first class constraints. The quantization of such systems has been
considered first by Dirac \cite{17} and was later refined by a number of 
authors.
It is now known under the name refined algebraic quantization (RAQ).
We will briefly sketch the main ideas following \cite{25}.
\begin{itemize}
\item[i)] {\it Phase Space and Constraints}\\
The starting point is a phase space $({\cal M},\{.,.\})$ together with a set
of first class constraints $C_I$ and possibly a Hamiltonian $H$.
\item[ii)] {\it Choice of Polarization}\\
In order to quantize the phase space we must choose a polarization, that is,
a Lagrangean submanifold $\cal C$ of $\cal M$ which is called configuration 
space. The coordinates of $\cal C$ have vanishing Poisson brackets among 
themselves. If $\cal M$ is a cotangent bundle, that is, 
${\cal M}=T^\ast {\cal Q}$ then it is natural to choose ${\cal Q}={\cal C}$
and we will assume this to be the case in what follows. For more general
cases, e.g. compact phases spaces one needs ideas from geometrical 
quantization, see e.g. \cite{26}.
The idea is that (generalized, see below) points of $\cal C$ serve as 
arguments of the vectors of the Hilbert space to be constructed.
\item[iii)] {\it Preferred Kinematical Poisson Subalgebra}\\
Consider the space $C^\infty({\cal C})$ of smooth functions on $\cal C$
and the space $V^\infty({\cal C})$ of smooth vector fields on $C$.
The vertical polarization of $\cal M$, that is, the space of fibre 
coordinates called momentum space, generates preferred elements of 
$V^\infty({\cal C})$ through $(v_p[f])(q):=(\{p,f\})(q)$ where we have 
denoted configuration and momentum coordinates by $q,p$ respectively
and $v[f]$ denotes the action of a vector field on a function. 
The pair $C^\infty(\cal C)\times V^\infty({\cal C})$ forms a Lie algebra
defined by $[(f,v),(f',v')]=(v[f']-v'[f],[v,v'])$ of which the algebra 
$\cal B$ generated by elements of the form $(f,v_p)$ forms a subalgebra.
We assume that $\cal B$ is closed under complex conjugation which becomes 
its $^\ast-$operation (involution).
\item[iv)] {\it Representation Theory of the Corresponding Abstract
$^\ast-$Algebra}\\
We are looking for all irreducible $^\ast-$representations
$\pi:\;{\cal B}\to {\cal L}({\cal H}_{kin})$ of $\cal B$ as linear 
operators 
on a kinematical Hilbert space ${\cal H}_{kin}$ such that the 
$^\ast-$relations becomes 
the operator adjoint and such that the canonical commutation relations 
are implemented, that is, for all $a,b\in {\cal B}$
\ba \label{1.3.1}
\pi(a)^\dagger &=& \pi(a^\ast) \nonumber\\
{[}\pi(a),\pi(b)] &=& i\hbar \pi([a,b])
\ea  
Strictly speaking, (\ref{1.3.1}) is to be supplemented by the domains on 
which the operators are defined. In order to avoid this one will work
with the subalgebra of $C^\infty({\cal C})$ formed by bounded functions, 
say of compact support and one will deal with exponentiated vector fields 
in order to obtain bounded operators. Irreducibility is a physically
meaningful requirement because we are not interested in Hilbert spaces 
with superselection sectors and the reason for why we do not require 
the full Poisson algebra to be faithfully represented is that this is 
almost always impossible in irreducible representations as stated in 
the famous Groenewald -- van Hove theorem. The Hilbert space that
one gets can usually be described in the form 
$L_2(\overline{{\cal C}},d\mu)$ where $\overline{{\cal C}}$ is a 
distributional extension of $\cal C$ and $\mu$ is a probability 
measure thereon. A well-known example is the case of free scalar fields
on Minkowski space where $\cal C$ is some space of smooth scalar fields
on $\Rl^3$ vanishing at spatial infinity while $\overline{{\cal C}}$ is 
the space of tempered distributions on $\Rl^3$ and $\mu$ is a 
normalized Gaussian measure on $\overline{{\cal C}}$.   
\item[v)] {\it Selection of Suitable Kinematical Representations}\\
Certainly we want a representation which supports also the constraints 
and the Hamiltonian as operators which usually will limit the number of 
available 
representations to a small number, if possible at all. The constraints 
usually are not in $\cal B$ unless linear in momentum and the expressions
$\hat{C}_I:=\pi(C_I), \hat{H}=\pi(H)$ will involve factor ordering 
ambiguities as well 
as regularizationand renormalization processes in the case of field theories.
In the generic case, $\hat{C}_I,\hat{H}$ will not be bounded and 
$\hat{C}_I$ will not be symmetric. We will require that $\hat{H}$ is 
symmetric and that the constraints are at least closable, that is, they 
are densely defined together with their adjoints. It is then usually 
not too difficult to find a dense domain 
${\cal D}_{kin}\subset {\cal H}_{kin}$ on which all these 
operators and their adjoints are defined and which they leave invariant.
Typically ${\cal D}_{kin}$ will be a space of smooth functions of rapid 
decrease so that arbitray derivatives and polynomials of the configuration
variables are defined on them and such spaces naturally come with their own 
topology which is finer than the subspace topology induced from 
${\cal H}_{kin}$ whence we have a topological inclusion
${\cal D}_{kin}\hookrightarrow {\cal H}_{kin}$.
\item[vi)] {\it Imposition of the Constraints}\\
The two step process in the classical theory of solving the 
constraints $C_I=0$ and looking for the gauge orbits is replaced by a one 
step process in the quantum theory, namely looking for solutions $l$ of 
the equations $\hat{C}_I l=0$. This is because it is obviously
solves the constraint at the quantum level (in the corresponding 
representation on the solution space the constraints are replaced by the 
zero operator) and it simultaneously looks for states that are gauge 
invariant because $\hat{C}_I$ is the quantum generator of gauge 
transformations.

Now, unless the point $\{0\}$ is in the common point spectrum of all the 
$\hat{C}_I$, solutions $l$ to the equations 
$\hat{C}_I l=0\;\forall\; I$ do not lie in ${\cal H}_{kin}$, rather 
they are distributions. Here one has several options, one could look
for solutions in the space ${\cal D}_{kin}'$ of continuous linear 
functionals on ${\cal D}_{kin}$ (topological dual) or in the space 
${\cal D}_{kin}^\ast$ of linear functionals on ${\cal D}_{kin}$ with the 
topology of pointwise convergence (algebraic dual). 
Since certainly ${\cal H}_{kin}\subset {\cal D}_{kin}'\subset {\cal 
D}_{kin}^\ast$ 
let us choose the latter option for the sake of more generality. The 
topology on ${\cal H}_{kin}$ is again finer than the subspace topology 
induced from ${\cal D}_{kin}^\ast$ so that we obtain a {\it Gel'fand triple}
or {\it Rigged Hilbert Space}
\be \label{1.3.1a} 
{\cal D}_{kin}\hookrightarrow {\cal H}_{kin} 
\hookrightarrow {\cal D}_{kin}^\ast
\ee
This a slight abuse of terminology since the name is usually reserved 
for the case that ${\cal D}_{kin}$ carries a nuclear topology 
(generated by a countable family of seminorms separating the points)
and that ${\cal D}_{kin}^\ast$ is its topological dual.

We are now looking for a subspace 
${\cal D}_{phys}^\ast\subset {\cal D}^\ast_{kin}$ such that for its 
elements $l$ holds
\be \label{1.3.2}
[\hat{C}_I' l](f):=l(\hat{C}_I^\dagger f)=0\;\;\forall\; f\in 
{\cal D}_{kin},\; \forall I
\ee
The prime on the left hand side of this eqution defines a dual, anti-linear
representation of the constraints on ${\cal D}_{kin}^\ast$. The reason 
for the adjoint on the right hand side of this equation is that if
$l$ would be an element of ${\cal H}_{kin}$ then (\ref{1.3.2}) would be 
replaced by
\be \label{1.3.3}
[\hat{C}_I' l](f):=<\hat{C}_I l,f>_{kin}=
<l,\hat{C}_I^\dagger f>_{kin}=:l(\hat{C}_I^\dagger f)\;\;\forall\; 
f\in {\cal D}_{kin},\; \forall I
\ee
where $<.,.>_{kin}$ denotes the kinematical inner product,
so that (\ref{1.3.2}) is the natural extension of (\ref{1.3.3}) from 
${\cal H}_{kin}$ to ${\cal D}_{kin}^\ast$.
\item[vii)] {\it Anomalies}\\
Since we have a first class constraint algebra, we know that classically 
$\{C_I,C_J\}=f_{IJ}\;^K C_K$ for some 
{\it structure functions} $f_{IJ}\;^K$ which depend in general on the 
phase space point $m\in {\cal M}$. The translation of this equation into
quantum theory is then plagued with ordering ambiguities, because 
the structure functions turn into operators as well. It may therefore 
happen that, e.g. 
\be \label{1.3.4}
[\hat{C}_I,\hat{C}_J]
=i\hbar \hat{C}_K \hat{f}_{IJ}\;^K
=i\hbar\{[\hat{C}_K,\hat{f}_{IJ}\;^K]
+\hat{f}_{IJ}\;^K \hat{C}_K \}
\ee
and it follows that any $l\in {\cal D}_{phys}^\ast$ also solves the 
equation $([\hat{C}_K,\hat{f}_{IJ}\;^K])'l=0$ for all $I,J$. If that
commutator is not itself a constraint again, then it follows that $l$
solves more than only the equations $\hat{C}_I' l=0$ and thus the quantum
theory has less physical degrees of freedom than the classical theory.
This situation, called an {\it anomaly}, must be avoided by all means.
\item[viii)] {\it Dirac Observables and Physical Inner Product}\\
Since generically ${\cal H}_{kin}\cap {\cal D}_{phys}^\ast=\emptyset$,
the space ${\cal D}_{phys}^\ast$ cannot be equipped with the scalar
product $<.,.>_{kin}$. It is here wehere Dirac observables come into play.
A {\it strong} Dirac observable is an operator $\hat{O}$ on ${\cal H}_{kin}$ 
which is, together with its adjoint, densely defined on ${\cal D}_{kin}$ and 
which commutes with all constraints, that is, $[\hat{O},\hat{C}_I]=0$
for all $I$. We require that $\hat{O}$ is the quantization of a real valued
function $O$ on the phase space and the condition just stated is the quantum 
version of the classical gauge invariance condition $\{O,C_I\}=0$ for all
$I$. A {\it weak} Dirac observable is the quantum version of the more
general condition $\{O,C_I\}_{|C_J}=0 \forall J=0\;\forall I$ and simply 
means that the space of solutions is left invariant by the 
natural dual action of the operator 
$\hat{O}'{\cal D}_{phys}^\ast\subset {\cal D}_{phys}^\ast$.  

A physical inner product on a subset ${\cal H}_{phys}\subset 
{\cal D}_{phys}^\ast$ is a positive definite sesquilinear form 
$<.,.>_{phys}$ with respect to which the $\hat{O}'$ become self-adjoint
operators, that is, $\hat{O}'=(\hat{O}')^\star$ where 
the adjoint on ${\cal H}_{phys}$ is denoted by $\star$. Notice that 
$[\hat{O}_1',\hat{O}_2']=([\hat{O}_1,\hat{O}_2])'$ 
so that commutation 
relations on ${\cal H}_{kin}$ are automatically transferred to 
${\cal H}_{phys}$ which then carries a proper $^\ast-$representation of 
the physical observables. The observables themselves will only be defined 
on a dense domain ${\cal D}_{phys}\subset {\cal H}_{phys}$ and we get a 
second Gel'fand triple
\be \label{1.3.5} 
{\cal D}_{phys}\hookrightarrow {\cal H}_{phys} 
\hookrightarrow {\cal D}_{phys}^\ast
\ee
In fortunate cases, for instance when the $\hat{C}_I$ are mutually commuting
self-adjoint operators on ${\cal H}_{kin}$, all we have said is just a 
fancy way of stating the fact that ${\cal H}_{kin}$ has a direct integral
decomposition
\be \label{1.3.6}
{\cal H}_{kin}=\int_S^\oplus d\nu(\lambda) {\cal H}_\lambda
\ee
over the spectrum $S$ of the constraint algebra with a measure $\nu$ and 
eigenspaces ${\cal H}_\lambda$ which are left invariant by the strong 
observables and therefore ${\cal H}_{phys}={\cal H}_0$.
In the more general cases that are of concern to us, more work is required.
\item[ix)] {\it Classical Limit}\\
It is by no means granted that the representation ${\cal H}_{phys}$
that one finally arrived at, carries semiclassical states, that is states
$\psi_{[m]}$ labelled by gauge equivalence classes $[m]$ of points 
$m\in {\cal M}$ with respect to which the Dirac observables have the correct
expectation values and with respect to which their relative fluctuations 
are small, that is, roughly speaking
\be \label{1.3.7}
|\frac{<\psi_{[m]},\hat{O}'\psi_{[m]}>_{phys}}{O(m)}-1|\ll 1
\mbox{ and }
|\frac{<\psi_{[m]},(\hat{O}')^2\psi_{[m]}>_{phys}}
{(<\psi_{[m]},\hat{O}'\psi_{[m]}>_{phys})^2}-1|\ll 1
\ee
Only when such a phase exists are we sure that we have not constructed some
completely spurious sector of the quantum theory which does not admit the 
correct classical limit.
\end{itemize}

\subsection{Selected Examples with First Class Constraints}
\label{s1.3.2}

In the case that a theory has only first class constraints, Dirac's
algorithm \cite{17} boils down to the following four steps:\\
1)\\
Define the momentum $p_a$ conjugate to the configuration variable 
$q^a$ by (Legendre transform)
\be \label{1.3.2.1}
p_a:=\partial S/\partial\dot{q}^a
\ee
where $S$ is the action.\\
2)\\
Equation (\ref{1.3.2.1}) defines $p_a$ as a function of $q^a,\dot{q}^a$
and if it is not invertible to define the $\dot{q}^a$ as a function
of $q^a,p_a$ we get a collection of so-called {\it primary constraints}
$C_I$, that is, identities among the $q^a,p_a$. In this situation one says
that $S$ or the Lagrangean is singular. 
\\
3)\\
Using that $q^a,p_a$ have canonical Poisson brackets,
compute all possible Poisson brackets $C_{IJ}:=\{C_I,C_J\}$. If 
some $C_{I_0 J_0}$ is not zero when all $C_K$ vanish, 
then add this $C_{I_0 J_0}$, called a {\it secondary constraint}, to
the set of primary constraints.\\
4)\\
Iterate 3) until the $C_I$ are in involution, that is, no new secondary
constraints appear.\\

In this report we will only deal with theories which have no second class
constraints, so this algorithm is all we need.
\begin{Exercise} \label{ex1.3.1} ~~~\\
Perform the quantization programme for a couple of simple systems in order 
to get a feeling for the formalism:
\begin{itemize}
\item[1.] {\it Momentum Constraint}\\
${\cal M}=T^\ast \Rl^2$ with standard Poisson brackets among 
$q^a,p_a;\;a=1,2$ and constraint $C:=p_1$. Choose 
${\cal H}_{kin}=L_2(\Rl^2,d^2x),\;
{\cal D}_{kin}={\cal S}(\Rl^2),\;
{\cal D}^\ast_{kin}={\cal S}'(\Rl^2)$ (spaces of functions of rapid 
decrease and tempered distributions respectively).\\
Solution:\\
Dirac observables are the conjugate pair $q^2,p_2$, ${\cal 
H}_{phys}=L_2(\Rl,dx_2)$.\\
Hint: Work in the momentum representation and conclude that the general
solution is of the form $l_f(p_1,p_2)=\delta(p_1)f(p_2)$ for 
$f\in {\cal S}'(\Rl)$.
\item[2.] {\it Angular Momentum Constraint}\\
${\cal M}=T^\ast \Rl^3$ with standard Poisson brackets among 
$q^a,p_a;\;a=1,2,3$ and constraints $C_a:=\epsilon_{abc} x^b p_c$. 
Check the first class property and choose the kinematical spaces as 
above with $\Rl^2$ replaced by $\Rl^3$.\\
Solution:\\
Dirac observables are the conjugate pair 
$r:=\sqrt{\delta_{ab} q^a q^b}\ge 0,\;
p_r=\delta_{ab} q^a p_b/r$, the physical phase space is $T^\ast \Rl_+$
and ${\cal H}_{phys}=L_2(\Rl_+,r^2 dr)$ where $\hat{r}$ is a 
multiplication 
operator and $\hat{p}_r=i\hbar\frac{1}{r}\frac{d}{dr} r$ with 
dense 
domain of symmetry given by the square integrable functions $f$ such that 
$f$ is regular at $r=0$.  \\
Hint:\\
Introduce polar coordinates and decompose kinematical wave functions 
into spherical harmonics. Conclude that the physical Hilbert space this 
time is just the restriction of the kinematical Hilbert space to the 
zero angular momentum subspace, that is, ${\cal H}_{phys}\subset
{\cal H}_{kin}$. The reason is of course that the spectrum of the 
$\hat{C}_a$ is pure point (discrete).
\item[3.] {\it Relativistic Particle}\\
Consider the Lagrangean $L=-m\sqrt{-\eta_{\mu\nu} \dot{q}^\mu 
\dot{q}^\nu}$ where $m$ is a mass parameter, $\eta$ is the Minkowski
metric and $\mu=0,1,..,D$. Verify that the Lagrangean is singular, that
is, the velocities $\dot{q}^\mu$ cannot be expressed in terms of the 
momenta $p_\mu=\partial L/\partial \dot{q}^\mu$ which gives rise to the 
{\it mass shell constraint} $C=m^2+\eta^{\mu\nu} p_\mu p_\nu$. Verify 
that this happens because the corresponding action is invariant under
Diff$(\Rl)$, that is, reparameterizations $t\mapsto \varphi(t),\;
\dot{\varphi}(t)>0$. Perform the Dirac analysis for constraints and 
conclude that the system has no Hamiltonian, just the Hamiltonian 
constraint $C$ which generates reparameterizations on the kinematical 
phase space ${\cal M}=T^\ast \Rl^{D+1}$ with standard Poisson brackets.
Now choose kinematical spaces as in 1. with $\Rl^2$ replaced by 
$\Rl^{D+1}$.\\
Solution:\\
Conjugate Dirac observables are 
$Q^a=q^a-\frac{q^0 p_a}{\sqrt{m^2+\delta^{ab}p_a p_b}}$ and 
${\cal H}_{phys}=L_2(\Rl^D,d^D p)$ on which $\hat{q}^0=0$.\\
Hint:\\
Work in the momentum representation and conclude that the general
solution to the constraints is of the form $l_f=\delta(C)f(p_0,\vec{p})$.
Now notice that the $\delta-$distribution can be written as a sum of 
two $\delta-$distribution corresponding to the positive and negative
mass shell and choose $f$ to have support in the former.

This example has features rather close to those of general 
relativity.
\item[4.] {\it Maxwell Theory}\\
Consider the action for free Maxwell-theory on Minkowski space and 
perform the Legendre transform. Conclude that there is a first 
class constraint $C=\partial_a E^a$ (Gauss constraint) with
Lagrange multiplier $A_0$ and a Hamiltonian 
$H=\int_{\Rl^3} d^3x [E^a E^b+B^a B^b]/2$ where 
$E^a=\dot{A}_a-\partial_a A_0$ is the electric 
field and $B^a=\epsilon^{abc} \partial_b A_c$ the magnetic one.
Verify that the Gauss constraint generates $U(1)$ gauge transformations 
$A\mapsto A-df$ while $E^a$ is gauge invariant. Choose ${\cal H}_{kin}$
to be the standard Fock space for three massless, free scalar fields 
$A_a$ and as ${\cal D}_{kin},\;{\cal D}_{kin}^\ast$ the finite linear span 
of $n-$particle states and its algebraic dual respectively.
\\
Solution:\\
Conjugate Dirac observables are the transversal parts of $A,E$ 
respectively, e.g.
$E_\perp^a=E^a-\partial_a\frac{1}{\Delta} \partial_b E^b$ 
where $\Delta$ is the Laplacian on $\Rl^3$. The physical Hilbert 
space is the standard Fock space for two free, masless scalar fields
corresponding to these transversal degrees of freedom.\\
Hint:\\
Fourier transform the fields and compute the standard annihilation and 
creation operators $\hat{z}_a(k),\hat{z}_a^\dagger(k)$ with canonical 
commutation relations. Express the Gauss constraint operator in terms 
of them and conclude that the gauge invariant part satisfies 
$\hat{z}_a(k) k^a=0$. Introduce $\hat{z}_I(k)=\hat{z}_a(k) e^a_I(k)$
where $\vec{e}_1(k),\vec{e}_2(k),\vec{e}_3(k):=\vec{k}/||k||$ form an 
oriented 
orthonormal basis. Conclude that physical states are states without 
longitudonal excitations and build the Fock space 
generated by the $\hat{z}^\dagger_1(k),\hat{z}^\dagger_2(k)$ from 
the kinematical vacuum state.
\end{itemize}
\end{Exercise}

\part{ Mathematical and Physical Foundations of Quantum 
General Relativity} 
\label{s2}

\section{Mathematical Foundations}
\label{s2.1}

\subsection{Polarization and Preferred Poisson Algebra $\cal B$}
\label{s2.1.1}

The first two steps of the quantization programme were already completed 
in section \ref{s1.2}: The phase space $\cal M$ is coordinatized by
canonically conjugate pairs $(A_a^j,E^a_j)$ where $A$ is an $SU(2)$
connection over $\sigma$ while $E$ is a $su(2)-$valued vector density of 
weight one over $\sigma$ and the Poisson brackets were 
displayed in (\ref{1.2.17}). Strictly speaking, since $\cal M$
is an infinite dimensional space, one must supply $\cal M$ with
a manifold structure modelled on some Banach space but we will skip these
functional analytic niceties here, see \cite{0} for further information.
Also we must specify the principal fibre bundle of which $A$ is the 
pull-back by local sections of a globally defined connection, and we must
specify the vector bundle associated to that principal bundle under the 
adjoint representation of which $E$ is the pull-back by local sections.
Again, in order not to dive too deeply into fibre bundle theoretic 
subtleties, we will assume that the principal fibre bundle is trivial
so that $A,E$ are actually globally defined. In fact, for the case of 
$G=SU(2)$ and $\dim(\sigma)=3$ one can show that the fibre bundle is 
necessarily trivial but for the generalization to the generic case we 
again refer the reader to \cite{0}.

With these remrks out of the way we may begin by defining a polarization.
The fact that GR has been casted into the language of a gauge theory 
suggests the choice ${\cal C}=\a$, the space of smooth $SU(2)$ connections 
over $\sigma$. 
  
The next question then is how to choose the space $C^\infty(\a)$. 
Since we are dealing with a field theory, it is not clear a priori
what smooth or even differentiable means. In order to give precise meaning
to this, one really has to equip $\a$ with a manifold structure modelled
on a Banach space. This is because one usually says that a function
$F:\a\to \Cl$ is differentiable at $A_0\in \a$ provided that
there exists a {\it bounded} linear functional 
$DF_{A_0}:\; T_{A_0}(\a)\to \Cl$
such that $F[A_0+\delta A]-F[A_0]-DF_{A_0}\cdot \delta A$ vanishes 
``faster than linearly'' for arbitrary tangent vectors 
$\delta A\in T_{A_0}(\a)$ at $A_0$. (The proper way of saying this is 
using the natural Banach norm on $T(\a)$.) Of course, in physicist's 
notation the differential $DF_{A_0}=(\delta F/\delta A)(A_0)$ is nothing
else than the functional derivative. Using this definition 
it is clear that polynomials in $A_a^j(x)$ are not differentiable
because their functional derivative is proportional to a 
$\delta-$distribution as it is clear from (\ref{1.2.17}). Thus we see
that the smooth functions of $A$ have to involve some kind of smearing 
of $A$ with test functions, which is generic in field theories.

Now this smearing should be done in a judicious way. The function\\
$F[A]:=\int_\sigma d^3x F^a_j(x) A_a^j(x)$ for some smooth test 
function $F^a_j$ of compact support is certainly smooth
in the above sense, its functional derivative being equal to $F_a^j$
(which is a bounded operator if $F$ is e.g. an $L_2$ function on $\sigma$
and the norm on the tangent spaces is an $L_2$ norm). However, this 
function does not transform nicely
under $SU(2)$ gauge transformations which will make it very hard to 
construct $SU(2)$ invariant functions from them. Here it helps to look
up how physicists have dealt with this problem in ordinary 
canonical quantum Yang-Mills
gauge theories and they found the following, more or less unique
solution \cite{27}: \\
Given a curve $c:\;[0,1]\to \sigma$ in $\sigma$ and a point $A\in \a$ we 
define the holonomy or parallel transport $A(c):=h_{c,A}(1)\in SU(2)$ as 
the unique 
solution to the following ordinary differential equation for 
functions $h_{c,A}:\;[0,1]\to SU(2)$
\be \label{2.1.1.1}
\frac{d h_{c,A}(t)}{dt}=h_{c,A}(t) A_a^j(c(t))\frac{\tau_j}{2} 
\dot{c}^a(t),\;\;h_{c,A}(0)=1_2      
\ee
\begin{Exercise} \label{ex2.1.1.1}
Verify that (\ref{2.1.1.1}) is equivalent with 
\be \label{ex2.1.1.1a}
A(c)={\cal P}\cdot\exp(\int_c A)=1_2+\sum_{n=1}^\infty
\int_0^t dt_1\int_{t_1}^1 dt_2\;..\;\int_{t_{n-1}}^1
A(t_1)..A(t_n)
\ee
where $\cal P$ denotes the path ordering symbol which orders the 
curve parameters from left to right according to their value 
beginning with the smallest one and 
$A(t):=A_a^j(c(t))\dot{c}^a(t)\tau_j/2$.
\end{Exercise}
With this definition it is not difficult to verify the following 
transformation behaviour of $A(c)$ under gauge transformations and 
spatial diffeomorphisms respectively (recall (\ref{ex1.2.1c})):
\be \label{2.1.1.2}
A^g(c)=g(b(c)) A(c) g(f(c))^{-1} \mbox{ and }
A^\varphi(c)=A(\varphi^{-1}(c))
\ee
where $b(c),f(c)$ denote the beginning and final point of a curve 
respectively.
Thus, the behaviour under gauge transformations is extremely simple
which makes it easy to construct gauge invariant functions, for instance
the Wilson loop functions Tr$(A(c))$ where $c$ is a closed curve, that is,
a {\it loop}. This is the reason why QGR is also denoted as {\it Loop
Quantum Gravity}. That holonomies also transform very naturally 
under spatial diffeomorphisms as depicted in the second equation of 
(\ref{2.1.1.2}) has the following mathematical origin: A connection 
is in particular a one-form, therefore it is {\it naturally integrated
(smeared) over one-dimensional submanifolds of $\sigma$}. Here natural 
means {\it without using a background metric}. Now the holonomy
is not really the exponential of $\int_c A$ but almost as shown
in (\ref{ex2.1.1.1a}). Thus, holonomies are precisely in accordance 
with our wish to construct a {\it background independent} quantum field 
theory. Moreover, the simple transformation behaviour under 
diffeomorphisms again makes it simple to construct spatially 
diffeomorphism invariant functions of holonomies: These will be 
functions only labelled by diffeomporphism invariance classes of 
loops, but these are nothing else than {\it knot classes}. Thus 
QGR has an obvious link with {\it topological quantum field theory (TQFT)}
\cite{28} which makes it especially attractive and was one of the major 
motivations for Jacobson, Rovelli and Smolin to consider Wilson loop 
functions for canonical quantum gravity \cite{29}. Finally one can show
\cite{30} that the holonomies separate the points of $\a$, i.e. they
encode all the information that is contained in a connection.

The fact that the holonomy smears $A$ only one-dimensionally is  
nice due to the above reasons but it is also alarming because
its functional derivative is certainly distributional and thus does not
exist in the strict mathematical sense. However, in order to obtain 
a well-defined Poisson algebra it is not necessary to have smooth 
functions of $A$, it is only sufficient. The key idea idea is that
if we smear also the electric fields $E$ then we might get a 
non-distributional Poisson algebra. By inspection from (\ref{1.2.17})
it is clear that $E$ has to be smeared in at least two dimensions 
in order to achieve this. Now again background independence comes to our 
help: Let $\epsilon_{abc}$ be the totally skew, background independent 
tensor density of weight $-1$, that is, 
$\epsilon_{abc}=\delta_{[a}^1\delta_b^2\delta_{c]}^3$ where $[..]$ denotes
total antisymmetrization. Then 
$(\ast E)^j_{ab}:=E^j_{ab}:=E^c_j\epsilon_{abc}$ 
is a 2-form of density weight $0$. Therefore {\it $E$ is naturally 
smeared in two dimensions}.
Notice that the smearing dimensions of momenta and configuration variables 
add up to the dimension of $\sigma$, they are dual to each other which is 
a generic phenomenon for any canonical theory in any dimension.
We are therefore led to consider the {\it electric fluxes}
\be \label{2.1.1.3}
E_j(S)=\int_S \ast E_j
\ee
where $S$ is a two-dimensional, open surface. It is easy to check that
$E(S):=E_j(S)\tau_j$ has the following transformation behaviour
\be \label{2.1.1.4}
E^g(S)=\int_S \mbox{Ad}_g(\ast E) \mbox{ and }
E^\varphi(S)=E(\varphi^{-1}(S))
\ee
Thus, while the transformation under spatial diffeomorphisms is again 
simple, the one under gauge transformations is not. However, the idea 
is that the $E_j(S)$ are the basic building blocks for more complicated 
functions of $E$ which are already gauge invariant. The prototype
of such a function is the area functional for a parameterized
surface $X_S:\; D\to \sigma,\;D\subset \Rl^2$ 
\be \label{2.1.1.5}
\mbox{Ar}(S):=\int_D d^2u \sqrt{\det(X_S^\ast q)}
\ee
\begin{Exercise} \label{ex2.1.1.2}
Define $n_a^S:=\epsilon_{abc} X^b_{S,u^1} X^c_{S,u^2}$ and verify that
(\ref{2.1.1.5}) coincides with 
\be \label{ex2.1.1.2a}
\mbox{Ar}(S):=\beta \int_D d^2u \sqrt{(E^a_j n_a^S)(E^b_j n_b^S)}
\ee
where $\beta$ is the Immirzi parameter.
\end{Exercise}
It is clear that $E_j(S)=\int_D d^2u E^a_j n_a^S$ so that the area 
functional can be written as the limit of a Riemann sum, over small
surfaces that partition $S$, of functions of the electric fluxes for 
those small surfaces.

\begin{figure}
\includegraphics[width=10cm,height=7cm]{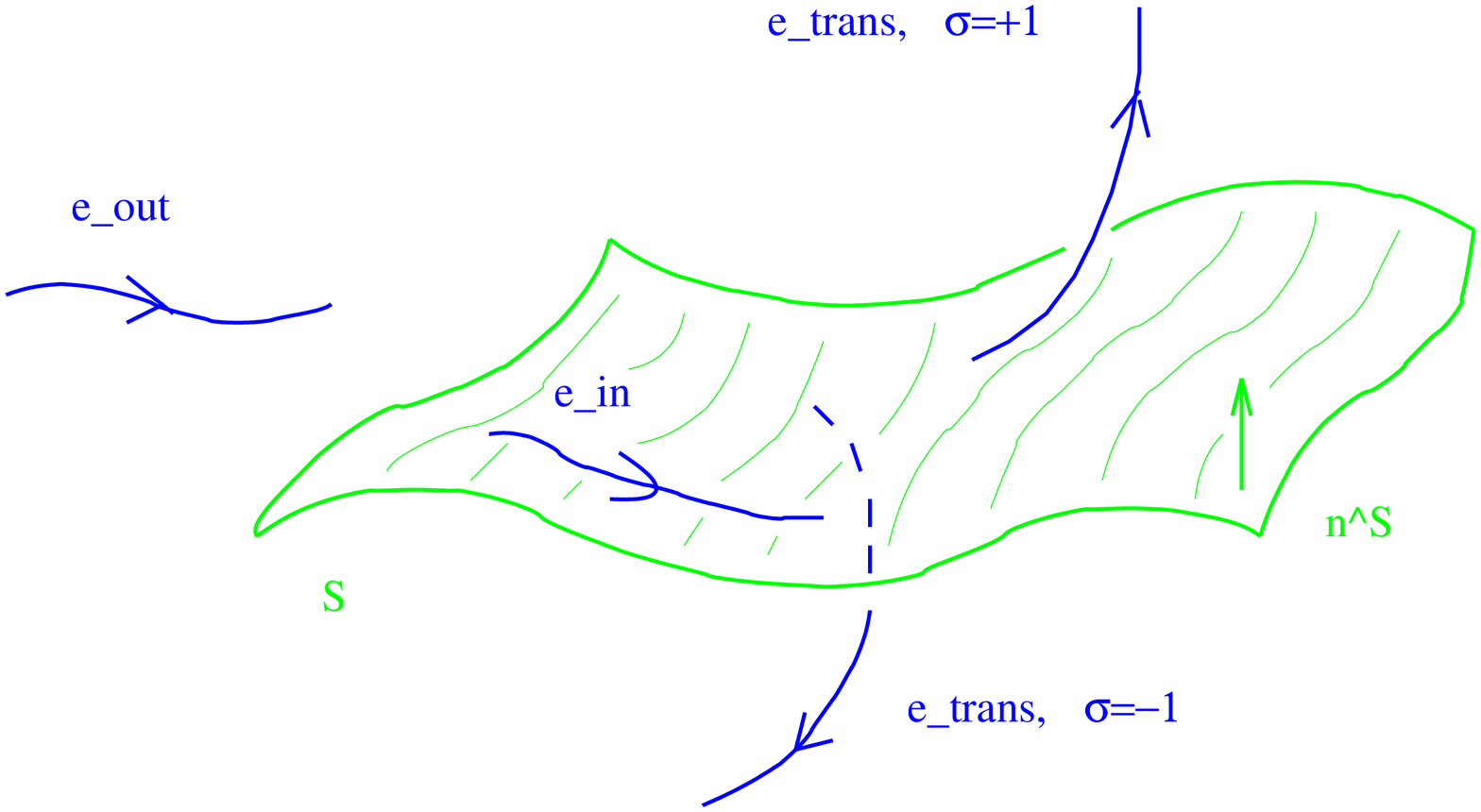}
\caption{Intersection structure of surfaces with edges}
\label{f7}
\end{figure}

Let us see whether the Poisson bracket between an electric flux 
and 
a holonomy is well-defined. Actually, let us be slightly more general and 
introduce the following notion: Let us losely think for the moment of
a graph $\gamma$ as a collection of a finite number of smooth, 
compactly supported, oriented curves, 
called edges $e$, which intersect at most in their end points, which are 
called vertices $v$. We denote by $E(\gamma),\;V(\gamma)$ the edge and 
vertex set of $\gamma$ respectively.
A precise definition will be given in section \ref{s2.1.2}. 
\begin{Definition} \label{def2.1.1.1} ~~\\
Given a graph $\gamma$ we define 
\be \label{2.1.1.6}
p_\gamma:\;\a\to SU(2)^{|E(\gamma)|};\;
A\mapsto (A(e))_{e\in E(\gamma)}
\ee
A function $f:\;\a\to \Cl$ is said to be cylindrical over a graph
$\gamma$, if there exists a function $f_\gamma:\;SU(2)^{|E(\gamma)|}\to\Cl$
such that $f=f_\gamma\circ p_\gamma$. We denote by 
Cyl$^n_\gamma,\;n=0,1,..,\infty$ the set of
$n-$times continuously differentiable cylindrical functions over $\gamma$ 
and by 
Cyl$^n$ the set functions which are cylindrical over {\it some} 
$\gamma$ with the same differentiability type. Here we say that 
$f=f_\gamma\circ p_\gamma\in \mbox{Cyl}^n_\gamma$ if and only if
$f_\gamma$ is $n-$times continuously differentiable with respect to the 
standard differential structure on $SU(2)^{|E(\gamma)|}$.
\end{Definition}
Our Poisson Algebra will be based on the set of functions Cyl$^\infty$
which certainly form an Abelean Poisson subalgebra. Our next task will
be to compute the Poisson bracket between a flux and an element of
Cyl$^\infty$. In order to compute this we will use the chain rule
($f\in\mbox{Cyl}^\infty_\gamma$)
\be \label{2.1.1.7}
\{E_j(S),f\}(A)=\sum_{e\in E(\gamma)} 
[\frac{\partial f_\gamma(\{h_{e'}\}_{e'\in E(\gamma)})}{\partial (h_e)_{AB}}
\{E_j(S),(h_e)_{AB}\}]_{|h_{e'}=A(e')}
\ee
so that the bracket will be well-defined once the bracket between a holonomy
and a flux is well-defined. To compute this the intersection structure 
of $e$ with $S$ is somewhat important. In order to simplify the notation,
we notice that we can always take $\gamma$ to be adapted to $S$, that is,
every edge $e$ belongs to one of the following three types:\\
a) $e\in E_{out}(\gamma)\;\;\Leftrightarrow\;\;e\cap S=\emptyset$.\\
b) $e\in E_{in}(\gamma)\;\;\Leftrightarrow\;\;e\cap S=e$.\\
c) $e\in E_{trans}(\gamma)\;\;\Leftrightarrow\;\;e\cap S=b(e)$.\\
This can be achieved by subdividing edges into a finite number of segments 
and inverting 
their orientation if necessary as depicted in figure \ref{f7}
(strictly speaking, this is true only if $S$ is compactly supported,
open, oriented and analytic).  
We also need to introduce the 
function $\sigma(S,e)$ which vanishes for 
$e\in E_{in}(\gamma)\cup E_{out}(\gamma)$ and which is $\pm 1$ 
for $e\in E_{trans}(\gamma)$ if the 
orientations of $S$ and $e$ agree or disagree respectively.
The easiest case is $e\in E_{trans}(\gamma),\;\sigma(S,e)=1$. 
We find 
\be \label{2.1.1.8}
\{E_j(S),A(e)\}=\kappa\int_D d^2u\;\; n_a^S(u)\int_0^1 ds \dot{e}^a(s)
\delta(X_S(u),e(s)) A(e_s)\frac{\tau_j}{2} A(e_s)^{-1} A(e)
\ee
where $e_s(t):=e(st)$. Noticing that the support of the $\delta-$distribution
is at $X_S(u)=e(0)$ which is an interior point of $S$ but a boundary 
point of $e$, a careful analysis reveals that (\ref{2.1.1.8}) reduces to
\be \label{2.1.1.9}
\{E_j(S),A(e)\}=\frac{\kappa}{4}\tau_j A(e)
\ee
With this result, (\ref{2.1.1.7}) can be written in the compact form
\be \label{2.1.1.10}
\{E_j(S),f\}(A)=\frac{\kappa}{4}\sum_{e\in E(\gamma)}
\sigma(S,e) [R_e^j f_\gamma]_{|h_{e'}=A(e')}
\ee
where we have defined the {\it right invariant vector fields}
\be \label{2.1.1.11}
(R_e^j f_\gamma)(\{h_{e'}\}_{e'\in E(\gamma)}\}):=
(\frac{d}{dt})_{t=0} f_\gamma(\{h_{e'}\}_{e'\not=e},e^{t\tau_j} h_e)
\ee
We can now define the vector fields $v_S^j$ on Cyl$^\infty$ by 
$v_S^j[f]:=\{E_j(S),f\}$ and arrive at the Poisson $^\ast-$algebra
${\cal B}$ generated by the $v^j_S,f\in \mbox{Cyl}^\infty$ with involution 
defined by complex conjugation through the general formula
$[(f,v),(f',v')]=(v[f']-v'[f],[v,v'])$.
\begin{Exercise} \label{ex2.1.1.3}  ~~~~\\
Fill the gaps between (\ref{2.1.1.7}) and (\ref{2.1.1.10}.\\
Hint:\\
Use formula \ref{ex2.1.1.1a} in order to derive (\ref{2.1.1.8}), then 
expand $X_S(u)-e(t)$ around $u=u_0$ defined by $X_S(u_0)=e(0)$ and $t=0$
to linear order in $u-u_0$ and sufficiently high order in $t$
to arrive at (\ref{2.1.1.9}). (Notice that
$e$ is only transversal, so $\dot{e}(0)$ may be tangential to $S$ in 
$e(0)$ !) Verify that the end result coincides with (\ref{2.1.1.10}).
\end{Exercise}
So we see that we arrive at a well defined algebra $\cal B$ by smearing 
the momenta in two dimensions. Could we also smear them in three 
dimensions ? The answer is negative: Consider a one-parameter family 
of surfaces $t\mapsto S_t$ and define $E_j(\{S\}):=\int dt\; E_j(S_t)$.
Then $f\mapsto \{E_j(S),f\}$ maps $f$ out of Cyl$^\infty$ because
it involves an integral over $t$ and thus depends on an uncountably 
infinite number of edges rather than a finite number. Thus this algebra
would not be closed so that if we would like to stick with at least
countably infinite graphs then we are {\it forced} to stick with 
two dimensional smearings of the electric fluxes !

\subsection{Representation Theory of $\cal B$ and 
Suitable Kinematical Representations}
\label{s2.1.2}

The representation Theory of $\cal B$ has been considered only rather
recently \cite{31} and the analysis is not yet complete. However, if one 
sticks to irreducible representations for which 1) the flux operators are 
well-defined and self-adjoint (in other words, the corresponding one 
parameter unitary groups are weakly continuous) and 2) the 
representation is spatially diffeomorphism invariant, then the unique
solution to the representation problem is the representation which we
describe in this section.

This representation is of the form ${\cal H}_0=L_2(\ab,d\mu_0)$ where
$\ab$ is a certain distributional extension of $\a$ and $\mu_0$
is a probability measure thereon. The most elegant description of this
Hilbert space uses the theory of $C^\ast-$algebras \cite{32} but 
fortunately there is a purely geometric description available \cite{33}
which is easier to access for the beginner. In what follows we assume 
for simplicity that $\sigma$ is an oriented, connected, simply connected
smooth manifold. One can show that each smooth manifold admits at least
one analytic structure (i.e. the atlas of charts consists of real analytic
maps) and we assume to have picked one once and for all.

\subsubsection{Curves, Paths, Graphs and Groupoids}
\label{s.2.1.2.1}

\begin{figure} 
\includegraphics[width=14cm,height=3.5cm]{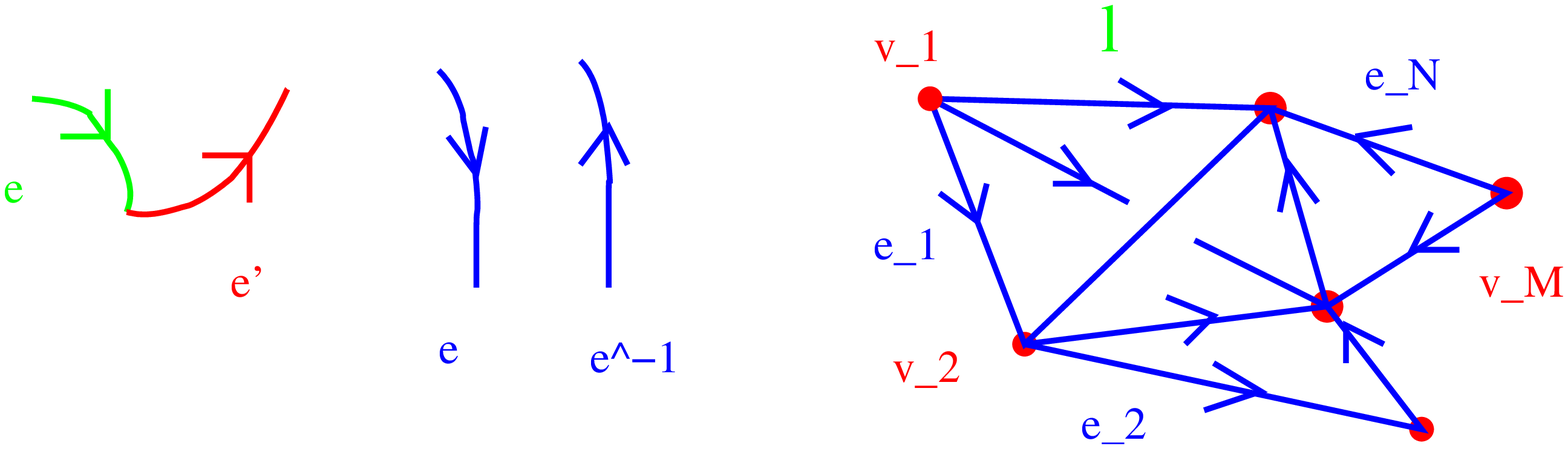}
\caption{Paths and graphs}
\label{f8}
\end{figure}

\begin{Definition} \label{def2.1.2.1} ~~~~~\\
i)\\
By a curve $c$ we mean a map $c:\;[0,1]\to \sigma$ which is piecewise 
analytic, continuous,
oriented and an embedding (does not come arbitrarily close to itself).
It is automatically compactly supported.
The set of curves is denoted $\cal C$ in what follows.\\
ii)\\
On $\cal C$ we define maps $\circ,\;(.)^{-1}$ called composition and 
inversion respectively by 
\be \label{2.1.2.1}
[c_1\circ c_2](t)= \left\{ \begin{array}{cc}
c_1(2t) &  t\in [0,\frac{1}{2}] \\
c_2(2t-1) &  t\in [\frac{1}{2},1] 
\end{array} \right.
\ee
if $f(c_1)=b(c_2)$ and 
\be \label{2.1.2.2}
c^{-1}(t)=c(1-2t)
\ee
iii)\\
By a path $p$ we mean an equivalence class of curves $c$ which differ
from each other by a finite number of reparameterizations and retracings, 
that is, $c\sim c'$ if there either exists a map   
$t\mapsto f(t),\;\dot{f}(t)>0$ with $c=c'\circ f$ or we may write 
$c,c'$ as compositions of segments in the form $c=s_1\circ s_2,\; 
c'=s_1\circ s_3\circ s_3^{-1}\circ s_2$ (and finite combinations of 
such moves). Notice that a curve induces its orientation and its end 
points on its 
corresponding path. The set of paths is denoted by $\cal P$. \\
iv)\\
By a graph $\gamma$ we mean a finite collection of elements of $\cal P$.
We may break paths into pieces such that $\gamma$ can be thought of
as a collection of edges $e\in E(\gamma)$, that is, paths which have an 
entire analytic
representative and which intersect at most in their end points $v\in 
V(\gamma)$ called vertices. The set
of graphs is denoted by $\Gamma$.
\end{Definition}
These objects are depicted in figure \ref{f8}.
\begin{Exercise} \label{ex2.1.2.1} ~~~\\
a)\\
Despite the name, composition and inversion does not equip $\cal C$
with a group structure for many reasons. Verify that composition is 
not associative and that the curve $c\circ c^{-1}$ is not simply 
$b(c)$ but rather a retracing. Moreover, contemplate that $\cal C$ 
does not have a unit and that not every two elements can be composed.\\
b)\\
Define composition and inversion of paths by taking the equivalence class 
of the compositions and inversions of any of their representatives and 
check that this definition is well defined. Check that then
composition of paths is associtive and that $p\circ p^{-1}=b(p)$.
However, $\cal P$ still does not have a unit and still not every
two elements can be composed.\\
c)\\
Let Obj$:=\sigma$ and for each $x,y\in\Sigma$ let 
$\mbox{Mor}(x,y):=\{p\in {\cal P}:\;b(p)=x,\;f(p)=y\}$. Recall the 
mathematical definition of a category and conclude that $\cal P$
is a category in which every morphism is invertible, that is, 
a {\it groupoid}.\\
d)\\
Define the relation $\prec$ on $\Gamma$ by saying that 
$\gamma\prec\gamma'$ if and only if every $e\in E(\gamma)$ is a finite 
composition of the $e'\in E(\gamma')$ and their inverses. Verify that
$\prec$ equips $\Gamma$ with the structure of a directed set, that is,
for each $\gamma,\gamma'\in\Gamma$ we find $\gamma^{\prime\prime}\in 
\Gamma$ such that $\gamma,\gamma'\prec \gamma^{\prime\prime}$. \\
Hint:\\
For this to work, analyticity of the curve representatives is crucial.
Smooth curves can intersect in Cantor sets and thus define graphs which 
are no longer finitely generated. Show first that this is not possible for 
analytic curves.\\
e)\\
Given a curve $c$ with path equivalence class $p$ notice that for 
the holonomy with respect to $A\in \a$ holds $A(c)=A(p)$.
Contemplate that, in particular, every group is a groupoid and that every 
connection $A\in \a$ qualifies as a groupoid homomorphism, that is,
$A:\;{\cal P}\to SU(2);\;p\mapsto A(p)$ with 
\be \label{ex2.1.2.1a}
A(p\circ p')=A(p)A(p') \mbox{ and } A(p^{-1})=(A(p))^{-1}
\ee
\end{Exercise}
The fact that holonomies are really defined on paths rather than curves 
and that holonomies are characterized algebraically by \ref{ex2.1.2.1a}
makes the following definition rather natural.
\begin{Definition} \label{def2.1.2.2}
The quantum configuration space is defined as the set
$\ab:=\mbox{Hom}({\cal P},SU(2))$ of all algebraic, arbitrarily 
non-continuous groupoid morphisms.
\end{Definition}
Here non-continuous means that in contrast to $A\in \a$ for an element
$A\in \ab$ it is possible that $A(p)=1$ varies discontinuously as we vary 
$p$ continuously. Thus, $\ab$ can be thought of as a distributional 
extension of $\a$.\\

\subsubsection{Topology on $\ab$}
\label{s2.1.2.2}

So far $\ab$ is just a set. We now equip it with a topology.
The idea is actually quite simple. Recall the maps (\ref{2.1.1.6})
which easily extend from $\a$ to $\ab$ and maps  
$\ab$ into $SU(2)^{|E(\gamma)|}$. Now $SU(2)^{|E(\gamma)|}$ is a compact 
Hausdorff topological group\footnote{Here it is crucial that $G=SU(2)$ is 
compact and thus for non-real Immirzi parameter all of what follows 
would be false.} in its natural manifold topology and we 
would like to exploit that.  Thus we are motivated to consider the spaces 
$X_\gamma:=\mbox{Hom}(\gamma,SU(2)^{|E(\gamma)|})$ where $\gamma$ is 
considered as a subgroupoid of $\Gamma$ with objects $V(\gamma)$ and 
morphisms generated by the $e\in E(\gamma)$. The map
\be \label{2.1.2.3} 
X_\gamma\to SU(2)^{|E(\gamma)|};\; x_\gamma\mapsto \{x_\gamma(e)\}_{e\in 
E(\gamma)}
\ee
identifies $X_\gamma$ with $G^{|E(\gamma)|}$ since $x_\gamma\in X_\gamma$
is already defined by which values it takes on the $e\in E(\gamma)$ and
we may thus use this identification in order to equip $X_\gamma$ with a 
compact Hausdorff topology. Now consider the uncountably infinite 
product
\be \label{2.1.2.4}
X_\infty:=\prod_{\gamma\in \Gamma} X_\gamma
\ee
A standard result from general topology, Tychonov's theorem, tells 
us that the smallest topology on $X_\infty$ such that all the maps
$p_\gamma:\;X_\infty\to X_\gamma;\;(x_\gamma)_{\gamma\in \Gamma}\mapsto 
x_\gamma$ are continuous is a compact Hausdorff topology\footnote{Recall
that we know the topology on a space when we know a base of open sets
from which we obtain all open sets by arbitrary unions and finite 
intersections. Since the preimages of open sets under continuous functions
are open by definition, we obtain a topology once we know which functions 
are continuous.}. Now we would like to identify $\ab$ with $X_\infty$ through
the restriction map
\be \label{2.1.2.5}
\Phi':\;\ab \to X_\infty;\;A\mapsto 
(x_\gamma:=A_{|\gamma}=p_\gamma(A))_{\gamma\in 
\Gamma}
\ee 
However, that map cannot be surjective because the points of $\ab$ satisfy
the following constraint which encodes the algebraic properties of a 
generalized connection: Let $\gamma\prec \gamma'$ and define the
graph restriction maps 
\be \label{2.1.2.6}
p_{\gamma'\gamma}:\;X_{\gamma'}\to X_\gamma;\;x_{\gamma'}\mapsto 
(x_{\gamma'})_{|\gamma}
\ee
which satisfy the compatibility condition
\be \label{2.1.2.6a}
p_{\gamma^{\prime\prime}\gamma}=
p_{\gamma'\gamma}\circ p_{\gamma^{\prime\prime}\gamma'} \mbox{ for } 
\gamma\prec\gamma'\prec\gamma^{\prime\prime}
\ee
Then automatically
\be \label{2.1.2.7}
p_{\gamma'\gamma}(A_{|\gamma'})=A_{|\gamma}
\ee
We are therefore forced to consider the subset of $X_\infty$ defined by
\be \label{2.1.2.8}
\overline{X}:=\{(x_\gamma)_{\gamma\in\Gamma}\in X_\infty;\;
p_{\gamma'\gamma}(x_\gamma')=x_\gamma\;\forall\;\gamma\prec\gamma'\}
\ee
\begin{Exercise} \label{ex2.1.2.2}   ~~~~\\
i)\\
Show that the maps (\ref{2.1.2.6}) are continuous surjections.\\
Hint:\\
Exploit the identification of the $X_\gamma$ with powers of $SU(2)$ and 
the continuity of multiplication and inversion in groups to establish 
continuity. To establish surjectivity use the fact that each edge $e$ of
$\gamma$ contains an edge $e'_e$ of $\gamma'$ as a segment such that 
the $e'_e$ do not overlap each other. Now given $x_\gamma$ set 
$x_{\gamma'}(e'_e)=x_\gamma(e)$ and extend trivially away from the $e'_e$.
Check that this defines an element of $X_{\gamma'}$.\\
ii)\\
Show that $\overline{X}$ is a closed subset of $X_\infty$.\\
Hint:\\
Since $\overline{X}$ is not a metric space we must work with nets and 
show that every net of points $x^\alpha\in \overline{X}$ which converges 
in $X_\infty$ actually converges in $\overline{X}$. Using the defintion of 
the topology on $X_\infty$, show that this is equivalent to showing that
the $p_\gamma(x^\alpha)=x_\gamma^\alpha$ converge to points $x_\gamma$
which satisfy (\ref{2.1.2.7}) and verify this using continuity of the 
$p_{\gamma'\gamma}$ just established.
\end{Exercise}
The surjectivity of the $p_{\gamma'\gamma}$ qualifies $\overline{X}$
as the so-called projective limit of the $X_\gamma$, a mathematical 
structure which is independent of our concrete context once we have 
a directed index set $\Gamma$ at our disposal and surjective projections
which satisfy the compatibility condition (\ref{2.1.2.6a}).

Now another standard result from topology now tells us that 
$\overline{X}$,
being the closed subset of a compact Hausdorff space, is a compact
Hausdorff space in the subspace topology and the question arises whether
\be \label{2.1.2.9}
\Phi:\;\ab \to \overline{X};\;A\mapsto 
(x_\gamma:=A_{|\gamma}=p_\gamma(A))_{\gamma\in 
\Gamma}
\ee 
is a bijection. Injectivity is fairly easy to see while surjectivity    
is a little bit tricky.
\begin{Exercise} \label{ex2.1.2.3} ~~~~~\\
Show that (\ref{2.1.2.9}) is a bijection.\\
Hint:\\
Given $x\in \overline{X}$ and $p\in {\cal P}$ choose any 
$\gamma_p\in\Gamma$ such that $p\in \gamma_p$ and define $A_x$ by
$A_x(p):=x_{\gamma_p}(p)$. Show that this definition is well defined 
using the directedness of $\Gamma$ and that $A_x$ is a groupoid 
homomorphism.
\end{Exercise}
Let us collect these results in the following theorem \cite{33a}.
\begin{Theorem}
\label{th2.1.2.1}  ~~~~\\
The space $\ab$ equipped with the weakest topology such that the 
maps $p_\gamma$ of (\ref{2.1.1.6}) are continuous, is a compact Hausdorff 
space.
\end{Theorem}
The value of this result is that it gives us a powerful tool for 
constructing measures on $\ab$.

\subsubsection{Measures on $\ab$}
\label{s2.1.2.3}

A powerful theorem due to Riesz and Markov, sometimews called the 
Riesz representation theorem, tells us that there is a one -- to -- one 
correspondence between the positive linear functionals $\Lambda$ on 
the algebra $C(\ab)$ of continuous functions on a compact Hausdorff space 
$\ab$ and (regular, Borel) probability measures $\mu$ thereon through the 
simple formula
\be \label{2.1.2.10}
\Lambda(f):=\int_{\ab} \;d\mu(A)\;f(A)
\ee
One says $\Lambda$ is represented by $f$. Here a linear functional is 
called positive if $\Lambda(|f|^2)\ge 0$ for any $f\in C(\ab)$. 
A function algebra on a compact space can be equipped with the 
sup -- norm $||f||:=\sup_{A\in \ab} |f(A)|$ which evidently 
has the so-called $C^\ast-$property $||f \overline{f}||=||f||^2$
so that (w.l.g. we may take $C(\ab)$ to be complete w.r.t. the norm)
$C(\ab)$ is a $C^\ast-$algebra. A standard
result in functional analysis reveals that positive linear functionals 
on $C^\ast-$algebras are automatically continuous, $|\Lambda(f)|\le
\Lambda(1)\;||f||$ and if we choose the normalization of $\Lambda$ 
to be $\Lambda(1)=1$ then $\mu$ is a probability measure.

\begin{figure} 
\includegraphics[width=16cm,height=8cm]{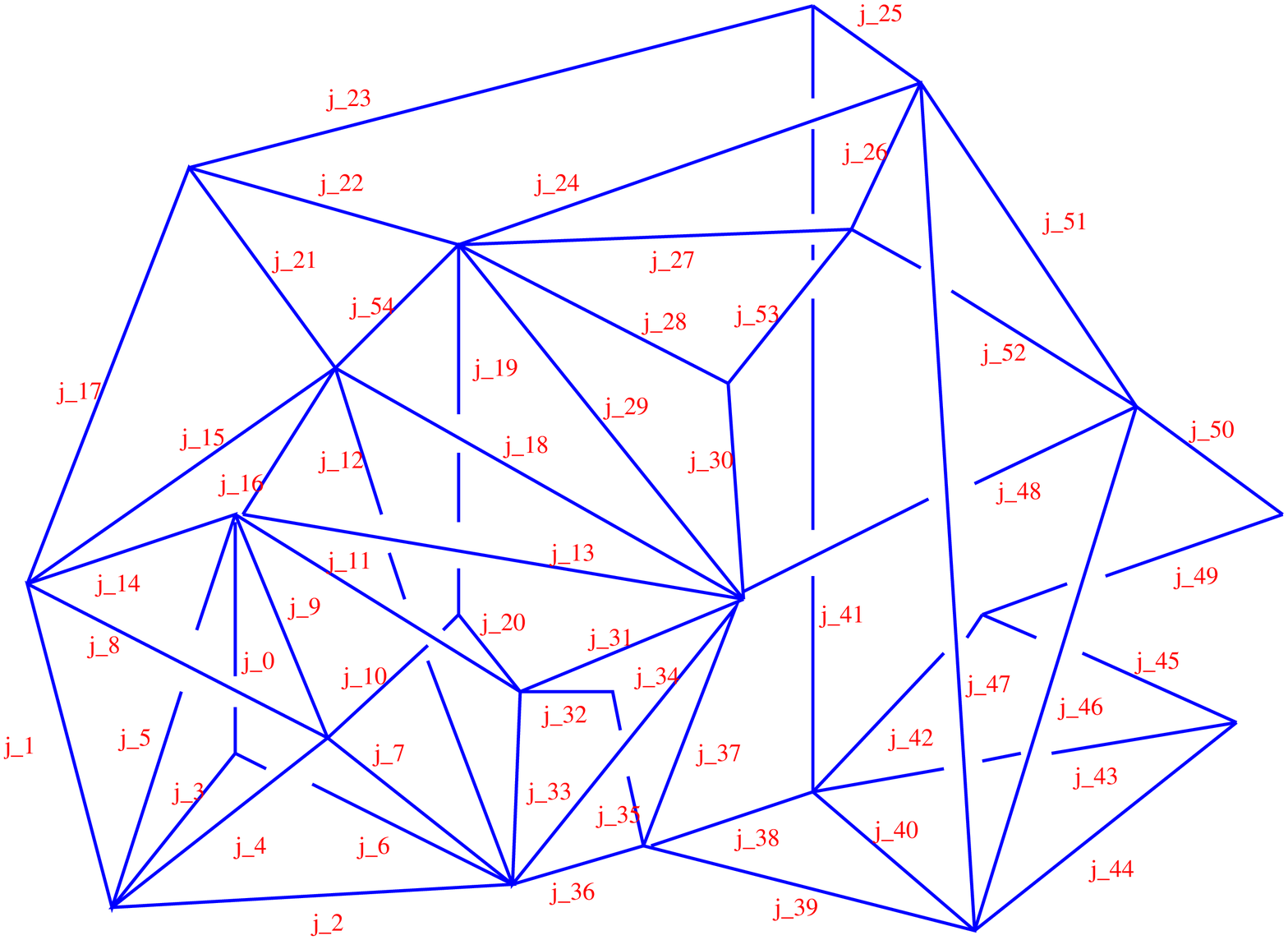}
\caption{A SNW. Orientations and magnetic quantum numbers are suppressed}
\label{f8a}
\end{figure}

In order to specify the measure $\mu_0$ that we are interested in, it is 
therefore enough to specify a positive linear functional $\Lambda_0$. The 
most elegant way of defining $\Lambda_0$ is through the following 
definition.
\begin{Definition} \label{def2.1.2.2a} ~~~~~\\
i)\\
Given a graph $\gamma$, label each edge $e\in E(\gamma)$ with a triple 
of numbers $(j_e,m_e,n_e)$ where 
$j_e\in\{\frac{1}{2},1,\frac{3}{2},2,..\}$ is a half-integral spin quantum 
number and $m_e,n_e\in\{-j_e,-j_e+1,..,j_e\}$ are magnetic quantum 
numbers. A quadruple 
\be \label{2.1.2.11}
s:=(\gamma,
\vec{j}:=\{j_e\}_{e\in E(\gamma)},
\vec{m}:=\{m_e\}_{e\in E(\gamma)},
\vec{n}:=\{n_e\}_{e\in E(\gamma)})
\ee
is called a spin network (SNW). We also write $\gamma(s)$ etc. for the 
entries of a SNW.\\
ii)\\
Choose once and for all one representative $\rho_j,\;j>0$ half integral, 
from 
each equivalence class of irreducible representations of $SU(2)$. Then 
\be \label{2.1.2.12}
T_s:\;\ab\to \Cl;\;A\mapsto
\prod_{e\in E(\gamma)} [\sqrt{2j_e+1} [\rho_{j_e}(A(e))]_{m_e n_e}]
\ee
is called the spin-network function (SNWF) of $s$. Here 
$[\rho_j(.)]_{mn}$
denotes the matrix elements of the matrix valued function $\rho_j(.)$.
\end{Definition}
An example of a SNW, which can be arbitrarily large and with vertices of
arbitrarily high valence, is given in figure \ref{f8a}.
The original motivation for the definition of spin network functions
\cite{34} 
in loop quantum gravity was the fact that they are linearly independent
in contrast to the Wilson loop functions which suffer from the so-called 
{\it Mandelstam identities}. For $SU(2)$ matrices $h,h'$ they are 
$\mbox{Tr}(h)\;\mbox{Tr}(h')=\mbox{Tr}(hh')+\mbox{Tr}(h(h')^{-1})$
and $\mbox{Tr}(h)=\mbox{Tr}(h^{-1})$ which leads to an infinite 
tower of identities of the 
form 
\be \label{2.1.2.13}
[\mbox{Tr}(A(\alpha_1))\;\mbox{Tr}(A(\alpha_2))]\mbox{Tr}(A(\alpha_1))
=\mbox{Tr}(A(\alpha_1))[\mbox{Tr}(A(\alpha_2))\;\mbox{Tr}(A(\alpha_1))]
\ee
depending on how we bracket the product of traces involving the three
loops $\alpha_1,\alpha_2,\alpha_3$ with a common base point. The  
SNWF's remove these cumbersome identities first by labelling functions 
by edges rather than loops and secondly by the simple observation
that a tensor product of (fundamental) representations can be uniquely
decomposed into irreducibles (Clebsh-Gordon decomposition).
\begin{Theorem}
\label{th2.1.2.1a} ~~~~~~~~~\\
The uniform (Ashtekar -- Lewandowski) measure $\mu_0$ is uniquely defined 
by the positive linear functional \cite{33b}
\be \label{2.1.2.14} ~~~~~\\
\Lambda_0(T_s):=\left\{ \begin{array}{cc} 
1 & s=(\emptyset,\vec{0},\vec{0},\vec{0})\\
0 & \mbox{otherwise}
\end{array} \right.
\ee
\end{Theorem}
\begin{Exercise} \label{ex2.1.2.4} ~~~~~\\
i)\\
Recall the representation theory of $SU(2)$ from the quantum mechanics of 
angular momentum and verify that the SNWF are indeed linearly 
independent.\\
ii)\\
Verify that $\Lambda_0$ is a positive linear functional.\\
Hint:\\
Using the Stone -- Weierstrass theorem, show first that the finite linear
combinations of SNWF are dense in $C(\ab)$. By continuity of $\Lambda_0$ 
it is therefore sufficient to check positivity on finite linear 
combinations 
\be \label{2.1.2.15} ~~~~~\\
f=\sum_{n=1}^N z_n T_{s_n},\;\;N<\infty,\;z_n\in\Cl
\ee
with $s_n$ mutually different SNW's. To see this, verify that 
$\Lambda_0(\overline{T_s} T_{s'})=0$ for $s\not=s'$ by using the 
Clebsh -- Gordon formula $j\otimes j'\equiv (j+j')\oplus (j+j'-1)\oplus ..
\oplus (|j-j'|)$.\\
iii)\\
For the representation theory of compact groups fundamental is a theorem 
due to Peter and Weyl \cite{35} which for $SU(2)$ amounts to saying that 
the functions 
\be \label{2.1.2.16}
T_{jmn}:\;SU(2)\to \Cl;\;h\mapsto \sqrt{2j+1} [\rho_j(h)]_{mn}
\ee
form an orthonormal basis for the Hilbert space $L_2(SU(2),d\mu_H)$
where $\mu_H$ is the normalized Haar measure on $SU(2)$ (the unique
normalized 
measure which invariant under inversion as well as left and right 
translation in $SU(2)$).
Based on this result, show that the SNWF form an orthonormal basis 
for the Hilbert space $L_2(\ab,d\mu_0)$.
\end{Exercise}
Let us summarize the results of the exercise in the following 
theorem \cite{34}.
\begin{Theorem}
\label{th2.1.2.2} 
~~~~~\\ The kinematical Hilbert space ${\cal H}_{kin}:=L_2(\ab,d\mu_0)$ 
defined by (\ref{2.1.2.14}) is non-separable and has the SNWF's
$T_s$ as orthonormal basis.
\end{Theorem}

\subsubsection{Representation Property}
\label{s2.1.2.4}

So far we did not verify that ${\cal H}_{kin}$ is a representation space 
for our 
$^\ast-$algebra $\cal B$ of basic operators. This will be done in the 
present section. Indeed, until today no other irreducible 
representation of the holonomy -- flux algebra has been found (except 
if one allows also infinite graphs \cite{36}).

By theorem (\ref{th2.1.2.3}) the subspace of finite
linear combinations of SNWF's is dense in ${\cal H}_{kin}$ with respect to 
the $L_2$ norm. On the other hand, we notice that the definition of 
Cyl$^\infty(\a)$ simply extends to Cyl$^\infty(\ab)$ and that finite linear
combinations of SNWF's form a subspace of 
$\mbox{Cyl}^\infty(\ab)$. Thus, we may choose 
${\cal D}_{kin}:=\mbox{Cyl}^\infty(\ab)$ and obtain a dense, invariant 
domain of $\cal B$ as we will see shortly. We define 
a representation of the holonomy -- flux algebra by 
($f'\in \mbox{Cyl}^\infty(\a),\; f\in \mbox{Cyl}^\infty(\ab),\;A\in \ab$) 
\ba \label{2.1.2.17}
{[}\pi(f)\cdot f'](A) &:=& (f'f)(A) \nonumber\\
{[}\pi(v_S^j)\cdot f](A) &=& 
{[}\pi(v_S^j)\pi(f)\cdot 1](A)
=[([\pi(v_S^j),\pi(f)]+\pi(f)\pi(E_j(S)))\cdot 1](A)
\nonumber\\
&:=& i\hbar[\pi(v_S^j[f]) \cdot 1](A)
=i\hbar (v_S^j[f])(A)
\ea
Thus $\pi(f)$ is a multiplication operator while $\pi(v_S^j)$ is a true 
derivative operator, i.e. it annihilates constants.
Notice that the canonical commutation relations are already obeyed by 
construction, thus we only need to verify the $^\ast-$relations and the 
fact that $\pi(v_S^j)$ annihilates constants will be crucial for that.

The $\pi(f)$ are bounded multiplication operators (recall that smooth, 
i.e. in particular continuous, functions on copmpact spaces are uniformly
bounded, that is, have a sup -- norm) so that the adjoint is 
complex conjugation, therefore there is nothing to check. As for 
$\pi(v_S^j)$ we notice that given two smooth cylindrical functions 
on $\ab$ we always find a graph $\gamma$ over which both of them are 
cylindrical and which is already adapted to $S$. 
\begin{Exercise} \label{ex2.1.2.5} ~~~\\
Let $f$ be cylindrical over $\gamma$. Verify that 
\be \label{2.1.2.18}
\Lambda_0(f)=\int_{SU(2)^{|E(\gamma)|}} \prod_{e\in E(\gamma)} d\mu_H(h_e)
f_\gamma(\{h_e\}_{e\in E(\gamma)})
\ee
Hint:\\
Write $f$ as a (Cauchy limit of) finite linear combinations of SNWF's
and verify that (\ref{2.1.2.18}) coincides with    
(\ref{2.1.2.14}).
\end{Exercise}
Using the explicit expression (\ref{2.1.1.10}) and the result of 
exercise \ref{ex2.1.2.5} it is easy to see that 
the symmetry condition $<f,\pi(v_S^j)f'>_{kin}=<\pi(v_S^j) f,f'>_{kin}$
is equivalent with the condition
\be \label{2.1.2.19} 
<F,R^j F'>_{L_2(SU(2),d\mu_H)}
=-<R^j F, F'>_{L_2(SU(2),d\mu_H)}
\ee
for any $F,F'\in C^\infty(SU(2))$ and $R^j$ is the right invariant vector 
field on $SU(2)$. However, $\mu_H$ is by definition 
invariant under left translations and $R^j$ is a generator of left 
translations in $SU(2)$ so the result follows. This shows that 
${\cal D}_{kin}$ is contained in the domain of $\pi(v_S^j)^\dagger$
and that the restriction of the adjoint to ${\cal D}_{kin}$ coincides
with $\pi(v_S^j)$. That ${\cal D}_{kin}$ is actually a domain of 
(essential) self-adjointness requires a little bit more work but is 
not difficult to see, e.g. \cite{0}. 

Finally, let us verify that the representation is irreducible.
By definition, a representation is irreducible if every vector is cyclic
and a vector $\Omega$ is cyclic if the set of vectors $\pi(a)\Omega,\;
a\in {\cal B}$ is dense. Now the vector $\Omega=1$ is cyclic because 
the vectors $\pi(f)\Omega=f,\;f\in \mbox{Cyl}^\infty$ are already dense.
Given an arbitrary element $\psi\in {\cal H}_{kin}$ we know that it is a 
Cauchy limit of finite linear combinations of spin network functions.
Thus, if we can show that we find a sequence $a_n\in {\cal B}$ such 
that $\pi(a_n)\psi$ converges to $\Omega$, then we are done. It is easy to 
see (exercise) that this problem is equivalent to showing that any $F\in 
L_2(F,d\mu_H)$ can be mapped by the algebra formed out of right 
invariant vector fields and smooth functions on $SU(2)$ to the constant 
function.  
\begin{Exercise} \label{ex2.1.2.6} ~~~~\\
Check that this is indeed the case.\\
Hint:\\
Show first that it is sufficient to establish that any polynomial
$p$ of the $a,b,c,d,\;ad-bc=1$ for 
$h=\left( \begin{array}{cc} a & b\\ c & d \end{array} \right)\in SU(2)$
can be mapped to the constant function. Show then that suitable linear
combinations of the $R^j,\,j=1,2,3$ with coefficients in $C^\infty(SU(2))$
produce the derivatives $\partial_a,\partial_b,\partial_c$ and convince 
yourself that $a^N p$ is a polynomial in $a,b,c$ for sufficiently large 
$N$.
\end{Exercise}
~\\
Let us collect these results in the following theorem \cite{36a}.
\begin{Theorem}
\label{th2.1.2.3} ~~~~~~\\
The relations (\ref{2.1.2.17}) define an irreducible representation of 
$\cal B$ on ${\cal H}_{kin}$.
\end{Theorem}
Thus, the representation space ${\cal H}_{kin}$ constructed 
satisfies all the requirements that qualify it as a good kinematical
starting point for solving the quantum constraints. Moreover, the 
measure $\mu_0$ is spatially diffeomorphism invariant as we will see 
shortly and together with the uniqueness result quoted at the beginning
of this section, this is the only representation with that property.
There are, however, doubts on physical grounds whether one should insist 
on a spatially diffeomorphism invariant representation because the 
smooth and 
even analytic structure of $\sigma$ which is encoded in the spatial 
diffeomorphism group should not play a fundamental role at short scales
if Planck scale physics is fundamentally discrete. In fact, as we will 
see later, QGR predicts a discrete Planck scale structure and therefore 
the fact that we started with analytic data and ended up with discrete 
(discontinuous) spectra of operators looks awkward. Therefore, on the one 
hand we should keep in mind that other representations are possibly 
better suited in the final picture, on the other hand there is no logical 
contradiction within the present formulation and in fact in 2+1 gravity 
one has a final combinatorical description while one started with 
analytical structures as well.

\section{Quantum Kinematics}
\label{s2.2}

In this section we discuss the complete solution of the Gauss and Vector 
constraint as well as the quantization of kinematical, geometrical operators 
that measure the length, area and volume of {\it coordinate} curves, 
surfaces and regions respectively. We call these results {\it kinematical}
because the Gauss and Vector constraint do not generate dynamics, this 
is the role of the Hamiltonian constraint which we will discuss in the 
third part. Moreover, the kinematical geometrical operators do not
commute with the Vector constraint or the Hamiltonian constraint and are 
therefore not Dirac observables. However, as we will show, one can turn
these operators easily into Dirac observables, at least with respect to the 
Vector constraint, and the fact that the spectrum is discrete is robust under
those changes. 

\subsection{The Space of Solutions to the Gauss and Spatial
Diffeomorphism Constraint}
\label{s2.2.1}

\begin{figure} 
\includegraphics[width=14cm,height=3cm]{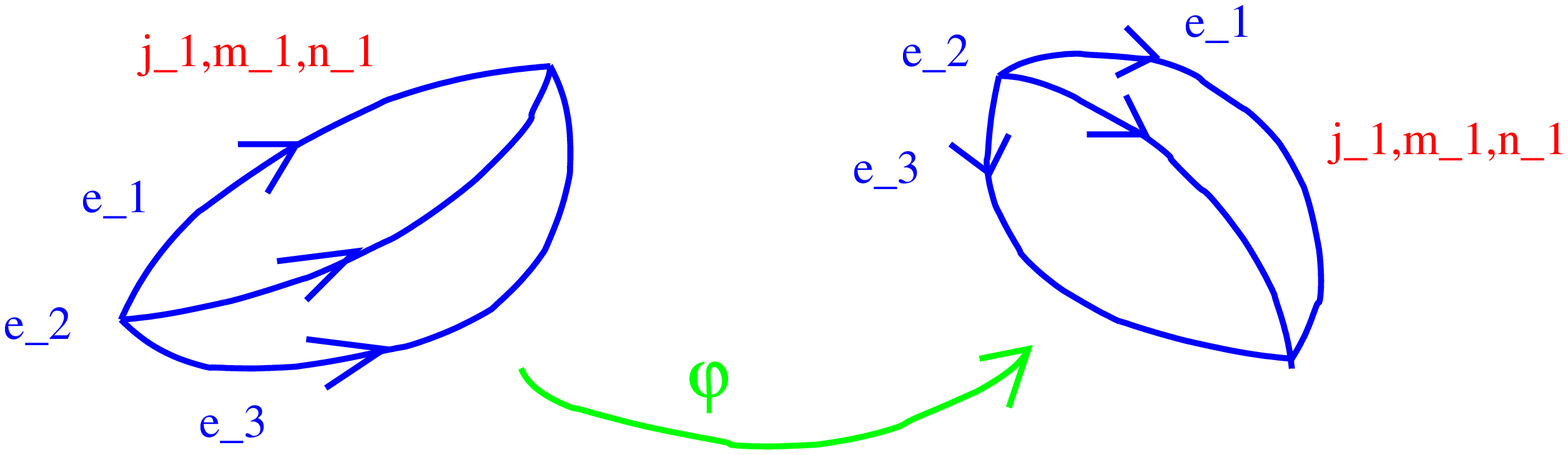}
\caption{Action of Spatial Diffeomorphisms on SNW's}
\label{f9}
\end{figure}

Recall the transformation behaviour of classical connections $A\in\a$ 
under $SU(2)$ gauge transformations and spatial diffeomorphisms
(\ref{2.1.1.2}). These equations trivially lift from $\a$ to $\ab$
and we may construct corresponding operators as follows: Let 
$\overline{{\cal G}}:=\mbox{Fun}(\Sigma,SU(2))$ be the set of local 
gauge transformations without continuity requirement and consider the set
$\mbox{Diff}^\omega(\Sigma)$ of {\it analytic} diffeomorphisms. We are 
forced to consider analytic diffeomorphisms as otherwise we would destroy 
the analyticity of the elements of $\Gamma$. These two groups have a natural
semi -- direct product structure that has its origin in the algebra 
(\ref{ex1.2.1a}) and is given by
\be \label{2.2.1.0}
[\overline{{\cal G}}\rhd \mbox{Diff}^\omega(\Sigma)]\times
[\overline{{\cal G}}\rhd \mbox{Diff}^\omega(\Sigma)]\to
[\overline{{\cal G}}\rhd \mbox{Diff}^\omega(\Sigma)];\;\;
[g,\varphi]\cdot[g',\varphi']=[(g'\circ\varphi^{-1})g,\varphi\circ\varphi']
\ee
\begin{Exercise} \label{ex2.2.1.1}  ~~~~\\
Verify (\ref{2.2.1.0}).\\
Hint:\\
Define $[g,\mbox{id}]\cdot A:=A^g,\;[\mbox{id},\varphi]\cdot A:=A^\varphi$
and $[g,\varphi]\cdot A:=[g,\mbox{id}]\cdot([\mbox{id},\varphi]\cdot A)$.
\end{Exercise}
We now define representations 
\ba \label{2.2.1.1}
&& \hat{U}:\;\overline{{\cal G}}\to {\cal B}({\cal H}_{kin});\;
g\mapsto \hat{U}(g) \nonumber\\
&& \hat{V}:\;\mbox{Diff}^\omega(\Sigma)\to {\cal B}({\cal H}_{kin});\;
\varphi\mapsto \hat{V}(\varphi)
\ea
densely on $f=p_\gamma^\ast f_\gamma\in {\cal D}_{kin}$ by
\ba \label{2.2.1.2}
[\hat{U}(g) f](A) &:=& 
f_\gamma(\{g(b(e))\;A(e)\;g(f(e))^{-1}\}_{e\in E(\gamma)}\})
\nonumber\\
{[}\hat{V}(\varphi) f](A) &:=& 
f_\gamma(\{A(\varphi^{-1}(e))\}_{e\in E(\gamma)}\})
\ea
Here ${\cal B}(.)$ denotes the bounded operators on a Hilbert space.
This definition of course comes precisely from the classical formula
(\ref{2.1.1.2}). The action of a diffeomorphism on a SNWF $T_s$ is therefore 
simply by mapping the graph $\gamma(s)$ to $\varphi^{-1}(s)$ while the 
labels $j_e,m_e,n_e$ are carried from $e$ to $\varphi^{-1}(e)$ as 
depicted in figure \ref{f9}. 
Then the following theorem holds \cite{36a}.
\begin{Theorem}
\label{th2.2.1.1} ~~~~\\
The relations (\ref{2.2.1.2}) define a unitary representation of the 
semi -- direct product kinematical group 
$\overline{{\cal G}}\times \mbox{Diff}^\omega(\Sigma)$.
\end{Theorem}
\begin{Exercise} \label{ex2.2.1.2}  ~~~~\\
Prove theorem (\ref{th2.2.1.1}).\\
Hint:\\
Check unitarity on the SNWF basis using the bi -- invariance of the Haar
measure. That (\ref{2.2.1.2}) holds can be traced back to exercise 
\ref{ex2.2.1.1}.
\end{Exercise}
The unitarity of the kinematical gauge group implies invariance of the 
measure $\mu_0$ and thus supplies additional motivation for the 
representation space ${\cal H}_{kin}$. Notice that the statement that 
(\ref{2.2.1.2}) defines a representation in particular means that the 
kinematical constraint algebra is {\it free of anomalies}. This should be 
contrasted with string theory where the anomly sits also in the spatial
diffeomorphism group (e.g. Diff$(S^1)$ for the closed string) unless one 
chooses the critical dimension $D=25(9)$ for the bosonic (supersymmetric)
string. \\
\\
Let us now solve the kinematical constraints. By definition, we are 
supposed to find algebraic distributions $l\in {\cal D}_{kin}^\ast$
which satisfy
\be \label{2.2.1.3}
l(\hat{U}(g)f)=l(\hat{V}(\varphi)f)=l(f)\;\;\forall\;
g\in\overline{{\cal G}},\;\varphi\in \mbox{Diff}^\omega(\Sigma),\;
f\in {\cal D}_{kin}
\ee
Now it is not difficult to see that any element of ${\cal D}_{kin}^\ast$
can be conveniently written in the form
\be \label{2.2.1.4}
l(.)=\sum_s\;\;c_s\;\;<T_s,.>_{kin}
\ee
where $c_s$ are complex valued coefficients and the uncountably infinite sum
extends over all possible SNW's. The general solution to (\ref{2.2.1.3})
is then easy to describe: Invariance under $\overline{{\cal G}}$ means that
for fixed $\gamma$ the coefficients $c_{\gamma,\vec{j},\vec{m},\vec{n}}$
have to be chosen, as $\vec{j},\vec{m},\vec{n}$ vary, in such a way that 
at each vertex of $\gamma$ the resulting function is gauge invariant.
That is, if $j_1,..,j_n$ are the labels of edges incident at $v$, then 
the $c_s$ have to arrange themselves to a projector on the trivial
representations contained in the tensor product $j_1\otimes..\otimes j_n$.
Such a projector is
also called intertwiner in the mathematical literature. For $SU(2)$ this 
leads to the theory of Clebsh-Gordon coefficients, $6j-$symbols etc.
As for Diff$^\omega(\Sigma)$ we see that 
$c_{\varphi(\gamma),\vec{j},\vec{m},\vec{n}}$ must be independent of 
$\varphi$, therefore $c_{\varphi,\vec{j},\vec{m},\vec{n}}$ depends only
on the {\it generalized knot class of $\gamma$} ! We say generalized 
because, as we will see later on, the physically relevant graphs are 
those with self-intersections while classical knot theory deals only with 
smooth curves.

One may ask whether one should already define a physical inner product 
with respect to the Gauss and spatial Diffeomorphism constraint and then 
solve the Hamiltonian constraint in a second, separate step on that 
already partly physical Hilbert space . While such a Hilbert space can 
indeed be constructed \cite{36a} it is of no use for QGR because the 
Hamiltonian constraint cannot leave that Hilbert space invariant as we 
see from the second equation in (\ref{1.2.9}) and we must construct 
the physical inner product from the full solution space to all constraints.
However, at least with respect to the kinematical constraints the 
full quantization programme including the question of 
observables has already been completed except for the analysis of the 
classical limit.

\subsection{Kinematical Geometrical Operators}
\label{s2.2.2}

We will restrict ourselves to the description of the area operator 
the classical expression of which we already wrote in (\ref{2.1.1.5})
and (\ref{ex2.1.1.2a}). \\
\\
In order to quantize $\mbox{Ar}(S)$ one starts from (\ref{ex2.1.1.2a})
and decomposes the analytical, compactly supported and oriented surface $S$ 
or, equivalently, its preimage $D$ 
under $X_S$ into small pieces $S_I$. Then the exact area functional is 
approximated by the Riemann sum 
\be \label{2.2.2.1}
\mbox{Ar}(\{S\})=\beta\sum_I \sqrt{E_j(S_I)^2}
\ee
This function is easily quantized because 
$\hat{E}_j(S_I)=i\hbar v_s^j$ is a self-adjoint operator so that 
the sum over $j$ of its squares is positive semi-definite, hence 
its square root is well-defined. Let us denote the resulting, partition 
dependent operator by $\widehat{\mbox{Ar}}(\{S\})$. Now one can show 
that the (strong) limit as the partition is sent to the continuum exists
\cite{37} and a partition independent operator 
$\widehat{\mbox{Ar}}(S)$ results \cite{37}.
\begin{Theorem}
\label{th2.2.2.1} ~~~\\
The area functional admits a well-defined quantization 
$\widehat{\mbox{Ar}}(S)$ on ${\cal H}_{kin}$ with the following properties:
\begin{itemize}
\item[i)] $\widehat{\mbox{Ar}}(S)$ is positive semidefinite, 
(essentially) self-adjoint with Cyl$^2(\ab)$ as domain of (essential)
self-adjointness.
\item[ii)] The spectrum Spec$(\widehat{\mbox{Ar}}(S))$ is pure point 
(discrete) with eigenvectors being given by finite linear combinations of 
spin network functions.
\item[iii)] The eigenvalues are given explicitly by
\ba \label{2.2.2.2}
\lambda_{j_1,j_2,j_{12}} &=&\frac{\beta\ell_P^2}{4}
\sqrt{2j_1(j_1+1)+2j_2(j_2+1)-j_{12}(j_{12}+1} 
\nonumber\\
J_{12} &\in &\{j_1+j_2,j_1+j_2-1,..,|j_1-j_2|\}
\ea
where $j_1,j_2$ are spin quantum numbers and $\ell_P^2=\hbar\kappa$ is 
the Planck area. The spectrum has an area gap (smallest non-vanishing
eigenvalue) given by
\be \label{2.2.2.3}
\lambda_0=\beta \ell_P^2\frac{\sqrt{3}}{4}
\ee
\item[iv)] Spec$(\widehat{\mbox{Ar}}(S))$ contains information about the 
topology of $S$, for instance it matters whether $\partial S=\emptyset$
or not.
\end{itemize}
\end{Theorem}
\begin{Exercise} \label{ex2.2.2.1} ~~~~~~~\\
Verify that the area gap is indeed given by (\ref{2.2.2.3}) and check that 
the distance between subsequent eigenvalues rapidly decreases as 
$j_1,j_2\to \infty$. Can one give an asymptotic formula for 
$N(A)$, the number of eigenvalues (discarding multiplicity) in the 
interval $[A-\ell_P^2,A+\ell_P^2]$ ? Thus, a 
correspondence principle, important for the classical limit is valid. 
If the spectrum would only consist of the {\it main series} 
$\lambda_j=\frac{\ell_P^2}{2}\sqrt{j(j+1)}$ which one obtains for 
$j_1=j_2=j,\;j_{12}=0$ then such a correspondence principle would 
certainly not hold which is, e.g., relevant for the black body 
spectrum of the Hawking radiation.
\end{Exercise}
Theorem \ref{th2.2.2.1} is an amazing result for several reasons:\\
A)\\
First of all, the expression for Ar$(S)$ depends non-polynomially, not 
even analytically on the product $E^a_j(x) E^b_j(x),\;x\in S$. Now 
$E^a_j(x)$ becomes an operator valued distribution in the quantum 
theory and products of distributions at the same point are usually
badly divergent. However, $\widehat{\mbox{Ar}}(S)$ is 
{\it perfectly well-defined} ! This is the first pay-off for sticking to 
a rigorous and background independent formalism !\\
B)\\
Although $S,\gamma,\Sigma,..$ are analytical, the spectrum 
Spec$(\widehat{\mbox{Ar}}(S))$ is discrete. In other words, suppose we 
are measuring the area of a sheet of paper with a spin-network state.
As long as the sheet does not cut an edge of the graph, the area 
eigenvalue is exactly zero {\it no matter how ``close" the edge and the 
sheet are}. We have put the word ``close" in inverted commas because 
this word has no meaning: Since there is no background metric, we do not 
know what close means, only diffeomorphism invariant notions have a meaning
such as ``the edge is cut" or ``the edge is not cut".
However, once the edge is cut the area eigenvalue jumps at least
by the area gap. This strongly hints that the microscopical 
geometry is really distributional (discontinuous) and that we have a 
discrete Planck scale structure, the role of the {\it atoms of geometry} 
being played by the one-dimensional (polymer-like) excitations labelled 
by SNW's. One may speculate that this discrete structure is  
fundamental and that the analyticity
assumptions that we began with should be unimportant, in the final 
picture everything should be only combinatorical. The smooth geometry that
we are familiar with at macroscopic scales is merely a result of coarse 
graining, for instance in order that a SNWF labelled with spin $j=1/2$
on every edge assigns to our sheet of paper its area of about 
$100$cm$^2$, an order of $10^{68}$ edges of the SNW have to cut the sheet !\\
C)\\
Qualitatively similar results apply to the volume operator 
$\widehat{\mbox{Vol}}(R)$ \cite{37,38} and the length operator 
$\widehat{\mbox{Len}}(c)$ \cite{39} whose classical expressions 
are given by
\be \label{2.2.2.4}
\mbox{Vol}(R)=\int_R d^3x \sqrt{\det(q)} \mbox{ and }
\mbox{Len}(c)=\int_c \sqrt{q_{ab} dx^a dx^b}
\ee
D)\\
These kinematical operators are certainly not Dirac observables because 
they are not even spatially diffeomorphism invariant (but $SU(2)$ invariant)
since the objects $R,S,c$ are just coordinate submanifolds of $\Sigma$.
Thus, one may wonder whether the properties of the spectrum just stated 
have any significance at all. The answer is believed to be affirmative as 
the following
argument shows: For instance, instead of $\mbox{Vol}(R)$ consider 
\be \label{2.2.2.5}
\mbox{Vol}_{EM}=\int_\Sigma d^3x \sqrt{\det(q)} \;\;
\theta(\frac{q_{ab}}{\sqrt{det(q)}}[E^a E^b+B^a B^b])
\ee
where we have coupled a Maxwell field to GR with electromagnetic fields
$E^a,E^b$ and $\theta$ is the step function. The physical meaning 
of (\ref{2.2.2.5}) is that it measures the volume of the region where 
the electromagnetic field energy density is non-vanishing and it is 
easy to check that (\ref{2.2.2.5}) is actually spatially diffeomorphism 
invariant ! Now in QGR the argument of the step function can be given a 
meaning as an operator (valued distribution) as we will see in the next 
section and the theta function of an operator can be defined through 
the spectral theorem. Since the spectrum of the $\theta-$function consists
only of $\{0,1\}$, the spectrum of (\ref{2.2.2.5}) should actually {\it 
coincide}
with that of $\widehat{\mbox{Vol}}(R)$ \cite{40}. A similar argument 
should also be valid with respect to Dirac observables commuting with the 
Hamiltonian constraint.\\ 
E)\\
The existence of the area gap is also at the heart of the finiteness of the 
Bekenstein -- Hawking entropy of black holes as we will see.

\part{Selected Areas of Current Research}
\label{s3}

\section{Quantum Dynamics}
\label{s3.1}

The Hamiltonian constraint $C$ of QGR is, arguably, {\it the holy grail}
of this approach to quantum gravity, therefore we will devote a 
substantial amount of space to this subject. In fact, unless one can 
quantize the Hamiltonian constraint, literally no further progress can be 
made so that it is important to know what its status is. From the 
explicit, non-polynomial
expression (\ref{1.2.16}) it is clear that a well-defined operator 
version of this object will be extremely hard to obtain and in fact
this had been the major obstacle in the whole approach until the mid 90's.
In particular, within the original ADM formulation only formal results 
were available. However, since with the new connection formulation also
the non-polynomial kinematical operators of the previous section could be 
constructed, chances might be better. 

At this point we include a brief account of the historical development
of the subject in order to avoid confusion as one looks at older 
papers: \\
Originally one chose the Immirzi parameter as $\beta=\pm i$ 
and considered $\tilde{C}=\sqrt{\det(q)}$ rather than $C$ because then 
$\tilde{C}$ is {\it actually a simple polynomial of only fourth order}
(the ``More" term disappears).
Polynomiality was considered as mandatory. There were three problems with 
this idea:\\ 
1) The non-polynomiality was shifted from $C$ into the 
reality conditions $A+\bar{A}=2\Gamma(E)$ where the spin connection $\Gamma$
is now a highly non-polynomial function of $E$. The operator version of 
this equation should be very hard to implement.\\
2) If $A$ is complex, then we are dealing with an $SL(2,\Cl)$ bundle rather
than an $SU(2)$ bundle. Since $SL(2,\Cl)$ is not compact, the mathematical
apparatus of section \ref{s2} is blown away.\\
3) Even formal trials to quantize $\tilde{C}$ resulted in either divergent, 
or background dependent operators.\\
In \cite{24} it was suggested to keep $\beta$ real which solves problems
1) and 2), however, then $\tilde{C}$ becomes even more complicated and anyway
problem 3) is not cured. Finally in \cite{41} it was shown that 
non-polynomiality
is not necessarily an obstacle, even better, it is actually {\it required} 
in 
order to arrive at a well-defined operator: It was established that the 
reason for problem 3)
is that $\tilde{C}$ is a scalar density of weight {\it two} while 
it was shown that only density weight {\it one} scalars have a chance to 
be quantized rigorously and background independently. Therefore the currently
accepted point of view is that $\beta$ should be real and that one uses the 
original unrescaled $C$ rather than $\tilde{C}$.

\subsection{A Possible New Mechanism for Avoiding UV Singularities 
in Background Independent Quantum Field Theories}
\label{s3.1.1}

Before we go into more details concerning \cite{41}, let us give a 
heuristic explanation {\it just why it happens that QGR may cure UV 
problems 
of QFT}, making the connection with the issue of the density weight
just mentioned. Consider classical Einstein -- Maxwell theory on 
$M=\Rl\times \sigma$ in its 
canonical formulation, then the Hamiltonian constraint gains an extra
matter piece given for unit lapse $N=1$ by 
\be \label{3.1.1}
H_{EM}=\frac{1}{2 e^2}\int_\sigma\; d^3x \; 
\overbrace{{\blue \frac{q_{ab}}{\sqrt{\det(q)}}}}^{\mbox{{\red Density 
weight -1}}}\;
\underbrace{{\green [E^a E^b+B^a B^b]}}_{\mbox{{\red Density weight +2}}} 
\ee
\begin{Exercise} \label{ex3.1.1} ~~~~\\
Starting from the Lagrangean 
\be \label{ex3.1.1a}
L=-\frac{1}{4e^2}\sqrt{|\det(g)|}F_{\mu\nu} F_{\rho\sigma} g^{\mu\rho}
g^{\nu\sigma}
\ee
where $F=2dA$ is the spacetime curvature of the Maxwell connection $A$ 
with unit cm$^{-1}$ and  
$e$ is the electric charge in units such that $\alpha=\hbar e^2$ is the 
dimensionfree Feinstrukturkonstante, perform the Legendre transform.
With the electric field $E^a$ being the momentum conjugate to the spatial
piece $A_a$ of $A$ verify that the ``Hamiltonian'' is given by 
$-A_0 G+N^a V'_a+N C'$ where $G=\partial_a E^a$ is the Gauss law,
$V'_a=F_{ab} E^b$ and $C'$ is the integrand of (\ref{3.1.1}) with 
$B^a=\epsilon^{abc} F_{bc}/2$ the magnetic field. Check that $G'$
generates $U(1)$ gauge transformations while $V'_a$ generates 
spatial diffeomorphisms where $A_a,E^a$ transform as a one -- form 
and a vector density of weight one respectively. Conirm that also 
$B^a$ is a vector density of weight one.
\end{Exercise}
As the exercise reveals, the geometry factor in (\ref{3.1.1}) is a 
symmetric covariant tensor of rank two of density weight $-1$ due to the 
factor $\sqrt{\det(q)}$ in the denominator while the matter part is a
symmetric contravariant tensor of rank two of density weight $+2$.
That the resulting scalar has net density weight is $+1$ is no coincidence 
but a direct 
consequence of the diffeomorphism invariance or background independence of 
any matter theory coupled 
to gravity: only the integral over $\sigma$ of a scalar density of weight 
$+1$ is spatially diffeomorphism invariant.

We can now quantize (\ref{3.1.1}) in two ways: \\
1) \\
In the first version we notice that if $g=\eta$ is the Minkowski metric,
that is, $q_{ab}=\delta_{ab}$ then (\ref{3.1.1}) reduces to the 
ordinary Maxwell Hamiltonian on Minkowski space. Thus we apply the 
formalism of QFT on a background spacetime, in this case 
Minkowski space, because we have fixed $q_{ab}$ to the non-dynamical 
$\Cl-$number field $\delta_{ab}$ which is not quantized at all. \\
2)\\
In the second version we keep $q_{ab}$ dynamical and quantize it as well.
Thus we apply QGR, a background independent quantization. Now $q_{ab}$
becomes a field operator $\hat{q}_{ab}$ and the statement that the metric 
is flat can at most have a semiclassical meaning, that is, the expectation 
value of $\hat{q}_{ab}$ in a gravitational state is close to 
$\delta_{ab}$.\\
Let us now sketch how these two different quantizations are performed and 
exactly pin-point how it happens that the first quantization is 
divergent while the second is finite.
\begin{itemize}
\item[1)] {\it QFT on a background spacetime}\\
As we have said, the metric $q_{ab}=\delta_{ab}$ is now no longer a 
dynamical entitity but just becomes a complex number. What we get is the
usual Maxwell Hamiltonian operator
\be \label{3.1.2}
\hat{H}_M=\frac{1}{2 e^2}\int_\Sigma\; d^3x \; 
\delta_{ab}
{\green [\hat{E}^a \hat{E}^b+\hat{B}^a \hat{B}^b]}
\ee
Notice the crucial difference with (\ref{3.1.1}): The net density weight
of the operator valued distribution in the integral is now $+2$ rather 
than $+1$ ! By switching off the metric as a dynamical field we have done 
a severe crime to the operator, because the net density weight $+2$ will
be remembered by the operator in any faithful representation of the 
canonical commutation relations and leads to the following problem:
The only coordinate density of weight one that one can construct 
is a $\delta-$distribution (and derivatives thereof), thus for instance 
the operator $\hat{E}^a(x)$ is usually represented as a functional 
derivative which one can rewrite formally as
\be \label{3.1.3}
\alpha \delta/\delta A_a(x)=\alpha\sum_{y\in \Sigma} 
\delta(x,y) 
\partial/\partial A_a(y)  
\ee
The right hand side of (\ref{3.1.3}) is a sum over terms each of which
consists of a well-defined operator $Y_a(y)=\partial/\partial A_a(y)$
and a distributional prefactor $\delta(x,y)$. It is for this reason
that expressions of the form $\hat{E}^a(x)\hat{E}^b(x)$ cannot be 
well-defined since we get products of distributions supported at the same 
point $x$ and which result in divergent expressions of the form
$\sum_{y,z} \delta(x,y)\delta(x,z)  Y_a(y) Y_b(z)
=\sum_y \delta(x,y)^2  Y_a(y) Y_b(y)$. The density weight two is correctly
encoded in the term $\delta(x,y)^2=\delta(0,0)\delta(x,y)$ which, however,
is meaningless. 

These heuristic arguments can of course be made precise: (\ref{3.1.2}) 
is quantized on the Fock space ${\cal H}_{Fock}$ and one obtains 
\be \label{3.1.4}
\hat{H}_M=\;:\hat{H}_M:\;+\;\hbar\int_\Sigma
\underbrace{[\sqrt{-\Delta_x} \delta(x,y)]_{x=y}}_{\mbox{{\red UV 
Singularity}}}
\ee
Here the colons stand for normal ordering. The UV (or short distance) 
singularity is explicitly identified as the coincidence limit $x=y$
of the integrand in the normal ordering correction. Therefore $\hat{H}_M$
is ill-defined on ${\cal H}_{Fock}$. Notice that even if the integrand 
would be finite, the integral suffers from an IR (or large volume)
singularity if $\sigma$ is not compact which comes from the fact that we 
are dealing with an infinite number of degrees of freedom. This 
singularity is, in contrast to the UV singularity, physical since it
captures the vacuum energy of the universe which is of course infinite if 
the volume is.
\item[2)] {\it QFT coupled to QGR}\\
This time we keep the metric as a dynamical variable and quantize it.
Thus instead of (\ref{3.1.2}) we obtain something of the form
\be \label{3.1.5}
\hat{H}_{EM}=\frac{1}{2 e^2}\int_\Sigma\; d^3x \; 
{\blue \widehat{\frac{q_{ab}}{\sqrt{\det(q)}}}}\;
{\green [\hat{E}^a \hat{E}^b+\hat{B}^a \hat{B}^b]}
\ee
This time the net density weight is still $+1$. Now while the expression
(\ref{3.1.3}) is still valid and implies that there will be a product
of $\delta-$distributions in the numerator coming from the matter operator
valued distributions, there is also a $\delta-$distribution {\it in the 
denominator} due to the factor $\sqrt{\det(q)}$ which comes about as 
follows: As we already mentioned in section \ref{s2.2.2} the volume 
functional in (\ref{2.2.2.4}) admits a well-defined quantization
of the form 
\be \label{3.1.6}
\widehat{\mbox{Vol}}(R) T_s=\ell_P^3
\sum_{v\in V(\gamma(s))\cap R} \hat{V}_v T_s
\ee
where $\hat{V}_v$ is a well-defined, dimensionfree operator (not an 
operator valued 
distribution !) built from the vector fields $v_S^j$. Since 
Vol$(R)$ is the integral over $R$ of $\sqrt{\det(q)}$ we conclude that 
$\sqrt{\det(q)}$ admits a quantization as an operator valued distribution,
namely
\be \label{3.1.7}
\widehat{\sqrt{\det(q)}}(x) T_s=\ell_P^3 \sum_{v\in V(\gamma(s))} 
\delta(x,v) \hat{V}_v T_s
\ee
Now certainly (\ref{3.1.5}) cannot be quantized on the Hilbert space 
${\cal H}_{kin}\otimes {\cal H}_{Fock}$ because ${\cal H}_{Fock}$
depends on a background metric (for instance through the Laplacian 
$\Delta$) which is not available to us. However, we may construct
a background independent Hilbert space ${\cal H}_{kin}'$ for Maxwell 
theory which is completely
identical to our ${\cal H}_{kin}$, just that $SU(2)$ is replaced by $U(1)$
\cite{41}. In ${\cal H}_{kin}'$ the role of spin network states is played
by charge network (CNW) states, that is, edges $e$ are labelled by 
integers $n_e$ (irreducibles of $U(1)$. Let us denote CNW's by 
$c=(\gamma,\vec{n}=\{n_e\}_{e\in E(\gamma)})$ and CNWF's by $T'_c$. Then a 
basis for the 
Einstein-Maxwell theory kinematical Hilbert space 
${\cal H}_{kin}\times {\cal H}'_{kin}$ is given by the states 
$T_s\otimes T'_c$. 

Now something very beautiful happens, which is not put in by hand but 
rather is a derived result: A priori the states $T_s,T'_c$
may live on different graphs, however, unless the graphs are identical,
the operator automatically (\ref{3.1.5}) annihilates $T_s\otimes T'_c$
\cite{42}.
This is the mathematical manifestation of the following deep physical
statement: {\it Matter can only exist where geometry is excited}. Indeed,
if we have a gravitational state which has no excitations in a 
coordinate region $R$
then the volume of that region as measured by the volume operator is 
identically zero. However, if a coordinate region has zero volume, then it 
is {\it physically simply not there, it is empty space}. Summarizing,
the operator (\ref{3.1.5}) is non-trivial only if $\gamma(s)=\gamma(c)$.

With this being understood, let us then sketch the action of 
(\ref{3.1.5}) on our basis. One finds heuristically
\ba \label{3.1.8}
&& \hat{H}_{EM} T_s\otimes T'_c
= m_P\sum_{v\in V(\gamma)}\sum_{e,e'\in E(\gamma),e\cap e'=v} \times
\nonumber\\
&\times & 
\int_\Sigma d^3x 
[{\blue \frac{\hat{q}_{e,e'}}{\hat{V}_v}}
\underbrace{ 
\underbrace{{\blue \frac{1}{\delta(x,v)}}}_{{\red \uparrow}} T_s]
\otimes
[\underbrace{{\green \delta(x,v)}}_{{\red \uparrow}}
}_{{\red Cancellation}}
{\green \delta(y,v) Y^e Y^{e'}} T'_c]_{x=y}]
\ea 
where $m_P=\sqrt{\hbar/\kappa}$ is the Planck mass.
Here $\hat{q}_{e,e'}$ and $Y^e$ are well-defined, dimensionfree  
operators (not 
distribution valued !) on ${\cal H}_{kin}$ and ${\cal H}'_{kin}$ 
respectively built from the right invariant vector fields $R^j_e,R_e$ that 
enter the definition of the flux operators as in (\ref{2.1.1.10}) and its
analog for $U(1)$. The product of $\delta-$distributions in the numerator
of (\ref{3.1.8}) has its origin again in the fact that the matter operator
has density weight $+2$ certainly also in this representation and 
therefore has to be there, so nothing is swept under the rug! The 
$\delta-$distribution in the denominator comes from (\ref{3.1.7})
and correctly accounts for the fact that the geometry operator has 
density weight $-1$. Again we have a coincidence limit $x=y$ which 
comes from a point splitting regularization and which in the background 
dependent quantization gave rise to the UV singularity. Now we see
what happens: One of the $\delta-$distributions in the numerator gets 
precisely cancelled by the one in the denominator leaving us with 
only one $\delta-$distribution correctly accounting for the fact that
the net density weight is $+1$. The integrand is then well-defined and 
the integral can be performed resulting in the {\it finite} expression
\ba \label{3.1.9}
&& \hat{H}_{EM} T_s\otimes T'_c
= \sum_{v\in V(\gamma)}\sum_{e,e'\in E(\gamma),e\cap e'=v} \times
\nonumber\\
&\times & 
[{\blue \frac{\hat{q}_{e,e'}}{\hat{V}_v}} T_s]
\otimes
[{\green Y^e Y^{e'}} T'_c]_{x=y}
\ea 
Notice that {\it finite} here means {\it non-perturbatively finite}, 
that is, not only finite order by order in perturbation theory
(notice that in coupling gravity we have a highly interacting theory in 
front of us). Thus, comparing our non-perturbative result to perturbation
theory the result obtained is comparable to showing that the perturbation 
series {\it converges} ! Notice also that for non-compact $\sigma$ 
the expression (\ref{3.1.9}) possibly has the physically correct
IR divergence coming from a sum over an infinite number of vertices.
\end{itemize}
\begin{Exercise} \label{ex3.1.2} ~~~~~\\
Recall the Fock space quantization of the Maxwell field and verify 
(\ref{3.1.4}).
\end{Exercise}
This ends our heuristic discussion about the origin of UV finiteness in 
QGR. The crucial point is obviously the density weight of the operator
in question which should be precisely $+1$ in order to arrive at a 
well-defined, background independent result: Higher density weight 
obviously leads to more and more divergent expressions, lower density 
weight ends in zero operators. 

\subsection{Sketch of a Possible Quantization of the Hamiltonian
Constraint}
\label{s3.1.2}

We now understand intuitively why the rescaled 
Hamiltonian constraint $\tilde{C}$ had no chance to be well-defined in the 
quantum theory: It is similar to (\ref{3.1.4}) due to its density weight 
$+2$. The same factor $1/\sqrt{\det(q)}$ that was responsible for making 
(\ref{3.1.5}) finite also makes the original, non-polynomial, unrescaled  
Hamiltonian constraint $C=\tilde{C}/\sqrt{\det(q)}$ finite. We will 
now proceed to some details how this is done, avoiding intermediate 
divergent expressions such as in (\ref{3.1.8}). 

The essential steps can already be explained for the first term in 
(\ref{1.2.16}) so let us drop the ``More" term and consider only
the integrated first term
\be \label{3.1.10}
C_E(N)= \frac{1}{\kappa} \int_\Sigma d^3x N
\frac{F_{ab}^j \epsilon_{jkl} E^a_j E^b_l}{\sqrt{|\det(E)|}}
\ee
Let us introduce a map 
\be \label{3.1.11}
R:\;\Sigma\to {\cal O}(\Sigma);\;x\mapsto R_x
\ee
where ${\cal O}(\Sigma)$ denotes the set of open, compactly 
supported, connected and simply connected subsets of $\Sigma$ and 
$R_x\in {\cal O}(\Sigma)$ is constrained by the requirement that $x\in 
R_x$. We define the volumes of the $R_x$ by
\be \label{3.1.12}
V(x):=\mbox{Vol}(R_x)=\int_{R_x} d^3y \sqrt{|\det(E)|}(y)
\ee
Then, up to a numerical prefactor we may write (\ref{3.1.10}) in 
the language of differential forms and in terms of a Poisson bracket
as
\be \label{3.1.13}
C_E(N)=\frac{1}{\kappa^2} 
\int_\Sigma N \mbox{Tr}(F\wedge \{A,V\})
\ee
\begin{Exercise} \label{ex3.1.3} ~~~~~\\
Verify that (\ref{3.1.12}) is really the volume of $R_x$ and 
(\ref{3.1.13}).
\end{Exercise}
The reasoning behind (\ref{3.1.13}) was to move the factor 
$1/\sqrt{\det(q)}(x)$ from the denominator into the numerator 
by using a Poisson bracket. This will avoid the $\delta-$distribution 
in the denominator as in (\ref{3.1.8}) and has the additional 
advantage that $\sqrt{\det(q)}$ now appears smeared over $R_x$
so that one obtains an operator, not a distribution.
Thus, the idea is now to replace the function $V(x)$ by the well-defined 
operator $\widehat{\mbox{Vol}}(R_x)$ and the Poisson bracket by a 
commutator divided by $i\hbar$. The only thing that prevents us from doing
this is that the operators $A_a^j, F_{ab}^j$ do not exist 
on ${\cal H}_{kin}$. However, they can be regularized in terms of 
holonomies as follows: \\
Given tangent vectors $u,v\in T_x(\Sigma)$ we define one parameter
homotopies of paths and loops of triangle topology 
\be \label{3.1.14}
\epsilon\mapsto p^u_{\epsilon,x},\; \alpha^{uv}_{\epsilon,x}
\ee
respectively with 
$b(p^u_{\epsilon,x})=b(\alpha^{uv}_{\epsilon,x})=x$ 
and $(\dot{p}^u_{\epsilon,x})_x=(\dot{\alpha}^{uv}_{\epsilon,x})_{x+}
=\epsilon u,\;
(\dot{\alpha}^{uv}_{\epsilon,x})_{x-}=-\epsilon v$ (left and right 
derivatives at $x$). Then for {\it 
smooth} connections $A\in \a$ the Ambrose -- Singer theorem tells us that 
\be \label{3.1.15}
\lim_{\epsilon\to 0} [A(p^u_{\epsilon,x}-1]/\epsilon=u^a A_a(x),\;\;\
\lim_{\epsilon\to 0} [A(\alpha^{uv}_{\epsilon,x}-1]/\epsilon^2=
u^a v^b F_{ab}(x)/2
\ee
\begin{Exercise} \label{ex3.1.4}
Verify (\ref{3.1.15}) by elementary means, using directly the differential
equation (\ref{2.1.1.1}).\\
Hint:\\
For sufficiently small $\epsilon$ we have up to $\epsilon^2$ corrections 
$p^u_{\epsilon,x}(t)=x+\epsilon t u$ and
$$
\alpha^{uv}_{\epsilon,x}(t)=x+\epsilon
\left\{ \begin{array}{cc}
tu & t\in [0,1] \\
u/3+(t-1)(v-u) & t\in [1,2] \\
(3-t)v & t\in [2,3] 
\end{array} \right.
$$
\end{Exercise}
Thus, given a triangulation $\tau_\epsilon$ of $\sigma$, 
that is, a decomposition of $\sigma$ into tetrahedra
$\Delta$ with base points $v(\Delta)$, edges 
$p_I(\Delta),\;I=1,2,3$ of $\Delta$ of the type $p^u_{\epsilon,v}$
starting at $v$ and triangular loops 
$\alpha_{IJ}(\Delta)=p_I(\Delta)\circ a_{IJ}(\Delta) p_J(\Delta)^{-1}$
of the type $\alpha^{uv}_{\epsilon,v}$  
where the {\it arcs} $a_{IJ}(\Delta)$ comprise the remaining three edges 
of $\Delta$, it is easy to show, using (\ref{3.1.15}), that up to a 
numerical factor 
\be \label{3.1.16}
C^{\tau_\epsilon}_E(N)=\frac{1}{\kappa^2} \sum_{\Delta\in \tau_\epsilon}
N(v(\Delta)) \sum_{IJK} \epsilon^{IJK} 
\mbox{Tr}(A(\alpha_{IJ}(\Delta) A(p_K(\Delta))\{A(p_K(\Delta))^{-1},V(v)\}
\ee
tends to $C_E(N)$ as $\epsilon\to 0$ (in this limit the triangulation gets 
finer and finer).
\begin{Exercise} \label{ex3.1.5} ~~~~\\
Verify this statement.
\end{Exercise}
The expression (\ref{3.1.16}) can now be readily quantized on 
${\cal H}_{kin}$ because holonomies and volume functionals are 
well-defined operators. However, we must remove the regulator $\epsilon$
in order to arrive at a quantization of (\ref{3.1.13}). Now the regulator
can be removed in many inequivalent ways because there is no unique way 
to refine a triangulation. Moreover, we must specify in which operator
topology $\hat{C}^{\tau_\epsilon}(N)$ converges. The discussion of these 
issues is very complicated and the interested reader is referred to 
\cite{41} for the detailed arguments that lead to the following 
solution:
\begin{itemize}
\item[i)] {\it Triangulation}\\
First of all we define the operator explicitly on the SNW basis $T_s$. In 
order
for the refinement limit to be non-trivial, it turns out that the 
triangulation must be refined in such a way that 
$\gamma(s)\subset\tau_\epsilon$ for sufficiently small $\epsilon$.
This happens essentially due to the volume operator which has non-trivial
action only at vertices of graphs.
Thus the refinement must be chosen depending on $s$. This is justified 
because classically all refinements lead to the same limit. One might 
worry that this does not lead to a linear operator, however, this is 
not the case  because it is defined on a basis.
\item[ii)] {\it Operator Topology}\\
The limit $\epsilon\to 0$ exists in the following sense:\\
 Let ${\cal D}^\ast_{Diff}\subset {\cal D}^\ast_{kin}$
be the space of solutions of the diffeomorphism constraint. We say
that a family of operators $\hat{O}_\epsilon$ converges to an operator
$\hat{O}$ on ${\cal H}_{kin}$ in the {\it uniform -- weak -- 
Diff$^\ast-$topology} provided that for each $\delta>0$ and for each 
$l\in {\cal D}^\ast_{Diff},\;f\in {\cal D}_{kin}$ there exists 
$\epsilon(\delta)>0$ {\it independent of $l,f$} such that
\be \label{3.1.17}
|l([\hat{O}_\epsilon-\hat{O}]f)|<\delta\;\;\forall \epsilon< 
\epsilon(\delta)    
\ee
This topology is of course motivated by physical considerations: Since
the operator is unbounded, the uniform (i.e. operator norm) topology is too 
strong. The strong or weak topologies (pointwise convergence in Hilbert 
space norm or as matrix elements) give a trivial (zero) limit (excercise !).
Thus one is naturally led to $^\ast$ topologies. The maximal dual
space on which to build a topology would be ${\cal D}^\ast_{kin}$ but 
one can check that the limit does not exist even pointwise in 
${\cal D}^\ast_{kin}$. Thus one is looking for suitable subspaces thereof.
The natural, physically motivated choice is, of course, the space 
${\cal D}^\ast_{Diff}$ which is singled out by the spatial diffeomorphism 
constraint. The reason for why have required uniform convergence in
(\ref{3.1.17}) is that this excludes the existence of the limit
for larger spaces 
${\cal D}^\ast_{Diff}\subset {\cal D}^\ast_\star \subset 
{\cal D}^\ast_{kin}$.
\end{itemize}
The end result is 
\ba \label{3.1.18}
&& \hat{C}^\dagger_E(N)=m_P\sum_{v\in V(\gamma(s))} N(v)
\;\; \sum_{\begin{array}{c} e,e',e^{\prime\prime}\in E(\gamma(s))\\
e\cap e'\cap e^{\prime\prime}=\{v\} \end{array}} 
\times \nonumber\\
&\times& 
\{\mbox{Tr}([A(\alpha_{\gamma(s),v,e,e'})-
(A(\alpha_{\gamma(s),v,e,e'}))^{-1}]A(p_{\gamma(s),v,e^{\prime\prime}})
[A(p_{\gamma(s),v,e^{\prime\prime}})^{-1},\hat{V}_v])
\nonumber\\
&& +\mbox{cyclic permutation in} \{e,e',e^{\prime\prime}\}\}\;\; T_s
\ea
The meaning of the loops $\alpha_{\gamma,v,e,e'}$ and paths 
$p_{\gamma,v,e^{\prime\prime}}$ that appears in this sum over vertices 
and triples of edges incident at them is maybe best explained in the 
following figure \ref{f10}. Their precise specification makes use 
of the axiom of choice and is diffeomorphism covariant, that is,
for $\varphi\in \mbox{Diff}^\omega(\Sigma)$, e.g. the loops 
$\alpha_{\gamma,v,e,e'}$ and
$\alpha_{\varphi(\gamma),\varphi(v),\varphi(e),\varphi(e')}$ are 
analytically diffeomorphic. Moreover, the {\it arcs} $a_{\gamma,v,e,e'}$ 
defined by 
\be \label{3.1.19}
\alpha_{\gamma,v,e,e'}=p_{\gamma,v,e}\circ a_{\gamma,v,e,e'}\circ
p_{\gamma,v,e'}^{-1}
\ee
are such that also $\gamma\cup a_{\gamma,v,e,e'}$ and 
$\varphi(\gamma)
\cup a_{\varphi(\gamma),\varphi(v),\varphi(e),\varphi(e')}$ are 
anlytically diffeomorphic.
\begin{figure}
\includegraphics[width=10cm,height=7cm]{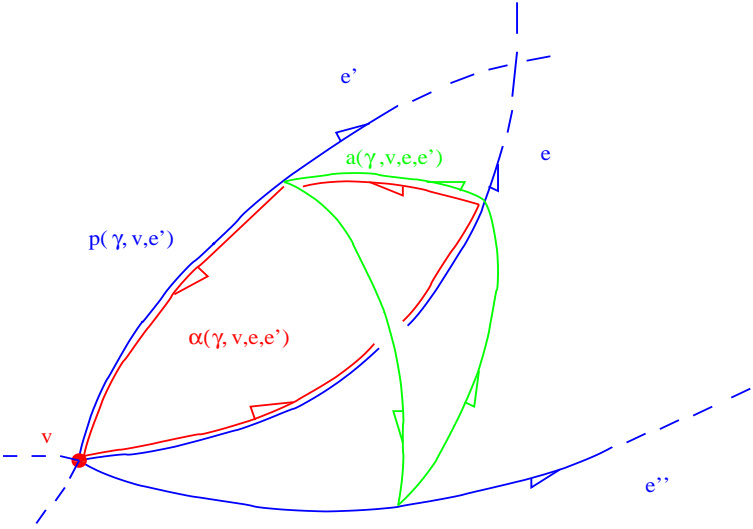}
\caption{Meaning of the loop, path and arc assignment of the Hamiltonian 
constraint. Notice how a tetrahedron emerges from those objects, making 
the link with the triangulation. The broken lines indicate possible 
other edges or continuations thereof.} 
\label{f10} 
\end{figure} 
The adjoint in (\ref{3.1.18}) is due to the fact that $C_E(N)$ is 
classically real-valued, so we are quantizing $\overline{C_E(N)}$ as well.
The operator (\ref{3.1.18}) is not symmetric, however, its adjoint 
is densely defined on ${\cal D}_{kin}$ and it is therefore closable.
Usually one requires real valued functions to become self-adjoint operators
because then by the spectral theorem the spectrum (possible measurement
values) is a subset of the real line. However, this argument is void when 
we are only interested in the kernel of the operator (``zero eigenvalue").
\begin{Exercise} \label{ex3.1.6} ~~~\\
Verify that $\hat{C}^\dagger_E(N)$ is not symmetric but it is, together with 
$\hat{C}_E(N)$, densely defined on ${\cal D}_{kin}$. Show that 
if real valued constraints $C_I$ form a Poisson algebra 
$\{C_I,C_J\}=f_{IJ}\;^K C_K$ with non-trivial, real valued structure 
functions such that $\{f_{IJ}\;^K,C_K\}_{\{C_L=0\}}\not=0$, 
then $\hat{C}_I,\hat{f}_{IJ}\;^K$ must not be both symmetric in order for 
the quantum algebra to be free of anomalies. Conclude that the failure 
of (\ref{3.1.18}) to be symmetric is likely to be required for reasons 
of consistency.
\end{Exercise}
The fact the loop $\alpha_{\gamma,v,e,e'}$ is not shrunk to $v$ as one 
would expect is of course due to our definition of convergence, in fact,
an arbitrary loop assignment $(\gamma,v,e,e')\mapsto \alpha_{\gamma,v,e,e'}$
that has the same diffeomorphism invariant characteristics is allowed, again
because in a diffeomorphism invariant theory there is no notion of 
``closeness" of $\alpha_{\gamma,v,e,e'}$ to $v$. Notice that the operator 
$\hat{C}_E(N)$ is {\it defined on ${\cal H}_{kin}$} using the axiom of 
choice and not on 
diffeomorphism invariant states as it is sometimes misleadingly stated in 
the literature \cite{43}. In fact, it cannot be because the dual operator 
$\hat{C}'_E(N)$ defined by
\be \label{3.1.20}
[\hat{C}'_E(N) l](f):=l(\hat{C}^\dagger_E(N) f)
\ee
for all $f\in {\cal D}_{kin},\;l\in {\cal D}^\ast_{kin}$ does not 
preserve ${\cal D}_{Diff}^\ast$ as is expected from the the classical
Poisson algebra $\{V,C\}\propto C\not= V$. If one wants to take this dual
point of view then one is forced to introduce a larger space 
${\cal D}^\ast_\star$ which is preserved but which does not solve the 
diffeomorphism constraint and is therefore unphysical. This has unnecessarily
given rise to a large amount of confusion in the literature and should be 
abandomed.

As we have said, the loop assignment is to a very large extent arbitrary 
at the level of ${\cal H}_{kin}$ and represents a serious {\it quantization
ambiguity}, it cannot even be specified precisely because we are using 
the axiom of choice. However, at the level of ${\cal H}_{phys}$ this 
ambiguity evaporates to a large extent because all choices that are 
related by a diffeomorphism result in the same solution space {\it to all
constraints} defined by elements $l\in {\cal D}^\ast_{Diff}$ which satisfy
in addition
\be \label{3.1.21}  
[\hat{C}'_E(N) l](f)=l(\hat{C}^\dagger_E(N) f)=0\;\;\forall\;\;
N\in C^\infty_0(\Sigma),\; f\in{\cal D}_{kin}
\ee
where $C^\infty_0(\Sigma)$ denotes the smooth functions of compact support.
Thus the solution space ${\cal D}^\ast_{phys}$ will depend only on the 
spatially diffeomorphism invariant characteristics of the loop assignment
{\it which can be specified precisely} \cite{41}, it essentially 
characterizes the amount by which the arcs knot the original edges of the 
graph. Besides this remaining ambiguity there are also factor ordering 
ambiguities {\it but no singularities} some of which are discussed in 
\cite{44}.

\begin{figure}
\includegraphics[width=10cm,height=7cm]{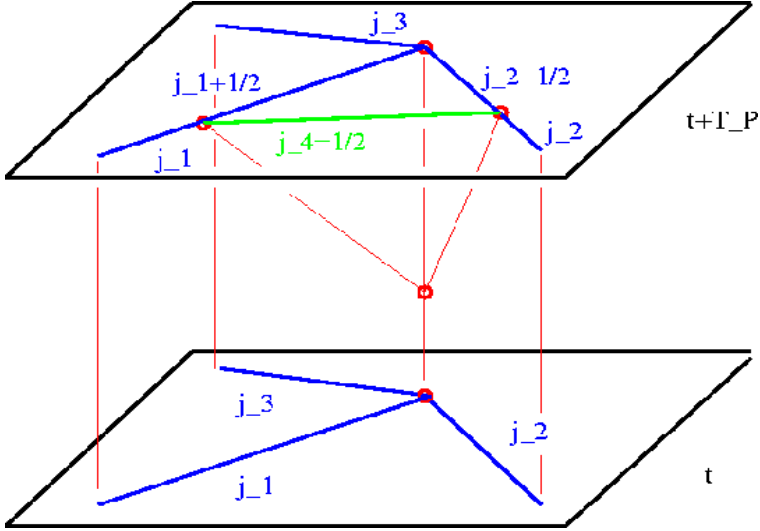}
\caption{Emergence of a spin foam from a SNW by the action of the 
Hamiltonian constraint} 
\label{f11}
\end{figure}

Let us list without proof some of the properties of this operator:
\begin{itemize}
\item[i)] {\it Matter Coupling}\\
Similar Techniques can be applied to the case of (possibly 
supersymmetric) matter coupled to GR \cite{41}.
\item[ii)] {\it Anomaly -- Freeness}\\
The constraint algebra of the Hamiltonian constraint with the spatial
diffeomorphism constraint and among each other is {\it mathematically
consistent}. From the classical constraint algebra $\{V,C\}\propto C$ 
we expect that 
$\hat{V}(\varphi)\hat{C}^\dagger_E(N)\hat{V}(\varphi)^{-1}=
\hat{C}_E(\varphi^\ast N)$ for all diffeomorphisms $\varphi$. However,
this is just the statement of the loop assignment being diffeomorphism 
covariant which can be achieved by making use of the axiom of choice.
Next, from $\{C,C\}\propto V$ we expect that the dual of 
$[\hat{C}^\dagger_E(N),\hat{C}^\dagger_E(N')]=
[\hat{C}_E(N'),\hat{C}_E(N)]^\dagger$ annihilates the elements 
of ${\cal D}_{Diff}^\ast$. This can be explicitly verified \cite{41}. 
We stress
that $[\hat{C}^\dagger_E(N),\hat{C}^\dagger_E(N')]$ is {\it not zero},
the algebra of Hamiltonian constraints is {\it not Abelean} as it is 
sometimes misleadingly stated in the literature. The commutator 
is in fact explicitly proportional to a diffeomorphism.
\item[iii)] {\it Physical States}\\
There is a rich space of rigorous solutions to (\ref{3.1.21}) and a precise 
algorithm for their construction has been developed \cite{41}.
\item[iv)] {\it Intuitive Picture}\\
The Hamiltonian constraint acts by annihilating and creating spin degrees 
of freedom and therefore the dynamical theory obtained could  be called 
``Quantum Spin Dynamics (QSD)" in analogy to ``Quantum 
Chromodynamics (QCD)" in which the Hamiltonian acts by creating and 
annihilating colour degrees of freedom. In fact we could draw a crude 
analogy to Fock space terminology as follows: The (perturbative) 
excitations of QCD carry a continuous label, the mode number $k\in \Rl^3$ 
and a discrete label, the occupation number $n\in\Nl$ (and others). In QSD 
the continuous labels are the edges $e$ and the discrete ones are spins $j$
(and others). So we have something like a non -- linear Fock 
representation in front of us. 

Next, when solving the Hamiltonian constraint, that is, when {\it integrating
the Quantum Einstein Equations}, one realizes that one is not dealing 
with a (functional) partial differential equation but rather with 
a (functional) partial difference equation. Therefore, when understanding 
coordinate time as measured how for instance volumes change, we conclude 
that also time evolution is necessarily {\it discrete}. Such discrete 
time evolution steps driven by the Hamiltonian constraint assemble 
themselves into what nowadays is known as a spin foam. A spin foam is a 
four dimensional complex of two dimensional surfaces where each surface is 
to be thought of as the world sheet of an edge of a SNW and it carries 
the spin that the edge was carrying before it was evolved\footnote{Thus, a 
spin foam model can be thought of as a background independent string theory !}.

Another way of saying this is that a spin foam is a complex of 
two-surfaces labelled by spins and when cutting a spin foam with a 
spatial three-surface $\Sigma$ one obtains a SNW. If one uses two such 
surfaces $\Sigma_t,\Sigma_{t+T_P}$ where $T_P=\ell_P/c$ is the Planck time
then one rediscovers the discrete time evolution of the Hamiltonian 
constraint. These words are summarized in figure \ref{f11}. 
\end{itemize}
While these facts constitute a promising hint that the Hilbert space 
${\cal H}_{kin}$ could in fact {\it support the quantum dynamics of GR}, 
there are well-taken concerns 
about the physical correctness of the operator $\hat{C}_E^\dagger(N)$:\\ 
\\
The 
problem is that one would like to see more than that the commutator of two
dual Hamiltonian constraints annihilates diffeomorphism invariant states,
one would like to see something of the kind
\be \label{3.1.22}
[\hat{C}^\dagger_E(N),\hat{C}^\dagger_E(N')]
=i\ell_P^2\widehat{[\int_\Sigma d^3x [N N'_{,a}-N_{,a} N'] q^ab V_b]}
\ee
The reason for this is that then one would be sure that 
$\hat{C}^\dagger_E(N)$ generates the correct quantum evolution.
While this requirement is not necessary, it is certainly sufficient and 
would be reassuring\footnote{Example: Suppose that $C_a$ are the angular 
momentum
components for a particle in in $\Rl^3$ with classical constraint algebra 
$\{C_a,C_b\}=\epsilon_{abc} C_c$. Introduce polar coordinates and 
define the non-self adjoint operators
$\hat{C}_1=i\hbar\partial/\partial\theta,\;
\hat{C}_2=i\hbar\partial/\partial\phi,\;\hat{C}_3=0$. Then the quantum 
constraint
algebra is Abelean and does not at all resemble the classical one, however,
the physical states are certainly the correct ones, functions that depend 
only on the radial coordinate.}. There are two obstacles that 
prevent us from rewriting the left hand side of (\ref{3.1.22}) in terms 
of the right hand side.\\
1)\\
The one parameter groups $s\mapsto \hat{V}(\varphi^u_s)$ of unitarities
where $\varphi_s^u$ are the one parameter groups of diffeomorphisms 
defined by the integral curves of a vector field $u$ are not weakly 
continuous, therefore a self-adjoint generator $\hat{V}(u)$ that we would
like to see on the right hand side of (\ref{3.1.22}) simply does not exist.
\begin{Exercise} \label{ex3.1.7}  ~~~~~~\\
Recall Stone's theorem about the existence of the self-adjoint generators
of weakly continuous one-parameter unitary groups and verify that 
$\hat{V}(\varphi^u_s)$ is not weakly continuous on ${\cal H}_{kin}$.
\end{Exercise}
2)\\
One can quantize the right hand side of (\ref{3.1.22}) by independent means
and it does annilate ${\cal D}^\ast_{Diff}$ \cite{41}, however, that 
operator does not resemble the left hand side in any obvious way. 
The reason for this is that even classically it takes a DinA4 page 
of calculation in order to rewrite the Poisson bracket $\{C_E(N),C_E(N')\}$
as in (\ref{3.1.22}) with $V_a$ given by (\ref{1.2.16}). The 
manipulations that must be performed in order to massage the Poisson bracket
into the desired form involve a) integrations by part, b) writing 
$F_{ab}$ in terms of $A_a$, c) derivatives of $\sqrt{\det(q)}$, 
d) multiplying fractions by functions in both numerator and denominator,
e) symmetry arguments in order to see that certain tems cancel etc.
(exercise !). These steps are obviously difficult to perform with 
operators.\\
\\
In summary, there is no mathematical inconsistency, however, there are 
doubts about the physical correctness of the Hamiltonian constraint operator
presently proposed although no proof exists so far that it is 
necessarily wrong. In order to make progress on this issue, it seems 
that we need to develop first a semiclassical calculus for the theory,
more precisely, we need coherent states so that expectation values of 
operators and their commutators can be replaced, up to $\hbar$ 
corrections, by their classical values and Poisson brackets respectively
for which then the manipulations listed in 2) above can be carried out.
If that is possible, and the outcome of these calculations is the 
expected one, possibly after changing the operator by
making use of the available quantization ambiguities, 
then one would be able to claim that one has indeed 
constructed a quantum theory of GR with the correct classical limit.
Only then can one proceed to solve the theory, that is, to construct 
solutions, the physical inner product and the Dirac observables.
The development of a semiclassical calculus is therefore one of the 
``hot" research topics at the moment.

Another way to get confidence in the quantization method applied to the 
Hamiltonian constraint is to study model systems for which the answer is 
known. This has been done for 2+1 gravity \cite{41} and for quantum cosmology
to which we turn in the next section.

\section{Loop Quantum Cosmology}
\label{s3.2}

\subsection{A New Approach To Quantum Cosmology}
\label{s3.2.1}

The traditional approach to quantum cosmology consists in a so-called 
mini -- superspace quantization, that is, one imposes certain spacetime 
Killing symmetries on the metric, plugs the symmetric metric into the 
Einstein Hilbert action and obtains an effective action which depends 
only on a finite number of degrees of freedom. Then one canonically 
quantizes this action. Thus one symmetrizes before quantization. These 
models are of constant interest and have 
natural connections to inflation. See e.g. \cite{45} for recent reviews.

What is not perfect about these models is that 1) not only do they switch off
all but an infinite number of degrees of freedom, but 2) also the 
quantization method applied to the reduced model usually is quite 
independent from that applied to the full theory. A fundamental approach to
quantum cosmology will be within the full theory and presumably involves 
the construction of semiclassical physical states whose probability 
amplitude is concentrated on, say a Friedmann -- Robertson -- Walker (FRW)
universe. This would cure both drawbacks 1) and 2). At the moment we 
cannot really carry out such a programme since the construction of the full
theory is not yet complete. However, one can take a more modest, hybrid
approach, where while dealing only with a finite number of degrees of 
freedom one {\it takes over all the quantization machinery from the full 
theory} ! Roughly speaking, one works on the space ${\cal H}_{kin}$ of 
the full theory but considers only states therein which satisfy the 
Killing symmetry. Hence one symmetrizes after quantization which amounts to
considering only a finite subset of holonomies and fluxes. This has the 
advantage of leading to a solvable model while preserving pivotal 
structures of the 
full theory, e.g. the volume operator applied to symmetric states will still
have a discrete spectrum as in the full theory while in the traditional
approaches it is continuous. Such a programme has been carried out in great
detail by Bojowald in a remarkable series of papers \cite{46} and his 
findings are indeed spectacular, should they extend to the full theory:
It turns out that the details of the quantum theory are {\it drastically 
different} from the traditional minisuperspace approach. In what follows 
we will briefly describe some of these results, skipping many of the 
technical details.

\subsection{Spectacular Results}
\label{s3.2.2}

Consider the FRW line element (in suitable coordinates)
\be \label{3.2.1}
ds^2=-dt^2+R(t)^2[\frac{dr^2}{1-kr^2}+r^2d\Omega_2^2]
=:-dt^2+R(t)^2 q^0_{ab} dx^a dx^b
\ee
The universe is closed/flat/open for $k=1/0/-1$. The only dynamical 
degree of freedom left is the so-called scale factor $R(t)$ which 
describes the size of the universe and its conjugate momentum.
The classical big bang singularity corresponds to the fact that the Einstein
equations predict that $\lim_{t\to 0} R(t)=0$ at which the metric
(\ref{3.2.1}) becomes singular and the inverse scale factor 
$1/R(t)$ blows up (the curvature will be $\propto 1/R(t)^2$ so this 
singularity is a true curvature singularity).

\begin{figure}
\includegraphics[width=10cm,height=7cm]{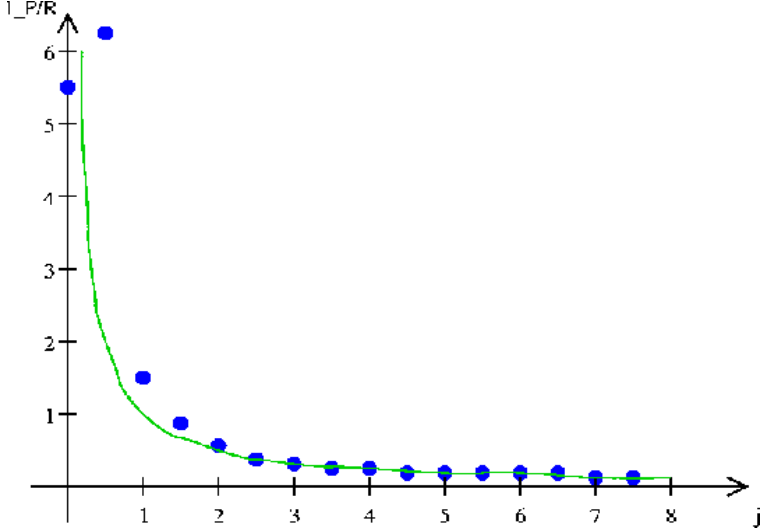}
\caption{Spectrum of the inverse scale factor}
\label{f12}
\end{figure}

We are interested in whether the curvature singularity $1/R\to\infty$ exists 
also in the quantum theory. To study this we notice that for 
(\ref{3.1.2}) $\det(q)=R^6\det(q^0)$. Hence, up to a numerical factor 
this question is equivalent to the question whether the operator 
corresponding to $1/\root{6} \of{\det(q)}$, when applied to symmetric states,
is singular or not. However, we saw in the previous section that 
one can trade a negative power of $\det(q)$ by a Poisson bracket with the 
volume operator. In \cite{46} precisely this, for the Hamiltonian 
constraint, {\it essential quantization technique} is applied which is why
{\it this model tests some aspects of the quantization of the 
Hamiltonian 
constraint}. Now it turns out that this operator, applied to symmetric 
states, leads to an operator $\widehat{\frac{1}{R}}$ which is diagonalized 
by (symmetric) SNWF's {\it and the spectrum is bounded} ! In figure 
\ref{f12} we plot the qualitative behaviour of the eigenvalues 
$\ell_P\lambda_j$ as a function of $j$ where $j$ is the spin label of a 
gauge invariant SNWF with a graph consisting 
of one loop only (that only such states are left follows from a
systematic analysis which defines what a symmetric SNWF is).
One can also quantize the operator $\hat{R}$ and one sees that its 
eigenvalues are essentially given by $j\ell_P$ up to a numerical factor. 
Thus the classical singularity corresponds to $j=0$ and one expects 
the points $\lambda_j\ell_p$ at the values $1/j$ on the curve $\ell_P/R$.
Evidently the spectrum is discrete (pure point) and bounded, at the 
classical singularity it is {\it finite}. In other words, the quantum 
universe never decreases to zero size. For larger $j$, in fact already for 
$R$ of the order of ten Planck lengths and above, the spectrum follows the 
classical curve rather closely hinting at a well-behaved classical 
limit (correspondence principle).

Even more is true: One can in fact quantize the Hamiltonian constraint by 
the methods of the previous section and solve it exactly. One obtains
an eighth order difference equation (in $j$). The solution 
therefore depends non-trivially on the initial condition. What is 
surprising, however, is the fact that only one set of initial conditions 
leads to the correct classical limit, thus in loop quantum cosmology 
initial conditions are {\it derived rather than guessed}. One can 
even propagate the quantum Einstein equations through the classical 
singularity and arrives at the picture of a bouncing universe.

Finally one may wonder whether these results are qualitatively affected 
by the operator ordering ambiguities of the Hamiltonian constraint. 
First of all one finds that these results hold only if one orders the 
loop in (\ref{3.1.18}) to the left of the volume operator as written there.
However, one is not forced to work with the holonomy around that loop 
in the fundamental representation of $SU(2)$, there is some flexibility
\cite{44} and one can choose a different one, say $j_0$. It turns out 
that the value $j_0$ influences the onset of classical behaviour, that is,
the higher $j_0$ the higher the value $j(j_0)$ from which on the 
spectrum in figure \ref{f12} lies on the curve $1/j$. Now this is important
when one copuples, say scalar matter because the operator 
$\widehat{\frac{1}{R}}$ enters the matter part of the Hamiltonian 
constraint and modifies the resulting effective equation for $R(t)$ in the 
very early phase of the universe and leads to a {\it quantum gravity driven
inflationary period} whose duration gets larger with larger $j_0$ !

Thus, loop quantum cosmology not only confirms aspects of the 
quantization of the Hamiltonian constraint but also predicts astonishing
deviations from standard quantum cosmology which one should rederive in 
the full theory.

\section{Path Integral Formulation: Spin Foam Models}
\label{s3.3}

\subsection{Spin Foams from the Canonical Theory}
\label{s3.3.1}

Spin Foam models are the fusion of ideas from topological quantum field 
theories and loop quantum gravity, see e.g. \cite{47} for a review, 
especially the latest, most updated one by Perez.
The idea that connects these theories is actually quite simple to explain
at an heuristic level:\\
\\
If we forget about 1) all functional analytic details, 2) the fact 
that the operator valued distributions corresponding to the Hamiltonian
constraint $\hat{C}(x)$ do not mutually commute for different $x\in\sigma$
and 3) that the Hamiltonian constraint operators $\hat{C}(N)$ are 
certainly not self-adjoint, at least as presently formulated, then 
we can formally write down the complete space of solutions to the 
Hamiltonian constraint as a so-called ``rigging map" (see e.g. \cite{0})
\be \label{3.3.1}
\bar{\eta}:\;{\cal D}_{kin}\to {\cal D}_{phys}^\ast;\;
f\mapsto \delta[\hat{C}]\;f:=[\prod_{x\in\Sigma}\;\delta(\hat{C}(x))\;f]
\ee
(where $\bar{\eta}=\mbox{c.c}\cdot\eta$ is the 
complex conjugate of the actual anti-linear rigging map).
Here the $\delta-$distribution of an operator is defined via the spectral
theorem (assuming the operator to be self-adjoint). Notice that we do not 
need to order the points $x\in\sigma$ as we assumed the $\hat{C}(x)$ to be 
mutually commuting for the moment and only under this assumption it is
true that, at least formally $\bar{\eta}[f](\hat{C}(N)f')=0$ 
(exercise)\footnote{At an even more formal level $\bar{\eta}[f]$
is also a solution in the non-commuting case if, as is the case with the 
currently proposed $\hat{C}$, the algebra with the spatial diffeomorphism 
constraint closes}.
Now we use the formula $\delta(x)=\int_\Rl \frac{dk}{2\pi} e^{ikx}$
to write the functional $\delta-$distribution $\delta[\hat{C}]$ as
a {\it path integral}
\be \label{3.3.2}
\delta[\hat{C}]=
\int_{{\cal N}'} [DN] e^{i\hat{C}(N)}
\ee
where we have neglected an infinite constant as usual in this formal
business. Here ${\cal N}'$ is the space of lapse functions at a fixed 
time. Let us introduce also the space of lapses with arbitray time 
dependence ${\cal N}_{t_1,t_2}$ in $t\in [t_1,t_2]$. Then, up to an 
infinite constant one can verify that
\be \label{3.3.2a}
\delta[\hat{C}]=
\int_{{\cal N}_{t_1}^{t_2}} [DN] 
e^{i\int_{t_1}^{t_2} dt \int_\Sigma d^3x N(x,t) \hat{C}(x)}
\ee
The rigging map machinery then tells us that the scalar product 
on the image of the rigging map is simply given by
\be \label{3.3.3}
<\bar{\eta}(f),\bar{\eta}(f')>_{phys}:=<f,\bar{\eta}(f')>_{kin}
=\int_{{\cal N}_{t_1}^{t_2}} [DN] <f,e^{i\int_{t_1}^{t_2} dt \hat{C}(N_t)} 
f'>_{kin}
\ee
This formula looks like a propagator formula, that is, like a transition 
amplitude between an initial state $f'$ on $\Sigma_{t_1}$ and a final 
state $f$ on $\Sigma_{t_2}$ after 
a multi-fingered time evolution generated by $\hat{C}(N_t)$.
In fact, if we use the Taylor expansion of the exponential function
and somehow regularize the path integral then the expansion coefficients
$<T_s,\hat{C}(N_t)^n T'{s'}>_{kin}$ can be interpreted as probability 
amplitude of the evolution of the SNW state $T'_{s'}$ to reach the SNW
state $T_s$ after $n$ time steps (recall figure \ref{f11}). 

Now by the usual formal manipulations that allow us to express 
a unitary operator $e^{i(t_2-t_1) \hat{H}}$ as a path integral over the 
classical pase space $\cal M$ (the rigorous version of which is the
Feynman -- Kac formula, e.g. \cite{49}) one can rewrite (\ref{3.3.3})
as 
\be \label{3.3.4}
<\bar{\eta}(f),\bar{\eta}(f')>_{phys}
=\int [DN\;D\vec{N}\;D\Lambda\;DA\;DE] <f,e^{i S} f'>_{kin}
\ee
where $S$ is the Einstein-Hilbert action written in canonical form in 
terms of the variables $A,E$, that is
\be \label{3.3.5}
S=\int_\Rl dt \int_\Sigma d^3x \{\dot{A}_a^j E^a_j-[-\Lambda^j G_j+
N^a V_a+NC]\}
\ee
and we have simultaneously included also
projections on the space of solutions to the Gauss and vector constraint.
Now the action (\ref{3.3.5}) is the $3+1$ split of the following 
{\it covariant action}
\be \label{3.3.6}
S=\int_M \{\Omega_{IJ}\wedge[\epsilon^{IJKL}-\beta^{-1}\eta^{IK}\eta^{JK}]
e_K\wedge e_L\}
\ee
discovered in \cite{50} where $\beta$ is the Immirzi parameter. 
Here $\Omega_{IJ}$ is the (antisymmetric) curvature two-form of an 
(antisymmetric) $SL(2,\Cl)$ connection 
one-form $\omega_{IJ}$ with Lorentz indices $I,J,K,..=0,1,2,3$, $\eta$ is 
the Minkowski metric and $e^I$ is the co-tetrad one-form. The first 
term in (\ref{3.3.6}) is called the Palatini action while the second
term is topological (a total differential modulo the equations of motion). 
The relationn between the four-dimensional fields 
$\omega^{IJ}_\mu,e^I_\mu$ (40 components) and 
the three-dimensional fields $A_a^j,E^a_j,\Lambda^j,N,N^a$ (25 components)
can only be established if certain so-called second class constraints
\cite{17} are solved.

\subsection{Spin Foams and BF -- Theory}
\label{s3.3.2}

Thus, it is formally possible to write the inner product between physical 
states as a {\it covariant} path integral for the classical canonical 
action and using only the kinematical inner product, thus providing a 
bridge between the covariant and canonical formalism. However, this 
bridge is far from being rigorously established as we had to perform
many formal, unjustified manipulations. Now rather than justifying the 
steps that lead from $\hat{C}$ to (\ref{3.3.4}) one can turn the logic 
upside down and {\it start} from a manifestly covariant formulation and 
{\it derive} the canonical formultion. This is the attitude that one 
takes among people working actively on spin foam models. Thus, let us 
forget about the topological term in (\ref{3.3.6}) and consider only the 
Palatini term. Then the Palatini action has precisely the form 
of a BF -- action
\be \label{3.3.7}
S_{BF}=\int_M \Omega_{IJ}\wedge B^{IJ}
\ee
just that the (antisymmetric) two-form field $B^{IJ}$ is not arbitrary
(it would have 36 independent components), it has to come from a 
tetrad with only 16 independent components, 
that is, it has to be of the form $\epsilon^{IJKL} e_K\wedge e_L$. 
\begin{Exercise} \label{ex3.3.1}  ~~~\\
Show that the condition that $B$ comes from a tetrad is 
almost\footnote{Another solution is $B^{IJ}=e^I\wedge e^J$ but 
this possibility is currently not discussed.} equivalent to the 
{\it simplicity constraint}
\be \label{3.3.8}
\epsilon_{IJKL} B^{IJ}_{\mu\nu} 
B^{KL}_{\rho\sigma}=c\epsilon_{\mu\nu\rho\sigma}
\ee
for some spacetime scalar density $c$ of weight one.
\end{Exercise}
The reasoning is now as follows: BF -- theory without the constraint
(\ref{3.3.8}) is a topological field theory, that is, it has no local
degrees of freedom. Therefore quantum BF -- theory is not really a QFT but
actually a quantum mechanical system and can therefore be handled 
much more easily than gravity. Let us now write an action equivalent
to the Palatini action given by
\ba \label{3.3.9}
S'_P[\omega,B,\Phi]&=&S_{BF}[\omega,B]+S_I[B,\Phi]\nonumber\\
S_I[B,\Phi] &:=& \int_M \Phi^{\mu\nu\rho\sigma}\epsilon_{IJKL}
B^{IJ}_{\alpha\beta} B^{KL}_{\gamma\delta}[
\delta^\alpha_\mu\delta^\beta_\nu\delta^\gamma_\rho\delta^\delta_\sigma
-\frac{1}{4!} 
\epsilon^{\alpha\beta\gamma\delta}\epsilon_{\mu\nu\rho\sigma}]
\ea
where the Lagrange multiplier $\Phi^{\mu\nu\rho\sigma}$ \cite{51}
is a four 
dimensional tensor density of weight one, symmetric in the index pairs 
$(\mu\nu)$ and $(\rho\sigma)$ and antisymmetric in each index pair.
Thus, $\Phi$ has $(6\cdot 7)/2=21$ independent components of which 
the totally skew component is projected out in (\ref{3.3.9}), leaving
us with $36-16=20$ independent components. Hence the Euler Lagrange
equations for $\Phi$ precisely delete the amount of unwanted degrees of 
freedom in $B$ and impose the simplicity constraint. Hence, 
classically $S'_P[\Omega,B,\Phi]$ and $S_P[\Omega,e]$ are equivalent. 
Thus, if we write a path integral for $S'_P$ and treat the Lagrange 
multiplier term $S_I$ in (\ref{3.3.9}) as an interaction Lagrangean  
(a perturbation) to BF -- theory, then
we can make use of the powerful techniques that have been developed for 
the path integral quantization for BF -- theory and its perturbation 
theory. 
\begin{Exercise} \label{ex3.3.2}  ~~~\\    
i) Write the Euler Lagrange equations for BF -- theory and conclude that
the solutions consist of flat connections $\omega$ and gauge invariant
$B-$ fields. Conclude that $\omega$ can be gauged to $zero$ by $SL(2,\Cl)$ 
transformations locally and that then $B$ is closed, that is, locally 
exact by Poincar\'e's theorem. Now, verify that the BF -- action is not 
only invariant under local $SL(2,\Cl)-$transformations but also under
\be \label{ex3.3.2a} 
B^{IJ}=\mapsto B^{IJ}+(D\wedge \theta)^{IJ}=
B^{IJ}+d\theta^{IJ}+\omega^I\;_K\wedge 
\theta^{KJ}+\theta^{IK}\wedge \omega_K\;^J
\ee
for some $sl(2,\Cl)$ valued one -- form $\theta$ and that therefore also
$B$ can be gauged to zero locally.\\
Hint:\\
Use the Bianchi identity for $\Omega$.\\
ii) Perform the Legendre transformation and conclude that there are as 
many first class constraints as canonical pairs so that again at most
a countable number of global degrees of freedom can exist.
\end{Exercise}
One may wonder how it is possible that a theory with less 
kinematical degrees of freedom has more dynamical (true) degrees of 
freedom. The answer is that BF -- theory has by far more symmetries than 
the Palatini theory, thus when constraining the number of degrees of 
freedom we are freezing more symmetries than we deleted degrees of 
freedom.  

Let us now discuss how one formulates the path integral corresponding
to the action (\ref{3.3.9}). It is formally given by
\be \label{3.3.10}
K_P(\Sigma_{t_1},\Sigma_{t_2})=\int [D\omega\;DB\;D\Phi] 
e^{iS'_P[\omega,B,\Phi]}
\ee
where $\Sigma_{t1},\Sigma_{t_2}$ denote suitable boundary conditions
specified in more detail below. Suppose we set $\Phi=0$, then 
(\ref{3.3.10}) is a path integral for BF -- theory and the integral 
over $B$ results in the functional $\delta-$distribution $\delta[\Omega]$
imposing the flatness of $\omega$. Now flatness of a connection is 
equivalent to trivial holonomy along contractable loops by the 
Ambrose -- Singer theorem. If one regularizes the path integral 
by introducing a triangulation $\tau$ of $M$, then $\delta[F]$ can be 
written as $\prod_\alpha \delta(\omega(\alpha),1)$ where the product
is over a generating system of independent, contractible loops in $\tau$ 
and $\delta(\omega(\alpha),1)$ denotes the $\delta-$distribution on 
$SL(2,\Cl)$ with respect to the Haar measure. Since $SL(2,\Cl)$ is a 
non-compact group, the $\delta-$distribution is a direct integral 
over irreducible, unitary representations rather than a direct sum as it 
would be the case for compact groups (Peter\&Weyl theorem). Such 
representations are infinite dimensional and are labelled by 
a continuous parameter $\rho\in\Rl_0^+$ and a discrete parameter 
$n\in\Nl_0^+$. Thus, one arrives at a triangulated a spin foam model:
For a fixed triangulation one integrates (sums) over all possible ``spins"
$\rho$ ($n$) that label the generating set of loops (equivalently: the 
faces that they enclose) of that triangulated four manifold. The analogy
with the state sum models for TQFT's is obvious.

Now what one does is a certain jump, whose physical implication is 
still not understood: Instead of performing
perturbation theory in $S_I$ one argues that formally integrating over 
$\Phi$ and thus imposing the simplicity constraint is equivalent to the 
restriction of the direct integral that enters the 
$\delta-$distributions to simple representations, that is, representations 
for which either $n=0$ or $\rho=0$. In other words, one says that 
the triangulated Palatini path integral is the same as the triangulated BF 
path integral restricted to simple representations. To motivate this 
argument, one notices that upon canonical quantization of BF theory 
on a triangulated manifold the 
$B$ field is the momentum conjugate to $\omega$ and if one quantizes on a 
Hilbert space based on $sl(2,\Cl)$ connections using the Haar measure
(similar as we have done for $SU(2)$ for a fixed graph), its 
corresponding flux
operator $\hat{B}_{IJ}(S)$ becomes a linear combination of right invariant 
vector fields $R^{IJ}$ on $SL(2,\Cl)$. 
Now the simplicity constraint becomes the condition that the second 
Casimir operator $R^{IJ} R^{KL}\epsilon_{IJKL}$ vanishes. However, on
irreducible representations this operator is diagonal with 
eigenvalues $n\rho/4$. While this is a strong motivation, it is certainly
not sufficient justification for this way of implementing the 
simplicity constraint in the path integral because it is not clear how 
this is related to integrating over $\Phi$.

In any case, if one does this then one arrives at (some version of) the
Lorentzian Barrett -- Crane model \cite{52}. Surprisingly, for a large 
class of triangulations $\tau$ the amplitudes 
\be \label{3.3.11}
K^\tau_P(\Sigma_{t_1},\Sigma_{t_2}):=
{[}\int [D\omega\;DB] e^{iS_{BF}[\omega,B]}]_{|{\rm simple reps.}}
\ee
actually converge although one integrates over a non-compact group !
This is a non-trivial result \cite{53}. The path integral is then over 
all possible representations that label the faces of a spin foam 
and the boundary conditions keep the representations on the boundary 
graphs, that is, spin networks fixed ($SL(2,\Cl)$ reduces to the $SU(2)$
on the boundary). This also answers the question of 
what the boundary conditions should be.  
 
There is still an open issue, namely how one should get rid of the 
regulator (or triangulation) dependence. Since BF -- theory is a 
topological QFT, the amplitudes are automatically triangulation 
independent, however, this is certainly not the case with GR. One 
possibility is to sum over triangulations and a concrete proposal of how 
to weigh the contributions from different triangulations comes from the 
so-called field theory formulation of the theory \cite{54}. Here 
one reformulates the BF -- theory path integral as the path integral for a 
scalar field on a group manifold which in this case is a certain power of 
$SL(2,\Cl)$. The action for that scalar field has a free piece and an 
interaction piece and performing the perturbation theory (Feynman graphs 
!) for that field theory is equivalent to the sum over BF -- theory 
amplitudes for all triangulated manifolds {\it with precisely defined 
weights}. This idea can be 
straightforwardly applied also to our context where the restriction
to simple representations is realized by imposing corresponding 
restrictions (projections) on the scalar field. \\
\\
Summarizing, spin foam models are a serious attempt to 
arrive at a covariant formulation of QGR but many issues are still 
unsettled, e.g.:\\
1)\\
There is no clean equivalence with the Hamiltonian formulation as we 
have seen. Without that it is unlcear how to interpret the spin foam
model amplitude and whether it has the correct classical limit. In order
to make progress on the issue of the classical limit, model independent
techniques for constructing ``causal spin foams" \cite{54a} with a built
in notion of quantum causality and renormalization methods \cite{54b}, 
which should allow in principle the derivation of a low energy effective 
action, have been developed.\\
2)\\
The physical correctness of the Barrett -- Crane model is unclear.
This is emphasized by recent results within the Euclidean formulation 
\cite{55} which suggest that the classical limit is far off GR since the 
amplitudes are dominated by spin values close to zero. This 
was to be expected because in the definition of the Barrett -- Crane 
model there is a certain flexibility concerning the choice of the 
measure that replaces $[D\omega\;DB]$ at the triangulated level and 
the result \cite{55} indicates that one must gain more control on that 
choice.\\
3)\\
It is not even clear that these models are four -- dimensionally
covariant: One usually defines that the amplitudes for a fixed 
triangulation are the same for any four - diffeomorphic triangulation.
However, recent results \cite{56} show that this natural definition 
could result nevertheless in anomalies. This problem is again related to 
the choice of the measure just mentioned.\\
\\
Thus, substantially more work is required in order to fill in the 
present gaps but the results already obtained are very promising indeed.

\section{Quantum Black Holes}
\label{s3.4}

\subsection{Isolated Horizons}
\label{s3.4.1}

Any theory of quantum gravity must face the question whether it can 
reproduce the celebrated result due to Bekenstein and Hawking \cite{57}
that a black hole in a spacetime $(M,g)$ should account for a quantum
statisical entropy given by
\be \label{3.4.1}
S_{BH}=\frac{\mbox{Ar}(H)}{4\ell_P^2}
\ee
where $H$ denotes the two-dimensional event horizon of the blak hole.
This result was obtained within the framework of QFT on Curved SpaceTimes
(CST) and should therefore be valid in a semiclassical regime in which 
quantum fluctuations of the gravitational field are neglible (large black 
holes). The most important question from the point of view of a 
microscopical theory of quantum gravity is, {\it what are the 
microscopical degrees of freedom that give rise to that entropy}. In 
particular, how can it be within a quantum field theory with an infinite 
number of degrees of freedom, that this entropy, presumably a measure for 
our lack of information of what happens behind the horizon, comes out 
finite. 

In \cite{58} the authors performed a bold computation: For
{\it any} surface $S$ and any positive number $A_0$ they asked the 
question how many SNW states there are in QGR such that the area operators
eigenvalues lie within the interval $[A_0-\ell_P^2,A_0+\ell_P^2]$.
This answer is certainly infinite because a SNW can intersect $S$ in an 
uncountably infinite number of different positions without changing the 
eigenvalues. This divergence can be made less severe by moding
out by spatial diffeomorphisms which we can use to map these different SNW
onto each other in the vicinity of the surface. However, since there are 
still an infinite number of non-spatially difeomorphic states which look 
the same in the vicinity of the surface but different away from it, the 
answer is still divergent. Therefore, one has to argue that one must 
not count information off the surface, maybe invoking the Hamiltonian
constraint or using the information that $S=H$ is not an arbitrary 
surface but actually the horizon of a black hole. Given this assumption, 
the result of the, actually quite simple counting problem came rather 
close to (\ref{3.4.1}) with the correct factor of $1/4$.

Thus the task left is to justify the assumptions made and to make the 
entropy counting water -- tight by invoking the information that 
$H$ is a black hole horizon. The outcome of this analysis created a whole
industry of its own, known under the name ``isolated horizons", which 
to large part is a beautiful new chapter in classical GR. In what follows 
we will focus only on a tiny fraction of the framework, mostly 
concentrating on the ingredients essential for the quantum formulation.
For reviews see \cite{59} which also contain a complete list of references
on the more classical aspects of this programme, the pivotal papers 
concerning the quantum applications are \cite{59a}. 

\begin{figure}
\includegraphics[width=10cm,height=7cm]{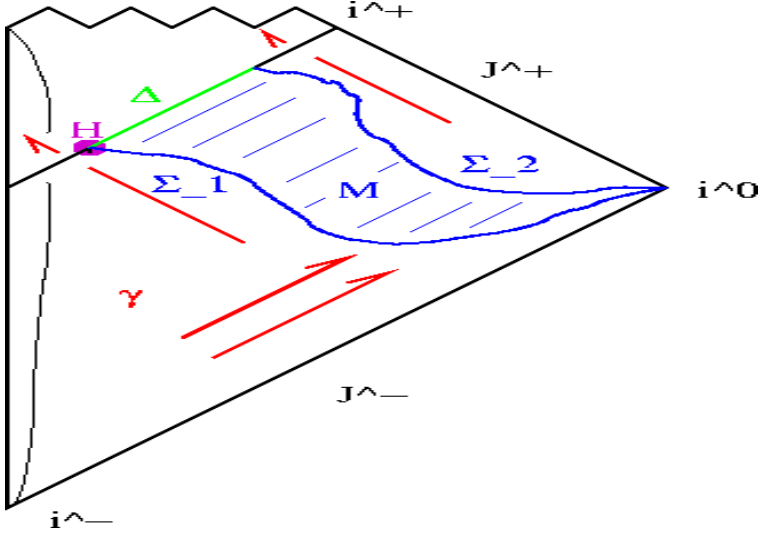}
\caption{An isolated horizon $\Delta$ boundary of a piece $M$ (shaded) of 
spacetime 
also bounded by spacelike hypersurfaces $\Sigma_1,\Sigma_2$. 
Radiation $\gamma$ may enter or leave $M$ and propagate into the 
singularity before or after the isolated horizon has formed but must 
not cross $\Delta$. An intersection of a spacelike hypersurface $\Sigma$
with $\Delta$ is denoted by $H$ which has the topology of a sphere.} 
\label{f13}
\end{figure}

By definition, an 
event horizon is the external boundary of the part of $M$ that does not
lie in the past of null future infinity $J^+$ in a Penrose 
diagramme. From an operational point of view, this definition makes little 
sense because in order to determine whether a candidate is an event 
horizon, one must know the whole spacetime $(M,g)$ which is never possible 
by measurements which are necessarily loacal in spacetime (what looks like 
an eternal black hole now could capture some dust later and the horizon 
would change its location). Thus one looks for some local substitute 
of the notion of an event horizon which captures the idea that the black 
hole has come to some equilibrium state at least for some amount of time.
This is roughly what an isolated horizon $\Delta$ is, illustrated 
in figure \ref{f13}.
More in technical details we have the following.
\begin{Definition} \label{def3.4.1} ~~~~\\
A part $\Delta$ of the boundary $\partial M$ of a spacetime $(M,g)$ is 
called an isolated horizon, provided that\\
1) $\Delta\equiv \Rl\times S^2$ is a null hypersurface and has zero shear
and expansion\footnote{Recall the notions of shear, expansion and twist
of a congruence of vector fields in connection with Raychaudhuri's 
equation.}.\\
2) The field equations and matter energy conditions hold at $\Delta$.\\
3) $g$ is Lie derived by the null generator $l$ of $\Delta$ at $\Delta$.   
\end{Definition}
The canonical formulation of a field theory on a manifold $M$ with 
boundary $\Delta$ must involve boundary conditions at $\Delta$ in order 
that the variation principle be well-defined (the action must be 
functionally differentiable). Such boundary conditions usually give birth
to boundary degrees of freedom \cite{60} which would normally be absent but 
now 
come into being because (part of the) gauge transformations are forced to
become trivial at $\Delta$. In the present situation what happens is that 
the boundary term is actually a {\it $U(1)$ Chern-Simons 
action}\footnote{It was observed first in \cite{60a} that general 
relativity in terms of connection variables and in the presence of 
boundaries leads to Chern -- Simons boundary terms.} 
\be \label{3.4.2}
S_{CS}=\frac{A_0}{\pi\beta} 
\int_{\Delta} W\wedge dW=\int_\Rl dt \int_{H} d^2y 
\epsilon^{IJ}[\dot{W}_I W_J+W_t (dW)_{IJ}]
\ee
where $W$ is a $U(1)$ connection one form and $H=S^2=\Sigma\cap \Delta$ 
is a sphere.
The relation between the bulk fields $A_a^j,E^a_j$ and the boundary fields
$W_I,\;I=1,2$ is given by
\be \label{3.4.3}
X_H^\ast A^j=W r^j\mbox{ and } 
[X_H^\ast(\ast E)_j]r_j=-\frac{A_0}{2\pi\beta} dW
\ee
where $X_H:\;H\to \Sigma$ is the embedding of the boundary $H$ of 
$\Sigma$ into $\Sigma$ and $r^j$ is an arbitrary but fixed unit vector in 
$su(2)$ which is to be preserved under $SU(2)$ gauge transformations at 
$\Delta$ and therefore reduces $SU(2)$ to $U(1)$. The number $A_0$ is 
the area of $H$ as measured by $g$ which turns out to be a constant of the 
motion as a consequence of the field equations. The existence of $r^j$ 
is a consequence of definition (\ref{def3.4.1}) and $\ast E$ is the 
natural metric independent two -- form dual to $E$. 

\subsection{Entropy Counting}
\label{s3.4.2}

\begin{figure}
\includegraphics[width=10cm,height=7cm]{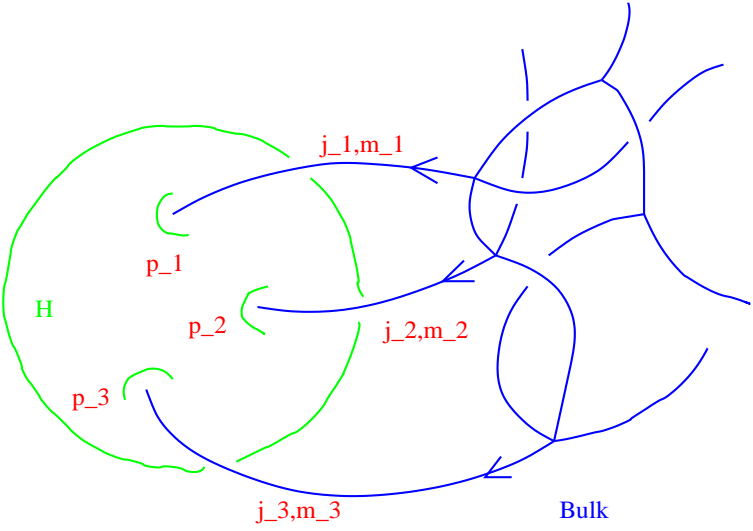}
\caption{Punctures, spins, magnetic quantum numbers and entropy counting.
Only the relevant boundary data are shown, the bulk information is traced 
over.}
\label{f14}
\end{figure}

One now has to quantize the system. This consists of several steps whose 
details are complicated and which we will only sketch in what follows.
\begin{itemize}
\item[i)] {\it Kinematical Hilbert Space}\\
The bulk and boundary degrees of freedom are independent of each other,
therefore we choose ${\cal H}_{kin}={\cal H}^\Sigma_{kin}
\otimes {\cal H}^H_{kin}$ where both spaces are of the form  
$L_2(\ab,d\mu_0)$ just that the first factor is for an $SU(2)$ bundle
over $\Sigma$ while the second is for an $U(1)$ bundle over $H$. 
\item[ii)] {\it Quantum Boundary conditions}\\
Equation (\ref{3.4.3}) implies, in particular, that in quantum theory
we must have schematically
\be \label{3.4.4}
[\widehat{[X_H^\ast(\ast E)_j]r_j}]\otimes 
\mbox{id}_{{\cal H}^H_{kin}}=
\mbox{id}_{{\cal H}^\Sigma_{kin}}\otimes 
[-\frac{A_0}{2\pi\beta} \widehat{dW}]
\ee
Now we have seen in the bulk theory that we have discussed in great detail
throughout this review, that $\ast E$ is an operator valued distribution 
which must be smeared by two -- surfaces in order to arrive at the 
well-defined electric fluxes. Since (\ref{3.4.4}) is evaluated at 
$H$, this flux operator will non-trivially act only on SNWF's $T_s$ which 
live in the bulk 
but intersect $H$ in punctures $p\in H\cap \gamma(s)$. Now the 
distributional character of the electric fluxes implies that the left
hand side of (\ref{3.4.4}) is non-vanishing only at those punctures.
Thus the curvature of $W$ is flat everywhere except for the punctures 
where it is distributional. 

Consider now SNWF's $T_s$ of the bulk theory and those of the boundary 
theory $T'_c$. Then $\widehat{[X_H^\ast(\ast E)_j]r_j}]$ acts on $T_s$
like the $z-$component of the angular momentum operator and will have 
distributional eigenvalues proportional to the magnetic quantum numbers
$m_e$ of the edges with punctures $p=e\cap H$ and spin $j_e$ where
$m_e\in \{-j_e,-j_e+1,..,j_e\}$.   
\item[iii)] {\it Implementation of Quantum Dynamics at $\Delta$}\\
It turns out that    
$\widehat{X_H^\ast(\ast E)_j]r_j}$ and $\widehat{dW}$ are the generators
of residual $SU(2)$ gauge transformations close to $X_H(H)$ and of 
$U(1)$ on $H$ respectively. Now these residual $SU(2)$ transformations 
are frozen to $U(1)$ transformations by $r_j$ and the most general 
situation 
in order for a state to be gauge invariant is that these residual $SU(2)$
transformations of the bulk theory and the $U(1)$ transformations of the 
boundary theory precisely cancel each other. It turns out that this
cancellation condition is precisely given by the quantum boundary 
condition (\ref{3.4.4}). Thus the states that solve the Gauss constraint
are linear combinations of states of the form $T_s\otimes T'_c$ 
where the boundary data of these states are punctures $p\in {\cal P}$ 
where 
$p\in \gamma(s)\cap H$, the spins $j_p=j_{e_p}$ of edges $e\in 
E(\gamma(s))$ with $e_p\cap H=p$ and their magnetic quantum 
numbers $m_p=m_{e_p}$. However, due to the specific features of the 
geometrical
quantization of Chern-Simons theories \cite{61} the $m_p$ cannot be 
specified freely, they have to satisfy the constraint 
\be \label{3.4.5}
\sum_{p\in {\cal P}} 2m_p=0 \mbox{ mod }k,\;\;
k=\frac{A_0}{4\pi \ell_P^2}
\ee
where $k$ is called the level of a Chern Simons theory which is
constrained to be an integer due to Weil's quantization obstruction
cocycle criterion of geometric quantization \cite{26}
and comes about as follows: The $T'_c$ are actually fixed to be 
$\Theta-$functions of level $k$ labelled by integers $a_p$ which satisfy
the gauge invariance condition
\be \label{3.4.6}
2m_p=-a_p \mbox{ mod }k,\;\;\sum_p a_p=0\mbox{ mod }k
\ee

Next, the spatial diffeomorphism constraint
of the bulk theory tells us that the position of the punctures on $H$ 
are not important, important is only their number. 

Finally, one of 
the boundary conditions at $\Delta$ implies that the lapse becomes 
trivial $N=0$ at $H$ if $\hat{C}(N)$ is to generate an infinitesimal 
time reparameterization\footnote{This does not mean that the lapse of a 
classical
isolated horizon solution must vanish at $S$, rather there is a subtle
difference between gauge motions and symmetries for field theories with 
boundaries \cite{60} where in this case symmetries map solutions to gauge 
inequivalent or equivalent ones respectively, if 
$N_{|H}\not=0$ or $N_{|H}=0$ respectively.}. Thus, luckily we can escape 
the open issues with the Hamiltonian constraint as far as the quantum
dynamics at $H$ is concerned. 
\end{itemize}
We can now come to the issue of entropy counting. First of all we notice
that Ar$(H)$ is a Dirac observable because $H$ is invariant under 
Diff$(H)$ and $N=0$ at $H$. Given $n$ punctures 
with spins $j_l,\;l=1,..,N$ the area eigenvalue for $H$ is 
\be \label{3.4.7}
\lambda(n,\vec{j})=8\pi \ell_P^2\beta \sum_{n=1}^n \sqrt{j_l(j_l+1)}
\ee
Now the physical Hilbert space is of the form 
\be \label{3.4.8}
{\cal H}_{phys}=\oplus_{n,\vec{j},\vec{m},\vec{a}=_k-2\vec{m}}\;\;
{\cal H}^B_{n,\vec{j},\vec{m}}\otimes {\cal H}^{BH}_{n,\vec{j},\vec{m}}
\otimes {\cal H}^H_{n,\vec{a}}
\ee
where ${\cal H}^{BH}_{n,\vec{j},\vec{m}}$ describes bulk degrees of 
freedom
at $H$ corresponding to the black hole (finite dimensional),
${\cal H}^B_{n,\vec{j},\vec{m}}$ describes bulk degrees of freedom 
away from $H$ and finally
${\cal H}^H_{n,\vec{a}}$ describes Chern -- Simons degrees of freedom 
which are completetly fixed in tems of $\vec{m}$ due to reasons of gauge
invariance (\ref{3.4.6}). The situation is illustrated in figure 
\ref{f14}.
Let $\delta>0$ and let $S_{A_0,\delta}$ be the set of eigenstates 
$\psi_{n,\vec{j},\vec{m}}\in {\cal H}^{BH}_{n,\vec{j},\vec{m}}$
of the area operator such that the eigenvalue lies in the interval
$[A_0-\delta,A_0+\delta]$ and $N_{A_0,\delta}$ their number.
Define the density matrix
\be \label{3.4.9}
\hat{\rho}_{BH}=\mbox{id}_B\otimes[\frac{1}{N}_{A_0,\delta}
\sum_{\psi\in S_{A_0,\delta}} |\psi><\psi|]\otimes \mbox{id}_H
\ee
The quantum statistical entropy from this microcanonical ensemble is given 
by
\be \label{3.4.10} 
S_{BH}=-\mbox{Tr}(\hat{\rho}_{BH}\ln(\hat{\rho}_{BH}))=\ln(N_{A_0,\delta})
\ee
Thus we just need to count states and the answer {\it will be finite 
because the area operator has an area gap}. 
\begin{Exercise} \label{ex3.4.1}  ~~~~\\
Estimate $N_{A_0,\delta}$ from above and below taking into account the 
constraint (\ref{3.4.5}) and that $k$ is an integer (purely combinatorical 
problem !). 
\end{Exercise}
The result of the counting problem is that $S_{BH}$ is indeed 
given by (\ref{3.4.1}) to leading order in $A_0$ (there are logarithmic 
corrections) for $\delta\approx \ell_P^2$ provided that 
\be \label{3.4.11}   
\beta=\frac{\ln(2)}{\pi\sqrt{3}}
\ee
Here the numbers $\ln(2),\sqrt{3}$ comes from the fact that the 
configurations with lowest
spin $j_l=1/2$ make the dominant contribution to the entropy 
with eigenvalue $\propto \beta n \sqrt{3}\approx A_0$ and number of states 
given by $N_{A_0,\delta}\approx 2^n$ that is, {\it two Boolean degrees of 
freedom per puncture} \cite{62}. This provides an explicit 
explanation for the origin of the entropy. Now fixing $\beta$ at
the value (\ref{3.4.11}) would make little sense would it be different for 
different types of 
black hole (that is, in presence of different matter, charges, rotation,
other hair,..). However, this is not the case !\\
\\
In summary, the analysis sketched above provides a self-contained 
derivation of $S_{BH}$ within QGR. The result is {\it highly non-trivial} 
because it was not to be expected from the outset that Loop Quantum 
Gravity, classical GR and Chern Simons theory would interact in such a 
harmonic way as to provide the expected result: 
Chern -- Simons theory is {\it very different} from QGR and still 
they have an interface at $H$. The result applies to
astrophysically interesting balck holes of the Schwarzschild type and does  
not require supersymmetry. Nevertheless, the calculation still has a
semiclassical input because the presence of the isolated horizon is fed 
in at the classical level already. It would be more satisfactory to have 
a {\it quantum definition of an (isolated) horizon} but this is a hard 
task and left for future research. Another unsolved problem then is the 
calculation of the Hawking effect from first principles.

\section{Semiclassical Analysis}
\label{s3.5}

\subsection{The Complexifier Machinery for Generating Coherent States}
\label{s3.5.1}

Let us first specify what we mean by semiclassical states.
\begin{Definition} \label{def3.5.1} ~~~~\\
Let be given a phase space ${\cal M},\;\{.,.\}$ with preferred Poisson 
subalgebra $\cal O$ of $C^\infty({\cal M})$ and a Hilbert space 
${\cal H},\;[.,.]$ together with an operator subalgebra $\hat{{\cal O}}$
of ${\cal L}({\cal H})$.
The triple ${\cal M},\;\{.,.\},{\cal O}$ is said to be a classical 
limit of the triple ${\cal H},\;[.,.],\hat{{\cal O}}$ provided that there
exists an (over)complete set of states $\{\psi_m\}_{m\in {\cal M}}$
such that for all $O,O'\in {\cal O}$ the 
{\it infinitesimal Ehrenfest property}
\ba \label{3.5.1}
&& 
|\frac{<\hat{O}>_m}{O(m)}-1|\ll 1 
\nonumber\\
&& 
|\frac{<[\hat{O},\hat{O}']>_m} {i\hbar\{O,O'\}(m)}-1|\ll 1 
\ea
and the {\it small fluctuation property}
\be \label{3.5.1a}
|\frac{<\hat{O}^2>_m}{<\hat{O}>_m^2}-1|\ll 1 
\ee
holds at generic\footnote{The set of points where (\ref{3.5.1}), 
(\ref{3.5.1a}) are 
violated should have small Liouville measure.} points in $\cal M$.
Here $<.>_m:=<\psi_m,.\psi_m>/||\psi_m||^2$ denotes the expectation value 
functional.
\end{Definition}
For systems with constraints, strictly speaking, semiclassical states 
should be physical states, that 
is, those that solve the constraints because we are not interested in 
approximating gauge degrees of freedom but only physical observables.
Only then are the predictions ($\hbar$ corrections to the classical
limit) of the theory reliable. In the present situation with QGR, however,
we are more interested in constructing {\it kinematical semiclassical
states} for the following reason: As we have shown, the status of the 
physical correctness of the Hamiltonian constraint operator $\hat{C}$
is unsettled. We would therefore like to test whether it has the correct
classical limit. This test is obviously meaningless on states which
the Hamiltonian constraint annihilates anyway. For the same reason it also 
does not make sense to construct semiclassical states which are 
at least spatially diffeomorphism invariant because the Hamiltonian 
constraint does not leave this space invariant.

The key question then is how to construct semiclassical states. 
Fortunately, for phase spaces which have a cotangent bundle structure 
as is the case with QGR, a rather general construction guideline is 
available \cite{63}, the so-called {\it Complexifier Method},  
which we will now sketch:\\

Let $({\cal M},\{.,.\})$ be a phase space with (strong) symplectic 
structure $\{.,.\}$ (notice that $\cal M$ is allowed to be infinite 
dimensional). We will assume that ${\cal M}=T^\ast {\cal C}$
is a cotangent bundle. Let us then choose a real polarization of $\cal M$, 
that is, a real Lagrangean submanifold $\cal C$ which will play the role of
our configuration space. Then a loose definition of a complexifier is as 
follows:
\begin{Definition} \label{def3.5.2}  
A complexifier is a positive definite function \footnote{For the rest
of this section $C$ will denote a complexifier function and not the 
Hamiltonian constraint.} $C$ on $\cal M$ with the 
dimension of an action, which is smooth
a.e. (with respect to the Liouville measure induced from $\{.,.\}$) and 
whose Hamiltonian vector field is everywhere non-vanishing on $\cal C$.
Moreover, for each point $q\in {\cal C}$ the function $p\mapsto 
C_q(p)=C(q,p)$ grows stronger than linearly with $||p||_q$
where $p$ is a local momentum coordinate and $||.||_q$ is a suitable norm 
on $T^\ast_q({\cal C})$.
\end{Definition}
In the course of our discussion we will motivate all of these 
requirements.

The reason for the name {\it complexifier} is that $C$ enables us to
generate a {\it complex polarization} of $\cal M$ from $\cal C$ as 
follows: If we denote by $q$ local coordinates of $\cal C$ (we do not 
display any discrete or continuous labels but we assume that local fields
have been properly smeared with test functions) then 
\be \label{3.5.2}
z(m):=\sum_{n=0}^\infty \frac{i^n}{n !} \{q,C\}_{(n)}(m)
\ee
define local complex coordinates of $\cal M$ provided we can invert 
$z,\bar{z}$ for $m:=(q,p)$ where $p$ are the fibre (momentum) 
coordinates of $\cal M$. This is granted at least locally by 
definition \ref{def3.5.2}. Here the multiple Poisson bracket
is inductively defined by $\{q,C\}_{(0)}=q,\;
\{q,C\}_{(n+1)}=\{\{q,C\}_{(n)},C\}$ and makes sense due to the required 
smoothness. What is interesting about 
(\ref{3.5.2}) is that it implies the following bracket structure
\be \label{3.5.3}
\{z,z\}=\{\bar{z},\bar{z}\}=0
\ee
while $\{z,\bar{z}\}$ is necessarily non-vanishing. The reason for this
is that (\ref{3.5.2}) may be written in the more compact form
\be \label{3.5.4}
z=e^{-i{\cal L}_{\chi_C}} q=([\varphi_{\chi_C}^t]^\ast q)_{t=-i}
\ee
where $\chi_C$ denotes the Hamiltonian vector field of $C$,
$\cal L$ denotes the Lie 
derivative and $\varphi^t_{\chi_C}$ is the one -- parameter family of 
canonical transformations generated by $\chi_C$. 
Formula (\ref{3.5.4}) displays the transformation (\ref{3.5.2})
as the analytic extension to imaginary values of the one parameter family
of diffeomorphisms generated by $\chi_C$ and since the flow generated by 
Hamiltonian vector fields leaves Poisson brackets invariant, (\ref{3.5.3})
follows from the definition of a Lagrangean submanifold. The fact that
we have continued to the negative imaginary axis rather than the 
positive one is important in what follows and has to do with the required
positivity of $C$.

The importance of 
this observation is that either of $z,\bar{z}$ are coordinates of a 
Lagrangean submanifold of the complexification ${\cal M}^\Cl$, i.e. a 
complex polarization and thus 
may serve to define a Bargmann-Segal representation of the quantum theory
(wave functions are holomorphic functions of $z$). The diffeomorphism 
${\cal M}\to {\cal C}^\Cl;\;m\mapsto z(m)$ shows that we may think of 
$\cal M$ either as a symplectic manifold or as a complex manifold 
(complexification of the configuration space). Indeed, the 
polarization is usually a positive K\"ahler polarization with respect
to the natural $\{.,.\}$-compatible complex structure on a cotangent 
bundle 
defined by local Darboux coordinates, if we choose the complexifier to be 
a function of $p$ only.  These facts make the associated Segal-Bargmann 
representation especially attractive.\\
\\
We now apply the rules of canonical quantization: a suitable 
Poisson algebra $\cal O$ of functions $O$ on $\cal M$
is promoted to an algebra $\hat{{\cal O}}$ of operators $\hat{O}$ on a 
Hilbert 
space $\cal H$ subject to the condition that Poisson brackets turn
into commutators divided by $i\hbar$ and that reality conditions 
are reflected as adjointness relations, that is,
\be \label{3.5.5}
[\hat{O},\hat{O}']=i\hbar \widehat{\{O,O'\}}+o(\hbar),\;\;
\hat{O}^\dagger=\hat{\bar{O}}+o(\hbar)
\ee
where quantum corrections are allowed (and in principle unavoidable 
except if we restrict $\cal O$, say to functions linear in momenta).
We will assume that the Hilbert space can be represented as a space of 
square integrable 
functions on (a distributional extension $\overline{{\cal C}}$ of) 
$\cal C$ with respect to a positive, faithful probability measure
$\mu$, that is, ${\cal H}=L_2(\overline{{\cal C}},d\mu)$ as it is 
motivated by the real polarization.   

The fact that $C$ is positive motivates to quantize it in such a way 
that it becomes a self-adjoint, positive definite operator. We will assume 
this to 
be the case in what follows.
Applying then the quantization rules to the functions $z$ in 
(\ref{3.5.2}) 
we arrive at
\be \label{3.5.6}
\hat{z}=\sum_{n=0}^\infty \frac{i^n}{n^!} 
\frac{[\hat{q},\hat{C}]_{(n)}}{(i\hbar)^n}=e^{-\hat{C}/\hbar}\hat{q}
e^{\hat{C}/\hbar}
\ee
The appearence of $1/\hbar$ in (\ref{3.5.6}) justifies the requirement
for $C/\hbar$ to be dimensionless in definition \ref{def3.5.2}.
We will call $\hat{z}$ {\it annihilation operator} for reasons that
will become obvious in a moment. 

Let now $q\mapsto \delta_{q'}(q)$ be the $\delta$-distribution with 
respect to $\mu$ with support at $q=q'$. (More in mathematical terms,
consider the complex probability measure, denoted as  
$\delta_{q'}d\mu$, which is defined by $\int \delta_{q'} d\mu f=f(q')$ 
for measurable $f$).
Notice that since $C$ is non-negative and necessarily depends 
non-trivially on momenta (which will turn into (functional) derivative 
operators in the quantum theory), the operator $e^{-\hat{C}/\hbar}$
is a {\it smoothening operator}. Therefore, although $\delta_{q'}$
is certainly not square integrable, the complex measure (which
is probability if $\hat{C}\cdot 1=0$) 
\be \label{3.5.7}
\psi_{q'}:=e^{-\hat{C}/\hbar}\delta_{q'}
\ee
has a chance to be an element of $\cal H$. Whether or not it does depends 
on the details of ${\cal M},\{.,.\},C$. For instance, if $C$ as a function
of $p$ at fixed $q$ has flat directions, then the smoothening effect
of $e^{-\hat{C}/\hbar}$ may be insufficient, so in order to avoid this
we required that $C$ is positive definite and not merely non-negative.
If $C$ would be indefinite, then (\ref{3.5.7}) has no chance to make sense 
as an $L_2$ function. 

We will see in a moment that (\ref{3.5.7}) qualifies as a candidate {\it
coherent state} if we are able to analytically extend (\ref{3.5.7}) to 
complex 
values $z$ of $q'$ where the label $z$ in $\psi_z$ will play the role
of the point in $\cal M$ at which the coherent state is peaked. In 
order that this is possible (and in order that the extended function is 
still square integrable), (\ref{3.5.7}) should be entire analytic.
Now $\delta_{q'}(q)$ roughly has an integral kernel of the form 
$e^{i(k,(q-q'))}$ (with some pairing $<.,.>$ between tangential and 
cotangential vectors) which is analytic in $q'$ but the integral over $k$,
after applying $e^{-\hat{C}/\hbar}$,
will produce an entire analytic function only if there is a damping factor 
which decreases faster than exponentially. This provides the intuitive 
explanation for the growth requirement in definition \ref{def3.5.2}.
Notice that the $\psi_z$ are not necessarily normalized.

Let us then assume that 
\be \label{3.5.8}
q\mapsto \psi_m(q):=[\psi_{q'}(q)]_{q'\to z(m)}=
[e^{-\hat{C}/\hbar}\delta_{q'}(q)]_{q'\to z(m)}
\ee
is an entire $L_2$ function. Then $\psi_m$ is automatically an 
{\it eigenfunction of the annihilation operator $\hat{z}$ with eigenvalue
$z$} since 
\be \label{3.5.9}
\hat{z}\psi_m=[e^{-\hat{C}/\hbar}\hat{q}\delta_{q'}]_{q'\to z(m)} 
=[q' e^{-\hat{C}/\hbar}\delta_{q'}]_{q'\to z(m)}=z(m)\psi_m
\ee
where in the second step we used that the delta distribution is a 
generalized eigenfunction of the operator $\hat{q}$. But to be an 
eigenfunction of an annihilation operator {\it is one of the accepted 
definitions of coherent states} ! 

Next, let us verify that $\psi_m$ indeed has a chance to be peaked at
$m$. To see this, let us consider the self-adjoint (modulo domain 
questions) combinations 
\be \label{3.5.10}
\hat{x}:=\frac{\hat{z}+\hat{z}^\dagger}{2},\;\; 
\hat{y}:=\frac{\hat{z}-\hat{z}^\dagger}{2i}
\ee
whose classical analogs provide real coordinates for $\cal M$.
Then we have automatically from (\ref{3.5.9})
\be \label{3.5.11}
<\hat{x}>_m:=\frac{<\psi_m,\hat{x}\psi_m>}{||\psi_m||^2}=
\frac{z(m)+\bar{z}(m)}{2}=:x(m)
\ee 
and similar for $y$. Equation (\ref{3.5.11}) tells us that the operator
$\hat{z}$ should really correspond to the function $m\mapsto z(m),\;m\in 
{\cal M}$.

Now we compute by similar methods that
\be \label{3.5.12}
<[\delta\hat{x}]^2>_m
:=\frac{<\psi_m,[\hat{x}-<\hat{x}>_m]^2\psi_m>}{||\psi_m||^2}
=<[\delta\hat{y}]^2>_m
=\frac{1}{2}|<[\hat{x},\hat{y}]>_m|
\ee
so that the $\psi_m$ are automatically {\it minimal uncertainty states for
$\hat{x},\hat{y}$}, moreover the fluctuations are unquenched (equal each 
other). This is 
the second motivation for calling the $\psi_m$ coherent states. Certainly
one should not only check that the fluctuations are minimal but also that 
they are small as compared to the expectation value, at least at generic 
points of the phase space, in order that the quantum errors are small. 

The {\it infinitesimal Ehrenfest} property
\be \label{3.5.13}
\frac{<[\hat{x},\hat{y}]>_z}{i\hbar}=
\{x,y\}(m)+O(\hbar)
\ee
follows if we have properly implemented the canonical commutation 
relations and 
adjointness relations. The size of the correction, however, does not 
follow from these general considerations but the minimal uncertainty
property makes small corrections plausible. Condition (\ref{3.5.13}) 
supplies information about how well the symplectic structure is reproduced
in the quantum theory.  

For the same reason one 
expects that the peakedness property
\be \label{3.5.14}
|\frac{<\psi_m,\psi_{m'}>|^2}{||\psi_m||^2\;||\psi_{m'}||^2}
\approx \chi_{K_m}(m')
\ee
holds, where $K_m$ is a phase cell with center $m$ and Liouville volume  
$\approx\sqrt{<[\delta \hat{x}]^2>_m <[\delta \hat{y}]^2>_m}$ 
and $\chi$ denotes the characteristic function of a set.

Finally one wants coherent states to be overcomplete in order that every
state in $\cal H$ can be expanded in terms of them. This has to be checked
on a case by case analysis but the fact that our complexifier coherent 
states are for real $z$ nothing else than regularized $\delta$ 
distributions which in turn provide a (generalized) basis makes this 
property plausible to hold.
\begin{Exercise} \label{ex3.5.1} ~~~~\\
Consider the phase space: ${\cal M}=T^\ast \Rl=\Rl^2$ with standard 
Poisson brackets $\{q,q\}=\{p,p\}=0,\;\{p,q\}=1$ and
configuration space ${\cal C}=\Rl$. Consider the 
complexifier $C=p^2/(2\sigma)$ where $\sigma$ is a dimensionful
constant such that $C/\hbar$ is dimensionfree. Check 
that it meets all the requirements 
of definition \ref{def3.5.2} and perform the coherent state construction
displayed above. \\
Hint:\\
Up to a phase, the resulting, normalized coherent states are the usual 
ones for the 
harmonic oscillator with Hamiltonian $H=(p^2/m+m\omega^2 q^2)/2$ with
$\sigma=m\omega$. Verify that the states $\psi_m$ are Gaussian peaked in 
the configuration representation with width $\sqrt{\hbar/\sigma}$ 
around $q=q_0$ and in the momentum representation around $p=p_0$ with width 
$\sqrt{\hbar\sigma}$ where $m=(p_0,q_0)$.  
\end{Exercise}
As it has become clear from the discussion, the complexifier method 
gives a rough guideline, but no algorithm, in order to arrive at a 
satisfactory family of coherent states, there are things to be checked
on a case by case basis. On the other hand, what is nice is that given 
{\it only one input}, namely the complexifier $C$, it is possible to arrive 
at a definite and constructive framework for a semiclassical analysis.
It is important to know what the classical limit of $\hat{C}$ is, 
otherwise, if we have just an abstract operator $\hat{C}$ then 
the map $m\mapsto z(m)$ is unknown and an interprtation of the states
in terms of $\cal M$ is lost.

\subsection{Application to QGR}
\label{s3.5.2}

Let us now apply these ideas to QGR. Usually the choice of $C$ is stronly
motivated by a Hamiltonian, but in QGR we have none. Therefore, at the 
moment the best we can do is to play with various proposals for 
$\hat{C}$ and to explore the properties of the resulting states.
For the simplest choice of $\hat{C}$ \cite{64} those properties have 
been worked out more or less completely and we will briefly describe 
them below. 

The operator $\hat{C}$ is defined by its action on cylindrical functions 
$f=p_\gamma^\ast f_\gamma$ by 
\be \label{3.5.15}
\frac{\hat{C}}{\hbar} f=-
p_\gamma^\ast[\frac{1}{2}[\sum_{e\in E(\gamma)} l(e) 
[R^j_e/2]^2]\;f_\gamma]
=:p_\gamma^\ast [\hat{C}_\gamma f_\gamma]
\ee
where the positive numbers $l(e)$ satisfy $l(e\circ e')=l(e)+l(e')$
and $l(e^{-1})=l(e)$ and $R^j_e$ are the usual right invariant vector 
fields. 
\begin{Exercise} \label{ex3.5.2} ~~~~\\
Recall the definition of the maps $p_{\gamma'\gamma}$ for 
$\gamma\prec \gamma'$ from section \ref{s2.1.2} and check that the 
$\hat{C}_\gamma$ are consistently defined, that is,
$\hat{C}_{\gamma'}\circ p_{\gamma'\gamma}^\ast= 
p_{\gamma'\gamma}^\ast\circ \hat{C}_\gamma$.
\end{Exercise}
This choice is in analogy to the harmonic oscillator where the 
quantum complexifier is essentially the Laplacian $-(d/dx)^2$.
The classical limit of (\ref{3.5.15}) depends in detail on the 
function $l$ which is analogous to the parameter $\hbar/\sigma$
for the case of the harmonic oscillator. For 
instance \cite{63}, one can 
choose a) three families of foliations $s\mapsto H_s^I,\;I=1,2,3$
of $\sigma$ by two dimensional surfaces $H^I_s$
such that there is a bijection 
$(s^1,s^2,s^3)\mapsto x(\vec{s}):=[\cap_I H^I_{s^I}]\in\sigma$
and b) a partition $P^I_s$ of the $H^I_s$ into small surfaces 
$S$ and define 
\be \label{3.5.16}
C=\frac{1}{2 a^2 \kappa}\int_\Rl ds \sum_{I=1}^3 \sum_{S\in P^I_s} 
[\mbox{Ar}(S)]^2
\ee
where $\mbox{Ar}(S)$ is again the area functional and $a$ is a 
dimensionful constant of dimension cm$^1$. The function $l$ for this 
example is then roughly\footnote{This formula gets exact in the limit
of infinitely fine partition, at finite coarseness, it is an approximation 
to the exact, more complicated formula.}  
$l(e)=\frac{(\beta\ell_P)^2}{a^2}\int ds \sum_I\sum_{S\in P^I_s} \chi_S(e)$
where $\chi_S(e)=1$ if $S\cap e\not=\emptyset$ and vanishes otherwise.

The $\delta-$distribution with respect to the measure $\mu_0$ can be 
written as the sum over all SNW's (exercise !)
\be \label{3.5.17}
\delta_{A'}(A)=\sum_s T_s(A') \overline{T_s(A)}
\ee
with resulting coherent states
\be \label{3.5.18}
\psi_{A^\Cl}(A)=
\sum_s e^{-\frac{1}{2}\sum_{e\in E(\gamma(s))} l(e) j_e(j_e+1)} 
T_s(A^\Cl) \overline{T_s(A)}
\ee
where the {\it $SL(2,\Cl)$ connection} $A^\Cl$ is defined by 
\be \label{3.5.19}
A^\Cl[A,E]=\sum_{n=0}^\infty \frac{i^n}{n!} (\{A,C\}_{(n)})[A,E]
\ee
Thus we see that in this case the symplectic manifold
given as the cotangential bundle ${\cal M}=T^\ast \a$ over the space 
of $SU(2)$ connections is also naturally given as the complex manifold
$\a^\Cl$ of $SL(2,\Cl)$ connections. From the general discussion above 
it now follows that the classical interpretation of the {\it annihilation 
operators} 
\be \label{3.5.20}
\hat{A}^\Cl(e):=e^{-\hat{C}/\hbar} \hat{A}(e) e^{\hat{C}/\hbar}
\ee  
is simply the holonomy of the complex connection $A^\Cl(e)$.

In order to study the semiclassical properties of these states 
we consider their {\it cut-offs} $\psi_{A^\Cl,\gamma}$ for each graph
$\gamma$ defined on cylindrical functions $f=p_\gamma^\ast f_\gamma$
by 
\be \label{3.5.21}
<\overline{\psi_{A^\Cl}},f>_{kin}=:
<\overline{\psi_{A^\Cl,\gamma}},f>_{kin}
\ee
Now, (if we work at the non-gauge invariant level,) one can check 
that
\be \label{3.5.22}
\psi_{A^\Cl,\gamma}(A)=\prod_{e\in E(\gamma(s)} 
\psi^{l(e)}_{A^\Cl(e)}(A(e))
\ee
where for any $g\in SL(2,\Cl),\;h\in SU(2)$ we have defined  
\be \label{3.5.23}
\psi^t_g(h):=\sum_j (2j+1) e^{-tj(j+1)/2} \chi_j(g h^{-1})
\ee
\begin{Exercise} \label{ex3.5.3} ~~~~~\\
Verify, using the Peter\&Weyl theorem, that for $g\in SU(2)$ we have 
$\psi^0_g(h)=\delta_g(h)$, the 
$\delta-$distribution with respect to $L_2(SU(2),d\mu_H)$. Conclude that 
(\ref{3.5.23})
is just the analytic extension of the heat kernel $e^{-t\Delta/2}$ 
where $\Delta$ is the Laplacian on $SU(2)$.
Thus the states (\ref{3.5.23}) are in complete analogy with those  
for the harmonic oscillator, just that $\Rl$ was replaced by $SU(2)$
and the complexification $\Cl$ of $\Rl$ by the complexification 
$SL(2,\Cl)$ of $SU(2)$. In this form, coherent states on compact 
gauge groups were originally proposed by Hall \cite{65}.
\end{Exercise}
The analysis of the semiclassical properties of the states 
$\psi_{A^\Cl,\gamma}$
on ${\cal H}_{kin}$ can therefore be reduced to that of the states 
$\psi^t_g$ on $L_2(SU(2),d\mu_H)$. We state here without proof that 
the following properties could be proved \cite{64}: I) Overcompleteness, 
II) expectation value property, III) Ehrenfest property, 
IV) peakedness in phase space, V) annihilation operator eigenstate 
property, VI) minimal uncertainty property and VII) small fluctuation
property. Thus, these states have many of the desired properties
that one requires from coherent states. 

\begin{figure}
\includegraphics[width=10cm,height=7cm]{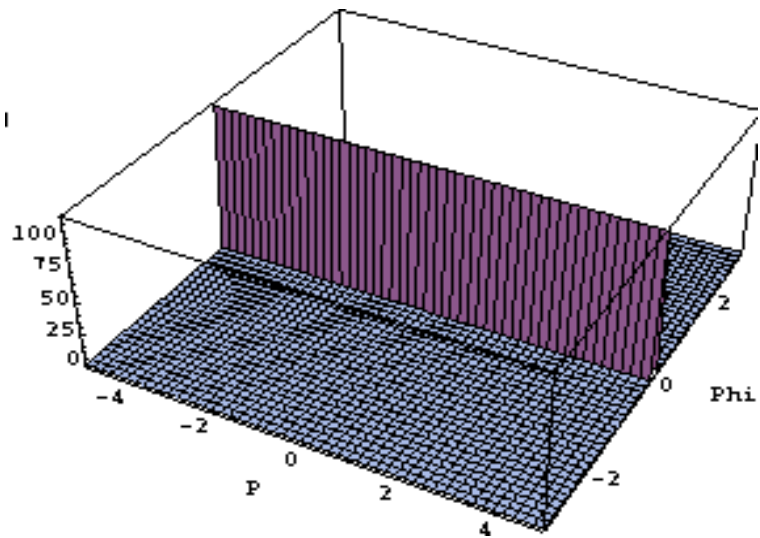}
\caption{Probability amplitude $u\mapsto j^t_{g_0}(u)$ at 
$p\in[-5,5], h_0=1$.}
\label{f15}
\end{figure}

\begin{figure}
\includegraphics[width=10cm,height=7cm]{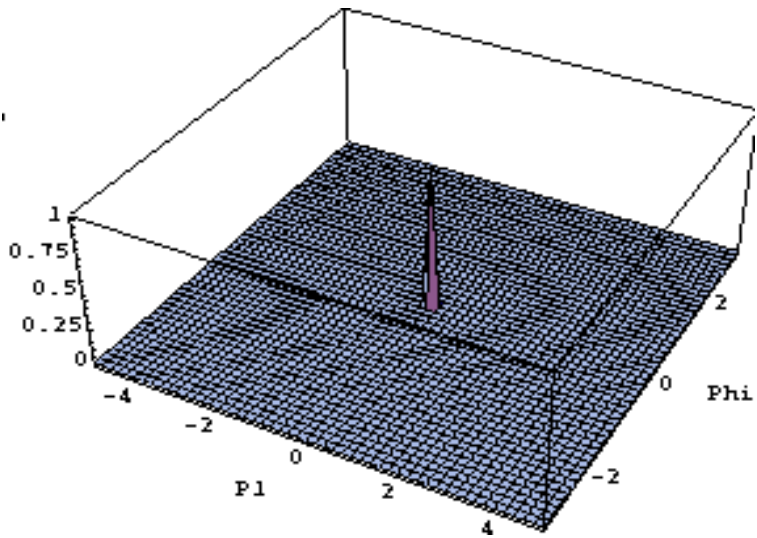}
\caption{Overlap function $g\mapsto i^t_{g_0}(g)$ at $p=0,h_0=1$
for $p_1\in [-5,5],u\in U(1)$.}
\label{f16}
\end{figure}

\begin{figure}
\includegraphics[width=10cm,height=7cm]{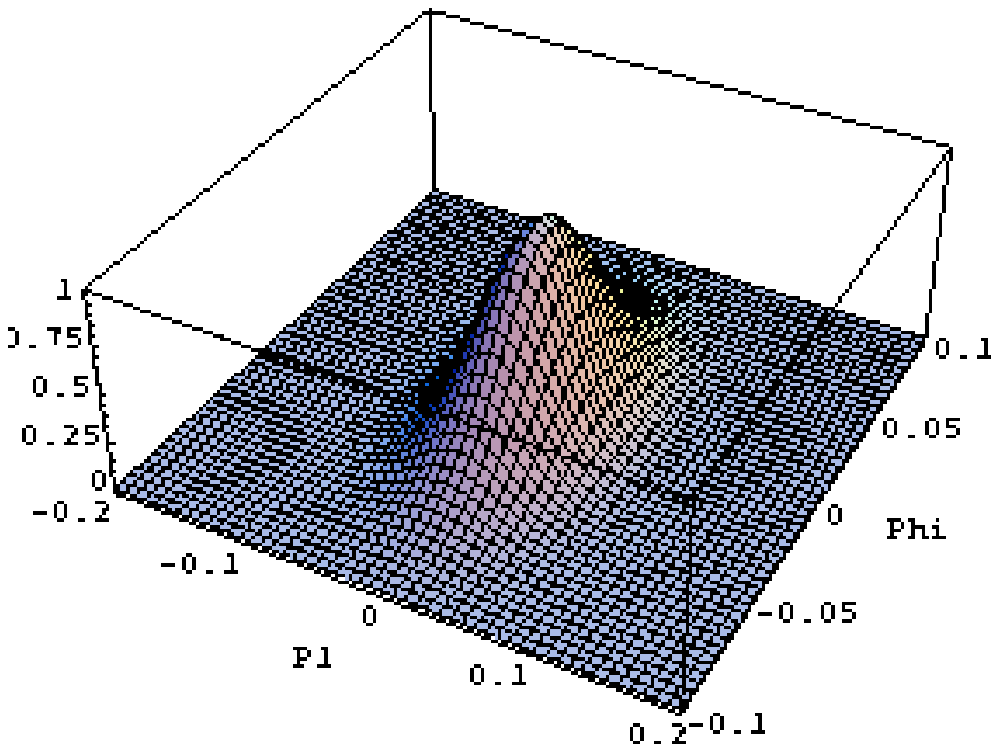}
\caption{Resolution of a neighbourhood of the peak of the 
function $g\mapsto i^t_{g_0}(g)$ at $p=0,h_0=1$.}
\label{f17}
\end{figure}

In the following graphic
we display as an example the peakedness properties of 
the analog of (\ref{3.5.23}) for the simpler case of the gauge 
group $U(1)$, the case of $SU(2)$ is similar but requires more plots 
because of the higher dimensionality of $SU(2)$. 
Thus $g_0=e^p h_0\in U(1)^\Cl=\Cl-\{0\},\;p\in\Rl,\;h_0\in U(1)$ and 
$u\in U(1)$ where we parameterize $u=e^{i\phi},\;\phi\in [-\pi,\pi)$.
Similarly, $g=e^{p_1} u,\;p_1\in\Rl$.
We consider in figure \ref{f15} the peakedness in the configuration 
representation given by the probability amplitude 
\be \label{3.5.24}
u=e^{i\phi}\mapsto j^t_{g_0}(u)=|\psi^t_{g_0}(u)|^2/||\psi^t_{g_0}||^2
\ee 
at $h_0=1, p\in [-5,5]$. 
In figure \ref{f16} the phase space peakedness expressed by the 
overlap function 
\be \label{3.5.25}
g=e^{p_1} u\mapsto i^t(g,g_0)
=\frac{|<\psi^t_g,\psi^t_{g_0}>|^2}{||\psi^t_g||^2\; ||\psi^t_{g_0}||^2}  
\ee
is shown at fixed $p=0,h_0=1$ for $p\in [-5,5],\;u\in U(1)$. We have 
made use of the fact (exercise !)
that $\psi_{g_0}(u)$ and $<\psi_g,\psi_{g_0}>$ respectively depend 
only on the combinations $g_0 u^{-1}=
e^{p_0} hu^{-1}$ and $\bar{g}g_0 =e^{p+p_1} u^{-1} h_0$ respectively.
Therefore, peakedness at $u=h_0$ or $g=g_0=e^p h_0$ respectively for any 
$h_0$ is 
equivalent to peakedness at $u=1$ or at $g=e^{p_0}$ respectively for 
$h_0=1$. Both plots are for the 
value $t=0.001$ and one clearly sees the peak width of 
$\sqrt{t}\approx 0.03$ when resolving those plots around the 
peak as in figure \ref{f17}, which has a close to Gaussian shape just like 
the harmonic oscillator coherent states have.
As a first modest application, these states have been used in order to 
analyze how one would obtain, at least in principle, the QFT's on CST's 
(Curved SpaceTime)
limit from full QGR in \cite{42}. In particular,
it was possible to perform a detailed calculation concerning the 
existence of Poincar\'e invariance violating dispersion relations of
photon propagation within QGR which were discussed earlier at a more
phenomenological level in the pioneering papers \cite{66}: The idea is that 
the metric field is a collection of  
quantum operators which are not mutually commuting. Therefore
it should be impossible to construct a state which is peaked on, say the 
Minkowski metric, and which is a simultaneous eigenstate of all 
the metric operator components, in other words, there should be no such
thing as a Poincar/'e invariant state in full QGR\footnote{This seems 
to contradict the fact that we are even interested in four dimensionally
diffeomorphism invariant states and the fact that the Poincar\'e group 
should be a tiny subgroup thereof. However, this is not the case because 
we require the states only to be invariant under diffeomorphisms which 
are pure gauge and those have to die off at spatial infinity. Poincar\'e
transformations are therefore not gauge transformation but symmetries and 
what we are saying is that there are no Poincar\'e symmetric, 
diffeomorphism gauge invariant states.}, already because such an object 
should be highly background dependent. The best one can construct is
a coherent state peaked on the Minkowski metric. The small fluctuations 
that are encoded in that state influence the propagation of matter and 
these tiny disturbances could accumulate to measurable sizes in so-called
$\gamma-$ray burst experiments \cite{67} where one measures the time delay
of photons of higher energy as compared to those of lower energy as they
travel over cosmological distances as a result of the energy dependence of 
the speed of light. If such an effect exists then it is a non-perturbative 
one because perturbatively defined QFT's on Minkowski space are by 
construction Poincar\'e invariant (recall e.g. the Wightman axioms from 
section \ref{s1.1}).\\ 
\\
These are certainly only first moderate steps. The development of the 
semiclassical analysis for QGR is still in its very beginning 
and there are many interesting and new mathematical and physical issues 
that have to be 
settled before one can seriously attack the proof that, for instance, 
the Hamiltonian constraint of section \ref{s3.1} has the correct 
classcical limit or that full QGR reduces to classical GR plus the 
standard model in the low energy regime.

\section{Gravitons}
\label{s3.6}

\subsection{The Isomorphism}
\label{s3.6.1}

The reader with a strong background in ordinary QFT and/or string theory
will have wondered throughout these lectures where in QGR the graviton, 
which plays such a prominent role in the perturbative, background dependent
approaches to quantum gravity, resides. In fact, if one understands the 
graviton, as usually, as an excitation of the quantum metric around   
Minkowski space, then there is a clear connection with the semiclassical
analysis of the previous section: One should construct a suitable 
coherent state which is peaked on the gauge invariant 
phase space point characterizing Minkowski space and identify suitable
excitations thereof as gravitons. It is clear that at the moment 
such graviton states from full QGR cannot be constructed, because we would 
need first to solve the Hamiltonian constraint. 

However, one can arrive at
an approximate notion of gravitons through the quantization of {\it 
linearized gravity}: Linearized gravity is nothing else than the expansion 
of the full GR action around the gauge variant initial data 
$(E^0)^a_j=\delta^a_j,\;(A^0)_a^j=0$ to second order in $E-E^0,A$
which results in a free, classical field theory with constraints. In fact, 
the usual notion 
of gravitons is {\it precisely the ordinary Fock space quantization}
of that classical, free field theory \cite{68}. In order to see whether
QGR can possibly accomodate these graviton states, Varadarajan in a 
beautiful series of papers \cite{69} has carried out a polymer like 
quantization of that free field theory on a Hilbert space 
${\cal H}_{kin}$ which is in complete analogy to that for full QGR,
the only difference being that the gauge group $SU(2)$ is 
replaced by the gauge group $U(1)^3$. While there are certainly large 
differences between the highly interacting QGR theory and linearized 
gravity, one should at least be able to gain some insight into the 
the answer to the question, how a Hilbert space in which the 
excitations are one dimensional can possibly describe the Fock 
space excitations (which are three dimensional).

The problem of describing gravitons within linearized gravity by 
polymer like excitations is mathematically equivalent to the simpler
problem of describing the photons of the ordinary Fock Hilbert space 
${\cal H}_F$ 
of Maxwell theory by polymer like excitations within a Hilbert space
${\cal H}_P=L_2(\ab,d\mu)$ where $\ab$ is again a space of generalized 
$U(1)$ connections with some measure $\mu$ thereon. Thus, we decribe 
the latter problem in some detail since it requires less space and has 
the same educational value.\\
\\
The crucial observation is the following isomorphism $\cal I$ between two 
different Poisson subalgebras of the Poisson algebra on the phase
space $\cal M$ of Maxwell theory coordinatized by a canonical pair 
$(E,A)$ defined by a $U(1)$ connection $A$ and a conjugate electric field 
$E$:
Consider a one-parameter family of
test functions of rapid decrease which are regularizations of
the $\delta-$distribution, for instance
\be \label{3.6.1}
f_r(x,y)=\frac{e^{-\frac{||x-y||^2}{2r^2}}}{(\sqrt{2\pi}r)^3}
\ee
where we have made use of the Euclidean spatial background
metric.
Given a path $p\in{\cal P}$ we denote its distributional {\it form factor}
by
\be \label{3.6.2}
X^a_p(x):=\int_0^1 dt \;\;\dot{p}^a(t) \delta(x,p(t))
\ee
The smeared form factor is defined by
\be \label{3.6.3}
X^a_{p,r}(x):=\int d^3y f_r(x,y) X^a_p(y)=\int_0^1 dt \;\;\dot{p}^a(t)
f_r(x,p(t))
\ee
which is evidently a test function of rapid decrease.
Notice that a $U(1)$ holonomy maybe written as
\be \label{3.6.4}
A(p):=e^{i\int d^3x X^a_p(x) A_a(x)}
\ee
and we can define a smeared holonomy by
\be \label{3.6.5}
A_r(p):=e^{i\int d^3x X^a_{p,r}(x) A_a(x)}
\ee
Likewise we may define smeared electric fields as
\be \label{3.6.6}
E^a_r(x):=\int d^3y f_r(x,y) E^a(y)
\ee
If we denote by $q$ the electric charge (notice that in
our notation $\alpha=\hbar q^2$ is the fine structure
constant), then we obtain the
following Poisson subalgebras:
On the one hand we have
smeared holonomies but unsmeared electric fields with
\be \label{3.6.7}
\{A_r(p),A_r(p')\}=\{E^a(x),E^b(y)\}=0,\;\;
\{E^a(x),A_r(p)\}=iq^2 X^a_{p,r}(x) A_r(p)
\ee
and on the other hand we have
unsmeared holonomies but smeared electric fields with
\be \label{3.6.8}
\{A(p),A(p')\}=\{E^a_r(x),E^b_r(y\}=0,\;\;
\{E^a_r(x),A(p)\}=iq^2 X^a_{p,r}(x) h_p
\ee
Thus the two Poisson algebras are isomorphic and also
the $^\ast$ relations are isomorphic, both
$E^a(x),E^a_r(x)$ are real valued while
both $A(p),A_r(p)$ are $U(1)$ valued. Thus, as abstract
$^\ast-$ Poisson algebras these two algebras are indistinguishable
and we may ask if we can find different representations
of it. Even better, notice that
$A_r(p) A_r(p')=A_r(p\circ p'),\;A_r(p)^{-1}=A_r(p^{-1})$
so the smeared holonomy algebra is also isomorphic to the
unsmeared one. Hence  there is an algebra 
$^\ast-$isomorphism ${\cal I}$ defined on the generators by
${\cal I}_r(h_p)=h_{p,r},\;\;{\cal I}_r(E_r)=E$. One must also 
show that the $A_r(p)$ are still algebarically independent as 
are the $A(p)$ \cite{69}.

\subsection{Induced Fock Representation With Polymer -- Excitations}
\label{s3.6.2}

Now we know that the unsmeared holonmy algebra is well represented
on the Hilbert space ${\cal H}_{kin}=L_2(\ab,d\mu_0)$ while the
smeared holonomy algebra is well represented on the
Fock Hilbert space ${\cal H}_F=L_2({\cal S}',d\mu_F)$ where
${\cal S}'$ denotes the space of divergence free, tempered distributions
and $\mu_F$ is the Maxwell-Fock measure. These measures are
completely characterized by their generating functional
\be \label{3.6.9}
\omega_F(\hat{A}_r(p)):=\mu_F(A_r(p))=
e^{-\frac{1}{4\alpha}\int d^3x X^a_{p,r}(x)\sqrt{-\Delta}^{-1}
X^b_{p,r}\delta_{ab}}
\ee
since finite linear combinations of the $h_{p,r}$ are dense
in ${\cal H}_F$ \cite{69}. Here $\Delta=\delta^{ab}\partial_a\partial_b$
denotes the Laplacian and 
we have taken a loop $p$ rather than an open path so that
$X_{p,r}$ is transversal.
Also unsmeared electric fields are represented through the Fock
state $\omega_F$ by
\be \label{3.6.10}
\omega_F(\hat{A}_r(p)\hat{E}^a(x)\hat{A}_r(p'))
=-\frac{\alpha}{2}[X^a_{p,r}(x)-X^a_{p',r}(x)]
\omega_F(\hat{h}_{p\circ p',r})
\ee
and any other expectation value follows from these and the
commutation relations.

Since $\omega_F$ defines a positive linear functional we may
define a new representation of the algebra $A(p),E^a_r$
by
\be \label{3.6.11}
\omega_r(\hat{A}(p)):=\omega_F(\hat{A}_r(p))
\mbox{ and }
\omega_r(\hat{A}(p)\hat{E}^a_r(x)\hat{A}(p'))
:=\omega_F(\hat{A}_r(p)\hat{E}^a(x)\hat{A}_r(p'))
\ee
called the $r-$Fock representation. In other words, we have 
$\omega_r=\omega_F\circ {\cal I}_r$.

Since $\omega_r$ is a positive linear functional on $C(\ab)$
by construction
there exists is a measure $\mu_r$ on $\ab$ that represents
$\omega_r$ in the sense of the Riesz 
representation theorem (recall \ref{2.1.2.10}). In \cite{70} Velhinho
showed that the one -- parameter family of measures
$\mu_r$ are expectedly mutually singular with respect to each
other and with respect to the uniform measure $\mu_0$ (that is, the 
support of one measure is a measure zero set with respect to the other 
and vice versa).\\
\\
Result 1: There is a unitary transformation between any of the Hilbert 
spaces ${\cal H}_r$ and their images under ${\cal I}_r$ in the usual Fock 
space ${\cal H}_F$. Since finite linear combinations of the 
$A_r(p)$ for fixed $r$ are still dense in ${\cal H}_F$ \cite{69}, there 
exists indeed a 
polymer like description of the usual $n$-photon states.\\
\\ 
Recall that the Fock vacuum $\Omega_F$ is defined to be the zero eigenvalue
coherent state, that is, it is annihilated by the annihilation operators
\be \label{3.6.12}
\hat{a}(f):=\frac{1}{\sqrt{2\alpha}}
\int d^3x f^a[\root 4\of{-\Delta} \hat{A}_a
-i(\root 4\of{-\Delta})^{-1}\hat{E}^a]
\ee
where $f^a$ is any transversal smearing field. We then have in
fact that $\omega_F(.)=<\Omega_F,.\Omega_F>_{{\cal H}_F}$. (For readers
familiar with $C^\ast-$algebras this means that 
$\Omega_F$ is the cyclic vector that is determined by $\omega_F$ through
the GNS construction.)
The idea is now the following: From (\ref{3.6.11}) we see that we can
easily answer any question in the $r-$Fock representation which has a
preimage in the Fock representation, we just have to replace
everywhere $A_r(p),E^a(x)$ by $A(p),E^a_r(x)$.
Since in the $r-$Fock representations only exponentials of connections
are defined, we should exponentiate the annihilation operators
and select the Fock vacuum through the condition
\be \label{3.6.13}
e^{i\hat{a}(f)}\Omega_F=\Omega_F
\ee
In particular, choosing $f=\sqrt{2\alpha}(\root 4\of{-\Delta})^{-1} X_{p,r}$
for some loop $p$ we get
\be \label{3.6.14}
e^{\int d^3x X^a_{p,r}[i\hat{A}_a+(\sqrt{-\Delta})^{-1}\hat{E}^a]}
\Omega_F
=\Omega_F
\ee
Using the commutation relations and the Baker -- Campell -- Hausdorff
formula one can write (\ref{3.6.14}) in terms of $\hat{A}_r(p)$
and the exponential of the electric field appearing in (\ref{3.6.14})
times a numerical factor. The resulting expression can then be
translated into the $r-$Fock representation. Denoting the translated 
expression by ${\cal I}_r^{-1}(e^{i\hat{a}(f)})$ we now ask 
the question, whether there exists a state $\Omega_r\in {\cal H}_{kin}
=L_2(\ab,d\mu_0)$ such that 
${\cal I}_r^{-1}(e^{i\hat{a}(f)})\Omega_r=\Omega_r$. Remarkably, expanding
$\Omega_r$ into the charge network basis introduced in section 
(\ref{s3.1}) one finds a (up to a multiplicative constant) 
unique solution given by
\be \label{3.6.15}
\Omega_r=\sum_c
e^{-\frac{\alpha}{2}\sum_{e,e'\in E(\gamma(c))} G^r_{e,e'} n_e(c) 
n_{e'}(c)} \overline{T_c}
\ee
where $c=(\gamma(c),\{n_e(c)\}_{e\in E(\gamma(c))})$ denotes
a charge network (the $U(1)$ analogue of a spin network) and
\be \label{3.6.16}
G^r_{e,e'}=\int d^3x X^a_{e,r}\sqrt{-\Delta}^{-1} X^b_{e',r}\delta^T_{ab}
\ee
where $\delta_{ab}^T=\delta_{ab}-\partial_a\Delta^{-1}\partial_b$
denotes the transverse projector.
\begin{Exercise} \label{ex3.6.1} ~~~~\\
Fill in the gaps that lead from (\ref{3.6.14}) to (\ref{3.6.16}).
\end{Exercise}
Let us discuss this result. First of all, (\ref{3.6.15}) is not 
normalizable with respect to the inner product on ${\cal H}_{kin}$
and neither are the images of $n-$photon states or coherent states
from ${\cal H}_F$. This seems to indicate that the space ${\cal H}_{kin}$
does not play any role for physically interesting states. However, in
\cite{63} it was shown that this is not the case: It turns out,
that, given a suitable regularization, that one can indeed obtain the 
expectation values such 
as $\omega_r(A(p))$ from the formal expression
\be \label{3.6.18}
\omega_r(A(p)):=\frac{<\Omega_r,A(p)\Omega_r>}{||\Omega_r||^2}
\ee
where both numerator and denominator are infinite but the fraction is 
finite.\\
\\
Result 2: The polymer images of photon states can be obtained as 
certain limits of states from ${\cal H}_{kin}$ which therefore is a valid 
starting point in order to obtain physically interesting representations.\\
\\
Moreover, as can be expected from the similarity between the formulas 
(\ref{3.6.16}) and (\ref{3.5.18}) (for $A^\Cl=0$ corresponding 
to vacuum $E=A=0$ in the present case), the states $\Omega_r$
also arise from a complexifier, given in this case by
\be \label{3.6.19}
C=\frac{1}{2q^2} \int_{\Rl^3} d^3x 
[E^a_r \sqrt{-\Delta}^{-1} E^b_r]\delta_{ab}
\ee
\\
Result 3: The complexifier framework is also able to derive images of 
$n-$photon states and usual Fock coherent states from the universal 
input of a complexifier.\\
\\
We conclude that at least for the linearized theory the question
posed at the beginning of this section could be answered affirmatively:
There is indeed a precise framework available for how to accomodate
graviton states into the framework of loop quantum gravity. 
This is a promising result and should have an analog in the full theory.

\part{Selection of Open Research Problems}
\label{s4}

Let us summarize the most important open research problems that have 
come up during the discussion in these lectures.
\begin{itemize}
\item[i)] {\it Hamiltonian Constraint and Semiclassical States}\\
The unsettled correctness of the quantum dynamics is the major roadblock
to completing the quantization programme of QGR. In order to make 
progress a better understanding of the kinematical semiclassical 
sector of the theory is necessary.
\item[ii)] {\it Physical Inner Product}\\
Even if we had the correct Hamiltonian constraint and the complete space 
of solutions, at the moment there is no really good idea available of 
how to construct a corresponding physical inner product  
because the constraint algebra is not a Lie algebra but
an open algebra in the BRST sense so that techniques from rigged 
Hilbert spaces are not available. A framework for such open algebras 
must be developed so that an inner product can be constructed at least
in principle.
\item[iii)] {\it Dirac Observables}\\
Not even in classical general relativity do we know enough Dirac observables.
For QGR they are mandatory for instance in order to select an inner 
product by adjointness conditions and in order to arrive at an 
interpretation of the final theory. A framework of how to define 
Dirac observables, at least in principle, even at the classical level,
would be an extremely important contribution. 
\item[iv)] {\it Covariant Formulation}\\
The connection between the Hamiltonian and the Spin Foam formulation is 
poorly understood. Without such a connection e.g. a proof of covariance of 
the canonical formulation on the one hand and a proof for the correct
classical limit of the spin foam formulation on the other cannot be 
obtained using the respective other formulation.
One should prove a rigorous Feynman -- Kac like formula that allows to 
switch between these complementary descriptions.
\item[v)] {\it QFT on CST's and Hawking Effect from First Principles}\\
The low energy limit of the theory in connection with the the construction 
of semiclassical states must be better understood. Once this is done, 
fundamental issues such as whether the Hawking effect is merely an 
artefact of an invalid description by QFT's on CST's while a quantum 
theory of gravity should be used or whether it is a robust result can be 
answered. 
Similar remarks apply to the information paradoxon associated with
black holes etc.
\item[vi)] {\it Combinatorial Formulation of the Theory}\\
The description of a theory in terms of smooth and even analytic 
structures curves, surfaces etc. at all scales in which the spectra of 
geometrical operators are discrete at Planck scales is awkward and 
cannot be the most adequate language. There should be a purely
combinatorical formulation in which notions such as topology, differential 
structure etc. can only have a semiclassical meaning.
\item[vii)] {\it Avoidance of Classical and UV Singularities}\\
That certain classical singularities are absent in loop quantum 
cosmology and that certain operators come out finite in the full theory
while in the usual perturbative formulation they would suffer from 
UV singularities are promising results, but they must be better 
understood. If one could make contact with perturbative formulations 
and pin -- point exactly why in QGR the usual perturbative UV singularities 
are absent
then the theory would gain a lot more respect in other communities of 
high energy physicists. There must be some analog of the renormalization 
group and the running of coupling constants that one usually finds
in QFT's and CST's.  Similar remarks apply to the generalization of the 
loop quantum cosmology result to the full theory.
\item[viii)] {\it Contact with String (M) -- Theory}\\
If there is any valid perturbative description of quantum gravity 
then it is almost certainly string theory. It is conceivable that
both string theory and loop quantum gravity are complementary descriptions 
but by themselves incomplete and that only a fusion of both can reach
the status of a fundamental theory. To explore these possibilities,
Smolin has launched an ambitious programme \cite{71} which to our mind 
so far did not raise the interest that it deserves\footnote{That we did 
not devote a section to this topic in this review is due to the fact that
we would need to include an introduction to M -- Theory into these 
lectures which would require too much space. The interested reader 
is referred to the literature cited.}. The contact arises through
Chern -- Simons theory which is part of both Loop Quantum Gravity and 
M -- Theory \cite{71a} (when considered as the high energy limit of 
11 dimensional Supergravity). Another obvious starting point
is the definition of M -- Theory as the quantum supermembrane in 
11 dimensions \cite{72}, a theory that could be obtained as the 
quantization of the classical supermembrane by our non-perturbative 
methods. Finally, a maybe even more obvious connection could be found 
through the so-called 
{\bf Pohlmeyer String} \cite{73} which appears to be a method to quantize 
the string non-perturbatively, without supersymmetry, anomalies or 
extra dimensions, by working 
{\it directly at the level of Dirac observables} which are indeed 
possible to construct explicitly in this case.    
\end{itemize}
We hope to have convinced the reader that Loop Quantum Gravity 
is an active and lively approach to a quantum theory of gravity which 
has produced already many non-trivial results and will continue to do so 
in the future. There are still a huge number of hard but fascinating 
problems to be solved of which the above list is at most the tip of an 
iceberg. If at least a tiny fraction of the readers  
would decide to dive into this challenging area and help in this 
endeavour, then these lectures would have been successful.\\
\\
\\
\\
\\
{\large Acknowledgements}\\
\\
We thank the Heraeus -- Stiftung and the organizers, Dominico Giulini,
Claus Kiefer and Claus L\"ammerzahl, for making this wonderful and 
successful meeting possible and the participants for creating a 
stimulating atmosphere through long and deep discussions, very often 
until early in the morning in the ``B\"urgerkeller".

\end{document}